\documentclass[aps,pra,reprint,superscriptaddress]{revtex4-2}
\usepackage{graphicx}
\usepackage{amsmath,amsfonts}
\usepackage{xr}   
\DeclareMathAlphabet{\mathbbold}{U}{bbold}{m}{n}
\begin{document}


\title{Modeling scattering matrix containing evanescent modes for wavefront shaping applications in disordered media}


\author{Michael Raju}
\email[]{michaelraju@umail.ucc.ie}
\affiliation{Tyndall National Institute, Lee Maltings Complex, Dyke Parade, Cork, Ireland, T12 R5CP}
\affiliation{School of Physics, University College Cork, College Road, Cork, Ireland, T12 K8AF}
\author{Baptiste Jayet}
\affiliation{Tyndall National Institute, Lee Maltings Complex, Dyke Parade, Cork, Ireland, T12 R5CP}
\author{Stefan Andersson-Engels}
\affiliation{Tyndall National Institute, Lee Maltings Complex, Dyke Parade, Cork, Ireland, T12 R5CP}
\affiliation{School of Physics, University College Cork, College Road, Cork, Ireland, T12 K8AF}



\begin{abstract}
We developed an open-source scalar wave transport model to estimate the generalized scattering matrix ($S$ matrix) of a disordered medium in the diffusion regime. The term \textit{generalized} refers to the incorporation of evanescent wave field modes alongside propagating modes in the estimation of the $S$ matrix. To achieve this, we employed the scalar Kirchhoff–Helmholtz boundary integral formulation together with the Green's function perturbation method, thereby extending the conventional Fisher-Lee relations to include evanescent modes. The estimated $S$ matrix, which satisfies the generalized unitarity and reciprocity relations, is modeled for a $2D$ disordered waveguide. The generalized transmission matrix contained within the $S$ matrix is utilized to estimate the optimal phase-conjugate wavefront for focusing onto an evanescent mode. The phenomenon of a universal transmission value of 2/3 for such an optimal phase conjugate wavefront is demonstrated in the context of evanescent wave mode focusing through a diffusive disorder. The presented code framework may be of interest to wavefront shaping researchers for visualizing and estimating wave transport properties in general. 
\end{abstract}


\maketitle


\section{Introduction \label{sec:genintro}}
Coherent control \cite{rotter2017light} of waves in disordered media through wavefront shaping has been an actively researched topic across various domains of physics, including electromagnetics and acoustics. Experimental demonstrations of coherent control using light, such as focusing outside a disordered medium \cite{vellekoop2007focusing}, enhancing total light transmission through a disorder \cite{popoff2014coherent,hsu2017correlation}, and focusing light within a region enclosed by disorder \cite{katz2019controlling}, have been reported. Preferential deposition \cite{ojambati2016coupling,sarma2016control,bender2022depth} of energy within a disordered medium has also emerged as an application in the field of coherent light control. The availability of wave-shaping devices, such as spatial light modulators, has significantly advanced research on coherent control experiments using light, particularly through the use of optical \cite{popoff2010measuring} transmission matrix. For a comprehensive study of the transport aspects in these wavefront shaping applications, modeling the scattering matrix ($S$ matrix) is particularly advantageous.

To estimate the regular $S$ matrix containing only the propagating wave modes for wavefront shaping applications, the Green's function perturbation method, combined with the conventional Fisher-Lee relations \cite{fisher1981relation,rotter2017light}, is used. This approach is commonly employed to validate wavefront shaping experiments \cite{goetschy2013filtering,popoff2014coherent,hsu2017correlation,yilmaz2019transverse} that involve only propagating wave modes. A waveguide like geometry, often associated with the Landauer formalism for electronic conduction \cite{akkermans2007mesoscopic}, is typically adopted for such validations. On the other hand, accounting for evanescent wave modes is experimentally significant in the context of wavefront shaping for achieving sub-wavelength resolution during focusing \cite{lerosey2007focusing,park2017time}. These experiments involving the focusing of evanescent wave modes are supported and motivated by the original proposal of the generalized $S$ matrix by Carminati et al. \cite{carminati2000reciprocity} and the associated description of its unitarity and reciprocity properties. Here, the term \textit{generalized} refers to the inclusion of evanescent modes alongside conventional propagating modes in the estimation of the $S$ matrix.  The overarching goals of this paper are $1)$ to describe the method used for numerically estimating the generalized $S$ matrix proposed by Carminati et al. \cite{carminati2000reciprocity} for a disordered medium in the Landauer-waveguide geometry, and 2) to combine the statistical analysis of scattered propagating and evanescent modes under this generalized modeling framework particularly targeting wavefront shaping applications. Additionally, the classical wave scattering methods presented in the paper are implemented in MATLAB\textsuperscript{\textregistered} and made available to readers as an open-source code package, accompanied by its user manual. Interested readers can use the code not only for the numerical estimation of the generalized $S$ matrix, but also for the visualization of wave transport within the medium, especially involving propagating eigenchannels. 

Several numerical approaches are already available to model electromagnetic scattering in disordered media and can in principle be used to estimate the scattering matrix. The Finite Difference Time Domain (FDTD) method, which solves Maxwell's equations on a spatial grid, has been used to numerically construct transmission matrices and extract eigenchannels in disordered slabs \cite{choi2011transmission}. However, FDTD requires a fine spatial grid, which becomes memory intensive for large three-dimensional domains. Therefore, the Pseudospectral Time-Domain (PSTD) method \cite{tseng2006pseudospectral} addresses this by replacing finite differences with Fourier-based spatial derivatives, reducing the required grid density to approximately two points per wavelength and enabling full three-dimensional solutions for macroscopic random media. Both FDTD and PSTD discretize the entire computational volume, which can increase the memory requirement. The discrete dipole approximation (DDA) offers an alternative by representing the scattering medium as a collection of polarizable dipoles rather than a uniform grid, and a systematic comparison of DDA and PSTD has shown that their relative performance depends strongly on the particle size parameter and refractive index 
contrast \cite{liu2012comparison}. Extending the dipole framework to large-scale inhomogeneous slabs, the coupled dipole approximation (CDA) treats the medium 
as a self-consistent system of polarizable dipoles interacting through a modified slab Green's function \cite{sukhov2008coupled}. When the medium consists of discrete identifiable scatterers rather than a continuous random permittivity, the superposition T-matrix method \cite{mishchenko2007multiple} provides an alternative route to determine numerically exact solutions by combining analytical single-particle T-matrices with inter-particle coupling to account for multiple scattering. All of the above methods simulate a specific realization of the disordered medium. A conceptually different strategy is offered by random matrix theory  \cite{byrnes2026random}, which instead characterizes the probability distribution of scattering matrix elements across an ensemble of disorder realizations by exploiting single-particle scattering properties and the asymptotic Gaussian statistics of random phasor sums in the limit of many scatterers. They obtain the scattering matrix of thicker systems by cascading transfer matrices of individually thin slabs, considering only propagating modes. Physical constraints such as reciprocity and unitarity are ensured within their statistical model, with extensions to polarized light \cite{byrnes2026random}. While all the methods discussed above scale favourably with the number of scatterers or discretization points, the electromagnetic propagation methods (FDTD, PSTD, DDA, CDA, and T-matrix) offer limited analytical tractability regarding the structure of the scattering matrix and its estimation. In particular, symmetry properties are typically checked numerically rather than derived at the level of the formulation. Furthermore, in all of the above approaches, although evanescent fields may be present within the medium during computation in some cases, the scattering matrices that can potentially be estimated are assembled predominantly in the basis of propagating modes. Evanescent mode contributions are typically discarded before or during the construction of the transmission matrix.

In the present work, we employ the Green's function perturbation method and generalize the conventional Fisher-Lee relations to incorporate evanescent modes. This approach is computationally more demanding than the methods above due to the necessity of storing and perturbing the Green's functions involved. For example, the computational complexity of the direct inversion of the Dyson's system of equations (Supplementary section \ref{supp:GreensfunPertb} A) scales as $\mathcal{O}(N_p^3)$ in the number of discretization points $N_p$. Even though the iterative method of solving the perturbation system of equations fully, which is the method we adopted in this paper (Supplementary section \ref{supp:GreensfunPertb} B), reduces the computational complexity to a sub-cubic level, the memory requirement to store the Green's function is relatively higher compared to the previously cited works. This makes our approach generally less suited to very large-scale three-dimensional disordered systems. However, we circumvent that issue by breaking a large medium into thinner slabs and using the $S$-matrix cascading rule involving the generalized $S$ matrix containing evanescent modes to estimate the effective $S$ matrix of the larger system. In addition to this, our method offers several analytical advantages that are central to the goals of this work. The generalized Fisher--Lee relations yield closed-form $S$-matrix elements including the evanescent modes, enabling analytical tractability and explicit analytical verification of symmetry relations such as reciprocity. It provides an accurate $S$-matrix for a given disorder realization as it is based on numerical perturbation method. The perturbed Green's function provides direct access to the internal field distribution throughout the medium for any arbitrary input, including the visualization and estimation of eigenchannels.

In this context, the key novelties and contributions of this paper are summarized as follows. First, the conventional Fisher-Lee relations are extended to include evanescent modes, enabling the construction of a generalized $S$ matrix that consistently incorporates both propagating and evanescent wave contributions. Second, this generalized framework provides a rigorous basis for modeling wavefront shaping scenarios involving sub-wavelength focusing enabled by evanescent waves. Third, it is shown that the universal $2/3$ optimal transmission rule, well established for propagating modes in diffusive media, remains valid for the background transmission associated with evanescent wave focusing. Finally, the generalized $S$ matrix formalism enables the cascading of $S$ matrices corresponding to thin layers of the disordered medium, allowing thicker systems to be constructed efficiently. By breaking the medium into slices, this approach makes it possible to compute wave transport for arbitrary sample thicknesses on standard desktop computational resources.

The paper is organized as follows. First, Sec.~\ref{Sec:definitionsBI} reviews the definitions of propagating and evanescent eigenmodes, along with the unperturbed and the disorder-perturbed Green's functions, which form the foundation for defining the $S$ matrix. For the disorder-perturbation of the Green's function, spatially correlated disorders are employed, where the correlation length can be used to tailor transport properties. Once the perturbed Green's function is obtained, the Kirchhoff–Helmholtz boundary integral formulation of the wave scattering problem is discussed. Such a boundary integral formulation is used to estimate the generalized $S$ matrix from the perturbed Green's function. Additionally, the boundary integral equation enables the visualization of the total field for a given incident wave from the left or the right side of the disorder and validates the numerical correctness of the estimated generalized $S$ matrix. Next, Sec.~\ref{sec:generalisedfisherlee} presents the direct estimation of the generalized $S$ matrix components from the perturbed Green's function, resulting in the generalization of the conventional Fisher-Lee relations. To ensure the numerical correctness of the code implementation, the estimated $S$ matrix is shown to satisfy the generalized unitarity and reciprocity relations in the Sec.~\ref{sec:uniandrecivalidation}. These relations are also crucial for discussing the properties of focusing onto an evanescent mode via phase conjugation of the transmitted propagating wave. Sec.~\ref{sec:truncno} discusses the role of the number of evanescent modes accounted, disorder correlation length, and their impact on numerical resolution and computational cost. Towards the final portion of the paper (Sec.~\ref{sec:focussing}), the scenario of wavefront shaping involving the focusing onto an evanescent mode in transmission and its associated properties, particularly the universal background transmission, is studied. The formation of a non-zero background intensity while focusing is well-studied \cite{davy2012focusing} in the case involving only propagating modes. To maintain the conciseness of the paper, a supplementary material \cite{supplementary} is provided, containing derivations of some of the results presented. 
 
 \begin{figure*}[t]
 	\includegraphics[width=1\textwidth]{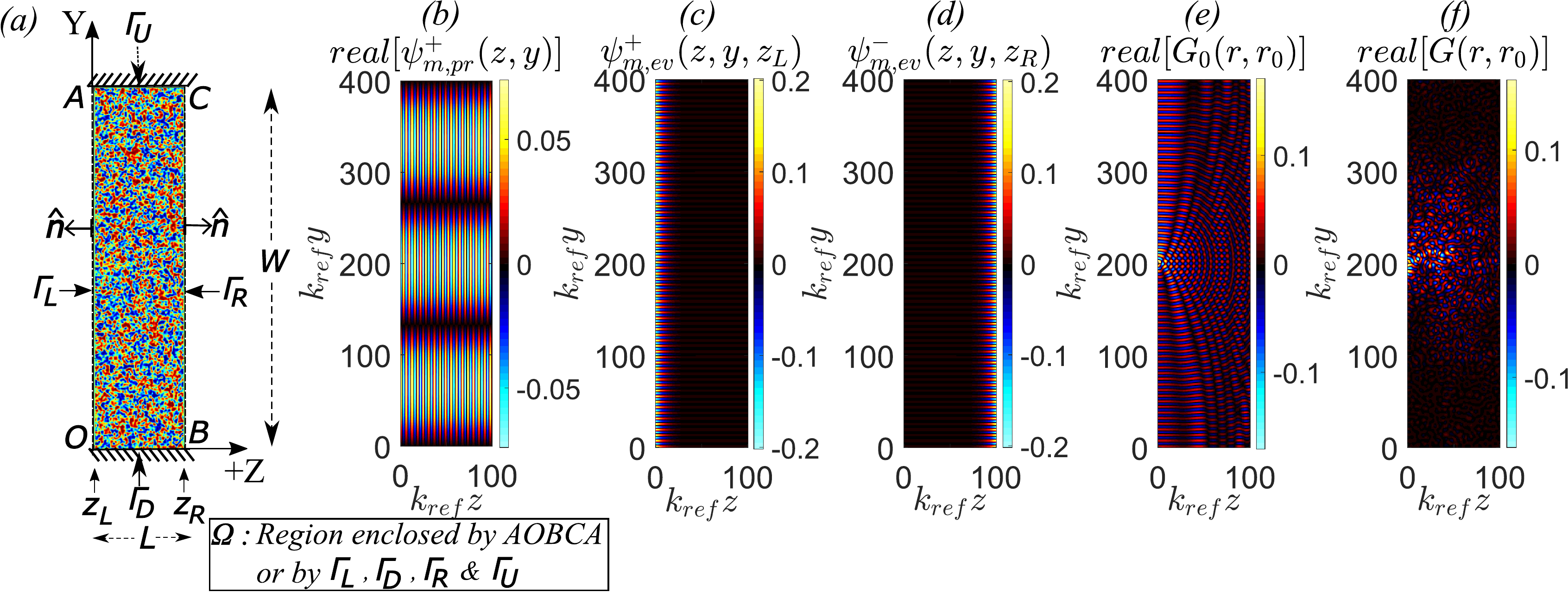}
 	\caption{\label{fig:geometryandeigmodes} \textbf{The waveguide-like geometry of the computational domain and the plots of the real parts of eigenmodes, unperturbed and perturbed Green's functions.} \textbf{(a)} Geometry of the computational domain. Boundary around the diffusive disorder ($AOBCA$) is denoted by $\Gamma$ which is the union of four boundary segments $\Gamma_{L}$, $\Gamma_{D}$, $\Gamma_R$ and $\Gamma_{U}$ where $L$ stands for left, $R$ stands for right, $U$ stands for up and $D$ for down sides of the disorder. The $2D$ region enclosed by $\Gamma$ is referred to as $\Omega$. Transverse boundaries $\Gamma_U$ and $\Gamma_{D}$ have reflection (mirror) boundary conditions. The transverse size of the slab is $W$ and the longitudinal thickness is $L$.  \textbf{(b)} Real part of $m^{th}$ eigenmode which is a propagating ($pr$) eigenmode. In this example, $m=3$, out of the $M_{pr}=130$ propagating modes possible for the given $k_{ref}W$. The $+$ symbol in the superscript denotes that the wave is propagating rightwards along $+z$ direction.  The discrete form (refer Supplementary Sec.~\ref{supp:discretization}) of the analytical equation given in Eq.~(\ref{eq:rlt_eig}) is used for plotting. \textbf{(c)} An evanescent eigenmode, plotted using Eq.~(\ref{eq:Levaneig}) for $m=132$, being the second evanescent mode out of the $M_{ev}=30$ evanescent modes taken. Here, the evanescent mode is incident rightwards (with a superscript $+$) originating from the boundary $\Gamma_{L}$ at $z_L$. Similarly, the left going scattered evanescent eigenmode exist at $\Gamma_{L}$ in presence of disorder, although not plotted here. \textbf{(d)} An evanescent eigenmode (plotted using Eq.~(\ref{eq:Revaneig}) for $m=132$) incident leftwards (with a superscript $-$) originating from $\Gamma_{R}$ at $z_R$. Similarly, the right going scattered evanescent mode exist at $\Gamma_{R}$ in presence of disorder, although not plotted here. \textbf{(e)} Real part of the disorder-less Green's function $G_0(z,y,z_0,y_0)$  for $(k_{ref}z_0,k_{ref}y_0)$ at $(0,201.46)$. Discrete form of Eq.~(\ref{eq:Greensfundef}) is used for plotting where 30 evanescent modes are used in addition to all the 130 propagating modes. \textbf{(f)} Similarly, plotting the perturbed Green's function $G(r,r_0)$ (corresponding to the disorder given in FIG.~\ref{fig:Smatrices}(b)) estimated by solving the Dyson's equation (Eq.~(\ref{eq:Dyson})) as given in Supplementary Sec.~\ref{supp:GreensfunPertb}.}
 \end{figure*}
 
\section{Definitions and notations used for the boundary integral formulation}\label{Sec:definitionsBI}
The paper addresses the $2D$ scalar Helmholtz wave equation without loss or gain, given as 
\begin{equation}
	\left[\nabla^2 + k_0^2\mu _r\tilde{\epsilon_r}(r)\right]\tilde{E}(r)=0,
	\label{eq:waveequationfinal}
\end{equation}
where $\tilde{E}(r)$ represents the complex-valued $2D$ scalar wave field as a function of the position $r=(z,y)$, $k_0=2\pi/\lambda_0$ is the free-space wave vector magnitude, $\lambda_0$ the free-space wavelength, $\tilde{\epsilon}_r(r)$ is the relative dielectric permittivity constant (as a function of the position) forming the scattering strength of the random media, and $\mu_r$ (taken as unity) is the relative permeability constant. The time-harmonic dependence of $\tilde{E}(r)$ is implicitly taken as $e^{-i\omega_0 t}$, where $i$ is the imaginary unit and $\omega_{0}$ is the angular temporal frequency.  The disorder dielectric constant $\tilde{\epsilon_r}(r)$ is defined as a spatially dependent perturbation $\delta \tilde{\epsilon_r}(r)$ added to a uniform reference (background) dielectric constant $\tilde{\epsilon}^{ref}_r$ as 
$\tilde{\epsilon_r}(r)=\tilde{\epsilon}^{ref}_r+ \delta\tilde{\epsilon_r}(r)$.
In this paper, $\delta\tilde{\epsilon_r}(r)$ is taken to be spatially correlated. A description of the generation of spatially correlated disorder used for the modeling can be found in the Supplementary Sec.~\ref{supp:spatialcorrdisordergeneration}. Although uncorrelated disorders with very small correlation length $l_c$ (approaching zero) are analytically and computationally simpler to implement, random media such as biological tissue samples involve non-zero spatial correlation lengths \cite{schmitt1996turbulent}. If $\delta\tilde{\epsilon_r}(r)=0$, the wave equation for the disorder-free region simplifies to $[\nabla^2 + k^2_{ref} ]\tilde{E}(r)=0$,
where $k^2_{ref}=k_0^2\tilde{\epsilon}^{ref}_r $ such that $k_{ref}=k_0 \eta_{ref}$. Here, $k_{ref}$ is the magnitude of the wave vector in a spatially uniform reference dielectric medium, and $\eta_{ref}=\sqrt{\tilde{\epsilon}^{ref}_r}$ is the refractive index of the reference medium. 

Completely reflecting (mirror) boundary conditions at the transverse boundaries (at $y=0$ and $y=W$ in FIG.~\ref{fig:geometryandeigmodes}(a)) were implemented to facilitate the accounting of the flux conservation just at the left and the right longitudinal boundaries (at $z=z_L$ and $z=z_R$). The subscripts $L$ and $R$ on the boundary $\Gamma$ denote the left and the right boundaries, respectively, as shown in the FIG.~\ref{fig:geometryandeigmodes}(a).
With respect to the geometry shown in FIG.~\ref{fig:geometryandeigmodes}(a), without any disorder, one can define both propagating eigenmodes (shown in FIG.~\ref{fig:geometryandeigmodes}(b)) and evanescent eigenmodes (shown in FIG.~\ref{fig:geometryandeigmodes}(c) and FIG.~(d)). These eigenmodes are the orthonormal basis wave states, forming the modal basis with which any arbitrary wave in the computational domain can be expressed as a basis expansion. The $S$ matrix is defined with respect to these eigenmode basis wave states, where the wave field modes can be propagating ($pr$) or evanescent ($ev$) in general. Conventionally, the right-traveling (along $+z$ direction) or the left-traveling (along $-z$ direction) propagating eigenmodes are combinedly defined as
\begin{equation}
	\psi^{\pm}_{m,pr}(z,y)=\phi^{\pm}_{m,pr}(z)\chi_m(y), 
	\label{eq:rlt_eig}
\end{equation}
where $\phi^{\pm}_{m,pr}(z)$ and $\chi_m(y)$ are given as

\begin{align}
&\phi^{\pm}_{m,pr}(z)=\frac{1}{\sqrt{k^{(m)}_z}} e^{\pm i k^{(m)}_z z},\\
&\chi_m(y)= \sqrt{\frac{2}{W}} \sin(k^{(m)}_y y),
\end{align}

such that $k^{(m)}_{y}=m \pi /W$ is the transverse wave vector component of the $m^{th}$ eigenmode along the $y$ direction, and  $k^{(m)}_{z}$ is the real longitudinal wave vector component of the $m^{th}$ eigenmode along the $z$ direction. The $``+"$ symbol in the superscript of $\phi^{+}_{m,pr}(z,y)$ and $\psi^{+}_{m,pr}(z,y)$ denotes wave propagation along the $+z$ direction (from the left to the right of the computational domain), while the $``-"$ symbol denotes the propagation along the $-z$ direction (from the right to the left). Here, the function $\chi_{m}(y)$ represents the transverse component of the eigenmode, satisfying the mirror(reflection) boundary conditions (at $y=0$ and $y=W$), with $W$ being the transverse size of the computational region, as shown in the FIG.~\ref{fig:geometryandeigmodes}(a). The index $(m)$ in the superscript of $k^{(m)}_y$ denotes the positive integer index for the $m^{th}$ wave vector. $(m)$ is a positive integer due to the discrete nature of $k^{(m)}_y$, which arises from the transverse reflecting boundary condition. An example for a propagating eigenmode is plotted in FIG.~\ref{fig:geometryandeigmodes}(b), where $m=3$ and the wave propagates rightward. $\chi_{m}(y)$  satisfies the orthogonality relationship $\langle \chi_{m}(y)| \chi_{n}(y) \rangle = \delta_{kron}(m,n)$, where $\delta_{kron}$ is the Kronecker delta function. There are approximately $2\eta_{ref} W/\lambda_0$ propagating eigenmodes. The wave vector components $k^{(m)}_{y}$ and $k^{(m)}_{z}$ obey the dispersion relation $\left(k^{(m)}_{y}\right)^2+\left(k^{(m)}_z\right)^2=k^2_{ref}$ where $k_{ref}= 2\pi \eta_{ref} /\lambda_0 = \omega_0 \eta_{ref}/c$. 

For the evanescent modes,
\begin{equation}
k^{(m)}_z=\sqrt{k^2_{ref}-\left(k^{(m)}_y\right)^2}=i k^{(m)}_{z,ev},
\end{equation}
where $k^{(m)}_z$ are purely imaginary and $k^{(m)}_{z,ev}$ are real numbers. The left- and right-going evanescent eigenmodes are defined to originate relative to the two boundaries $\Gamma_L$ and $\Gamma_{R}$ (boundaries shown in the FIG.~\ref{fig:geometryandeigmodes}(a) at $z_L$ and $z_R$), positioned close to the disorder.  These evanescent modes attenuate along the direction of incidence onto the disorder from either the left or right side of the disorder and also attenuate when scattered away. The evanescent modes are defined as the follows, with dependence on $z_L$ or $z_R$ :
\begin{subequations}
	\label{allequations1} 
	\begin{eqnarray}
		\psi^{\pm}_{m,ev}(z,y,z_L)&=&\phi^{\pm}_{m,ev}(z,z_L)\chi_m(y),\label{eq:Levaneig}\\
		\psi^{\pm}_{m,ev}(z,y,z_R)&=&\phi^{\pm}_{m,ev}(z,z_R)\chi_m(y),\label{eq:Revaneig} 
	\end{eqnarray}
\end{subequations}
where
\begin{subequations}
	\label{allequations2} 
	\begin{eqnarray}
		\phi^{\pm}_{m,ev}(z,z_L)&=&\frac{1}{\sqrt{|k^{(m)}_{z}|}} e^{- k^{(m)}_{z,ev} |z-z_L|}, \\
		\phi^{\pm}_{m,ev}(z,z_R)&=&\frac{1}{\sqrt{|k^{(m)}_{z}|}} e^{- k^{(m)}_{z,ev} |z-z_R|}.
	\end{eqnarray}
\end{subequations}
The symbol $``\pm"$ on the L.H.S of Eq.~(\ref{allequations2}) indicates the directions of incidence/scattering of the evanescent mode originating relative to $z_L$ or $z_R$, (with $``+"$ for rightward and $``-"$ for leftward emergence). Note that $``\pm"$ does not appear on R.H.S of Eq.~(\ref{allequations2}), as both the leftward- and rightward-emerging modes attenuate away from $z_L$ or $z_R$ along $z$ . Two examples of the incident evanescent modes are shown in FIG.~\ref{fig:geometryandeigmodes}(c) and FIG.~\ref{fig:geometryandeigmodes}(d), illustrating a right-going evanescent eigenmode at $\Gamma_{L}$ and a left-going evanescent eigenmode at $\Gamma_{R}$, respectively. Although not plotted, scattered evanescent modes from the disorder also exist, which are left-going at $\Gamma_{L}$ and a right-going at $\Gamma_{R}$. 

The eigenmodes, referred to as $\psi_m$ in general, include a normalization factor ${1}/{\sqrt{|k^{(m)}_z|}}$. For the propagating modes, ${1}/{\sqrt{k^{(m)}_z}}$ ensures that the flux (current) component flowing normal to a given boundary region $\Gamma_{L}$ or $\Gamma_{R}$ is unity. The flux density of the $m^{th}$ eigenmode is defined as $\vec{J}^{(m)}(z,y)=J^{(m)}_z(r)\hat{e}_z + J^{(m)}_y(r)\hat{e}_y=\mathfrak{Im}\left\{\psi^{*}_m \nabla \psi_m\right\}$, where $\hat{e}_z $ and $\hat{e}_z$ are unit vectors along the $Z$ and $Y$ directions, respectively, and $\mathfrak{Im}$ refers to the imaginary part. As there is no flux crossing the transverse boundaries $\Gamma_{U}$ and $\Gamma_{D}$, $J^{(m)}_y(r)$ is zero along the boundary normal. For the evanescent modes, as there is no flux crossing the boundaries, there is no physical necessity for defining a normalization factor. Nevertheless, a normalization factor ${1}/{\sqrt{|k^{(m)}_z}|} = {1}/{\sqrt{k^{(m)}_{z,ev}}}$ is included in the definition of the evanescent eigenmodes. This is primarily for the convenience of modeling, ensuring that the propagating modes and the evanescent modes are of the same numerical order of magnitude. This avoids precision errors when estimating the $S$ matrix. 

Next, using the propagating and evanescent eigenmodes, the analytical expression \cite{felsen2004wave}  for the $2D$ Green's function $G_0(r,r_0)$ (plotted in FIG.~\ref{fig:geometryandeigmodes}(e)) without any disorder perturbation is given as
\begin{equation}
	\begin{split}
		&G_0(z,y,z_0,y_0)=\\
		&\sum_{m=1}^{\infty}\frac{1}{2ik^{(m)}_z}e^{i k^{(m)}_z| (z-z_0)|}\sqrt{\frac{2}{W}} \sin(k^{(m)}_y y)\sqrt{\frac{2}{W}} \sin(k^{(m)}_y y_0), \\
	\label{eq:Greensfundef}
	\end{split}
\end{equation} 
where $r_0=(z_0,y_0)$ is the source location, $k_z^{(m)}$ is real for the propagating terms in the summation, and $k_z^{(m)}=i k_{z,ev}^{(m)}$ is imaginary for the evanescent components.  Although the summation in Eq.~(\ref{eq:Greensfundef}) runs from $m=1$ to infinity, it is truncated to include all $M_{pr}$ propagating modes, where $M_{pr}\approx2\eta_{ref}W/\lambda_0$, and a certain truncation number of evanescent modes,  $M_{ev}$. As an example, in this paper, $M_{ev}$ is set as 30, $M_{pr} = 130$, and $M_{total} = M_{pr}+M_{ev}$. Considerations to be taken for choosing this truncation number for the evanescent modes is described in Sec.~\ref{sec:truncno} for a given spatially correlated disorder realization. Thus, the summation in Eq.~(\ref{eq:Greensfundef}) runs from $m=1$ to $m=M_{total}$ for the numerical implementation of $G_0(r,r_0)$ in the code. 

Using $G_0(r,r_0)$ and the dielectric disorder perturbation $\delta\tilde{\epsilon_r}(r)$ (shown in FIG.~\ref{fig:Smatrices}(b)), the perturbed Green's function $G(r,r_0)$ (plotted in FIG.~\ref{fig:geometryandeigmodes}(f)) is estimated using Dyson's equation \cite{akkermans2007mesoscopic,carminati2021principles},
as given below :
\begin{equation}
	G(r,r_0)=G_0(r,r_0)+ \int_{r_1}  G_0(r,r_1) V(r_1) G(r_1,r_0) dr_1, \label{eq:Dyson}
\end{equation}
where $V(r)=-k_0^2\delta\tilde{\epsilon_r}(r)$ is the scattering potential at the position $r$. Dyson's integral equation is solved in this paper by generalizing the method given in~\cite{martin1994iterative,martin1995generalized}, solving for a block of scattering particles taken at once and iteratively updating the Green's function. For details on the method used to obtain $G(r,r_0)$, refer to Supplementary Sec.~\ref{supp:GreensfunPertb}. As the propagating and the evanescent eigenmodes $\psi^{\pm}_m(r)$ form a complete basis set, the perturbed Green's function $G(r,r_0)$ can also be expanded in terms of these eigenmodes. On the left boundary, $G(r,r_0)$ can be expanded in terms of left-going eigenmodes, and on the right boundary, $G(r,r_0)$ can be expanded in terms of right-going eigenmodes. Hence, $G(r,r_0)$ satisfies the homogeneous boundary conditions given in the  Supplementary Sec.~\ref{supple:homoBC}, which are also satisfied by the eigenmodes. These boundary conditions are useful for simplifying the analysis presented in the following section. Next section deals with the Kirchhoff–Helmholtz boundary integral formulation of the scattering problem, incorporating $G(r,r_0)$ and the total field around the boundary $\Gamma$. This is followed by the direct numerical estimation of the generalized $S$ matrix from $G(r,r_0)$. The goal of this section is to establish the relationship between the perturbed Green's function $G(z,y,z_0,y_0)$ and the total wave field $\tilde{E}(z,y)$ that exists inside and around a medium due to an arbitrary wave incident on the disorder. This relationship helps estimate the response of a random medium excited by an arbitrary incident wave. 

First, consider the case of wave incidence on the disorder from the left side only, where $E^{inc}(z,y)$ (traveling from left to right) is the incident wave. On the left boundary $\Gamma_L$, the total wave field due to wave incidence only from the left side is given as
\begin{equation}
	\tilde{E}(z,y)|_{(z,y)\in\Gamma_L}=\tilde{E}^{inc}(z,y)|_{(z,y)\in\Gamma_L} + \tilde{E}^{refl}(z,y)|_{(z,y)\in\Gamma_L}
\label{eq:totalwaveonleft}	
\end{equation}
where $\tilde{E}^{refl}(z,y)|_{(z,y)\in\Gamma_L}$ is the reflected wave on $\Gamma_L$. $\tilde{E}^{inc}(z,y)$ and $\tilde{E}^{refl}(z,y)$ can be expanded in terms of eigenmodes and their associated complex coefficients on $\Gamma_L$ as follows:
\begin{align}
	\tilde{E}^{inc}(z,y)|_{(z,y)\in\Gamma_L}&= \sum_{m=1}^{M_{total}} c^{+}_m \psi^{+}_m(z,y)|_{(z,y)\in\Gamma_L} & &  \label{eq:incwavegenmain}\\ 
         & =\sum_{m=1}^{M_{total}} c_m^+ \chi_m(y)\phi^+_m(z)|_{(z,y)\in\Gamma_L}\nonumber        &  &   \\
      \text{such that}~\phi^+_m(z)=&\begin{cases}
      	&\phi^+_{m,pr}(z)~\text{when}~m\leq M_{pr}\nonumber\\
      	&\phi^+_{m,ev}(z,z_L)~\text{when}~m > M_{pr}
      \end{cases}\nonumber
\end{align}
where $c^{+}_m$ is the $m^{th}$ complex coefficient used to synthesize the incident wave in terms of eigenmodes propagating along the $+z$ direction. Similarly, 
\begin{align}
\tilde{E}^{refl}(z,y)|_{(z,y)\in\Gamma_L}&=\sum_{n=1}^{N_{total}} r^{-}_n \psi^{-}_n(z,y)|_{(z,y)\in\Gamma_L}& & \label{eq:reflwavegenmain}\\
	&=\sum_{n=1}^{N_{total}}r^-_n\chi_n(y)\phi^-_n(z)|_{(z,y)\in\Gamma_L} \nonumber &&\\
	\text{such that}~\phi^-_n(z)=&\begin{cases}
	&	\phi^-_{n,pr}(z)~\text{when}~n\leq N_{pr}\\
	&	\phi^-_{n,ev}(z,z_L)~\text{when}~n > N_{pr}
	\end{cases}	\nonumber\\
	\text{and}~r^-_n&=\sum_{m=1}^{M_{total}} \mathbf{S}^{n m}_{11} c^{+}_m,\nonumber			
\end{align}
where $\mathbf{S}^{n m}_{11}$ are the elements of the generalized reflection matrix $\mathbf{S}_{11}$ for the wave incidence only from the left side of the disorder. Here, $m$ as the index denotes the incident eigenmode number, and $n$ denotes the scattered eigenmode number.

On the right boundary $\Gamma_R$, only the transmitted wave $\tilde{E}^{trans}(z,y)|_{(z,y)\in\Gamma_R}$ exists for the scenario of wave incidence only from the left side of the disorder. Therefore, the total field on $\Gamma_{R}$ is given as
\begin{align}
	\tilde{E}(z,y)|_{(z,y)\in\Gamma_R}&=\tilde{E}^{trans}(z,y)|_{(z,y)\in\Gamma_R}& &\label{eq:transtotalwave}\\
	&=\sum_{n=1}^{N_{total}} t_n^+\phi^+_n(z) \chi_n(y)|_{(z,y)\in\Gamma_{R}} \nonumber&&\\
	\text{such that}~\phi^+_n(z)=&\begin{cases}
		&\phi^+_{n,pr}(z)~\text{when}~n\leq N_{pr}\\
		&\phi^+_{n,ev}(z,z_R)~\text{when}~n > N_{pr}
	\end{cases}\nonumber\\
\text{and}~t^+_n&= \sum_{m=1}^{M_{total}} \mathbf{S}^{n m}_{21} c^{+}_m, 
\nonumber
\end{align}
where $\mathbf{S}^{n m}_{21}$ are the elements of the generalized transmission matrix $\mathbf{S}_{21}$ for wave incidence only from the left side of the disorder.

Inside the medium, the total field of the wave is denoted as $\tilde{E}(z,y)$ itself. Consider the Helmholtz equation for the Green's function and the total field as follows : 
\begin{align}
	&[\nabla^2 + k_0^2\tilde{\epsilon_r}(z,y)]G(z,y,z_0,y_0)=\delta(z,z_0)\delta(y,y_0),
	\label{eq:greensfun}\\
	&[\nabla^2 + k_0^2 \tilde{\epsilon_r}(z,y)]\tilde{E}(z,y)=0,
	\label{eq:totalfield}
\end{align}
where $\nabla^2$ is the Laplacian operator with respect to the field position $(z,y)$. In Eq.~(\ref{eq:greensfun}), $(z_0,y_0)$ is the Dirac-delta source location. Multiplying Eq.~(\ref{eq:greensfun}) by $\tilde{E}(z,y)$ and Eq.~(\ref{eq:totalfield}) by $G(z,y,z_0,y_0)$, followed by subtraction and the application of the divergence theorem, the following Kirchhoff–Helmholtz boundary integral equation is obtained, as shown in appendix Sec.~A of  \cite{domotor2015scattering} :
\begin{align}
		&\tilde{E}(z_0,y_0)|_{(z_0,y_0)\in\Omega}=\nonumber\\
		& \int \int \nabla\cdot\left\{\tilde{E}(r)\nabla G(r,r_0)-G(r,r_0)\nabla \tilde{E}(r)\right\}dz dy \nonumber \\
		&=\int_\Gamma \left\{\tilde{E}(r)\nabla G(r,r_0)-G(r,r_0)\nabla \tilde{E}(r)\right\}\cdot~\hat{n} d\Gamma,
	\label{eq:generalBoundInt}
\end{align}
where $r$ denotes $(z,y)$, and $r_0$ denotes $(z_0,y_0)$. Note that $(z_0,y_0)$ now represents the location where the field is visualized, and it is no longer the source location. This is consistent with the reciprocity property of the Green's function, $G(r,r_0)=G(r_0,r)$. Since there is no flux crossing $\Gamma_{U}$ and $\Gamma_{D}$, the boundary integral involves only the terms associated with $\Gamma_{L}$ and $\Gamma_{R}$. The Kirchhoff–Helmholtz boundary integral equation implies that the total field $\tilde{E}(z_0,y_0)$ at a point $(z_0,y_0)$ inside the closed region $\Omega$ bounded by $\Gamma$ can be estimated if the value of $\tilde{E}(z,y)$, and the Green's function $G(z,y,z_0,y_0)$ together with its spatial derivatives, are known at the boundaries $\Gamma_{L}$ and $\Gamma_R$.  The boundary integral can be simplified (with the derivation provided in Supplementary Sec.~\ref{supp:BIsimplification}) as follows. For the
wave incidence only from the left to the right of the disorder $(\text{incidence onto}~\Gamma_{L})$:
\begin{align}
&\tilde{E}\left(z_0,y_0\right)|_{(z_0,y_0)\in\Omega}= \nonumber\\
&\sum^{M_{total}}_{m=1}2i k_z^{(m)}\int \tilde{E}_m^{inc}(z,y)G\left(z,y,z_0,y_0\right)|_{(z,y)\in \Gamma_L}dy 
\label{eq:BIvisualfinal}
\end{align}
where $\tilde{E}_m^{inc}(z,y)|_{(z,y)\in\Gamma_L}=c^{+}_m \psi^{+}_m(z,y)|_{(z,y)\in\Gamma_L}$. Here, $c^{+}_m$ is the $m^{th}$ complex coefficient used to synthesize the incident wave in terms of eigenmodes propagating along the $+z$ direction. 

Similarly, for wave incidence only from the right side of the disorder $(\text{incidence onto}~\Gamma_{R})$, the equation is similar to that of Eq.~(\ref{eq:BIvisualfinal}), except that now ${(z,y)\in \Gamma_R}$ as follows :
\begin{align}
&\tilde{E}\left(z_0,y_0\right)|_{(z_0,y_0)\in\Omega}=	\label{eq:BIvisualfinalR2L}\\
&\sum^{M_{total}}_{m=1}2i k_z^{(m)}\int \tilde{E}_m^{inc}(z,y)G\left(z,y,z_0,y_0\right)|_{(z,y)\in \Gamma_R}dy\nonumber
\end{align}
where $\tilde{E}_m^{inc}(z,y)|_{(z,y)\in\Gamma_R}=c^{-}_m \psi^{-}_m(z,y)|_{(z,y)\in\Gamma_R}$. Similar to that of Eq.~(\ref{eq:incwavegenmain}),
\begin{align}
	\tilde{E}^{inc}(z,y)|_{(z,y)\in\Gamma_R}=\sum_{m=1}^{M_{total}} c^{-}_m \psi^{-}_m(z,y)|_{(z,y)\in\Gamma_R}.\label{eq:rightincidence}
\end{align}
The simplified boundary integral equations Eq.~(\ref{eq:BIvisualfinal}) or Eq.~(\ref{eq:BIvisualfinalR2L}), along with Eq.~(\ref{eq:totalwaveonleft}) and Eq.~(\ref{eq:transtotalwave}), can be used for two purposes. $1)$ To estimate the generalized reflection ($\mathbf{S}_{11}~\text{and}~\mathbf{S}_{22}$) and transmission ($\mathbf{S}_{21}$ and $\mathbf{S}_{12}$) matrices. $2)$  To visualize the total field $\tilde{E}(z_0,y_0)|_{(z_0,y_0)\in\Omega}$, for a given incident wave from the left side or the right side of the disorder. As an example, in Supplementary Sec.~\ref{supp:visualisingeigchannel}, the transmission eigenchannels associated with $\mathbf{S}^{pr,pr}_{21}$ and $\mathbf{S}^{pr,pr}_{12}$ are plotted using the boundary integral equations.

\section{Extended Fisher-Lee relations for estimating the generalized $S$ matrix}\label{sec:generalisedfisherlee}

This section explains how the boundary integral equations are used to estimate the generalized $S$ matrix where the traditional Fisher-Lee method \cite{fisher1981relation,datta1997electronic,domotor2015scattering,rotter2017light} is generalized to include the evanescent wave modes as well. The definition of $S$ matrix is given as

\begin{align}
	\mathbf{S}=\left[
	\begin{array}{cc}
		\mathbf{S}_{11} & \mathbf{S}_{12} \\
		\mathbf{S}_{21} & \mathbf{S}_{22} \\
	\end{array}
	\right],
	\label{eq:Smatrix}
\end{align} 
where $\mathbf{S}_{21}$ and $\mathbf{S}_{12}$ are the generalized transmission matrices for wave incidence from the left and right, respectively. Similarly, $\mathbf{S}_{11}$ and $\mathbf{S}_{22}$ are the generalized reflection matrices for wave incidence from the left and right, respectively. The size of the transmission matrix $\mathbf{S}_{21}$ is generally given as $[N_{total}\times M_{total}]$ where $M_{total}$ is the total number of incident eigenmodes and $N_{total}$ the total number of transmitted eigenmodes. For the quasi-$1D$ geometry shown in FIG.~\ref{fig:geometryandeigmodes}(a), $M_{pr}=N_{pr}$ and $M_{ev}=N_{ev}$. Due to the wave incidence only from the left side of the disorder onto the boundary $\Gamma_{L}$, there exists the total field composed of the incident wave and the reflected wave on $\Gamma_{L}$. On the transmitted side, the total wave field consists of the transmitted wave. One can match the total field to the boundary integral equation and solve for $\mathbf{S}_{11}$ and $\mathbf{S}_{21}$ to obtain the generalized Fisher-Lee relations. Similarly, the estimation of $\mathbf{S}_{22}$ and $\mathbf{S}_{12}$ follows the same logic, where the wave incidence is only from the right side of the disorder. 
First, the methodology adopted for deriving the generalized $\mathbf{S}_{11}$ matrix elements is presented as follows:  

The simplified boundary integral equation previously introduced in Eq.~(\ref{eq:BIvisualfinal}), where the wave incidence occurs only from the left side of the disorder, serves as the starting point for deriving the $\mathbf{S}_{11}$  matrix elements. As the L.H.S of Eq.~(\ref{eq:BIvisualfinal}) denotes the total field due to the wave incidence only from the left, one can estimate the total field on the left boundary.
Substituting Eq.~(\ref{eq:totalwaveonleft}) into the L.H.S of Eq.~(\ref{eq:BIvisualfinal}), the following is obtained : 

\begin{align}
	&\left\{\tilde{E}^{inc}(z_0,y_0)+ \tilde{E}^{refl}(z_0,y_0)\right\}|_{(z_0,y_0)\in\Gamma_L} =\sum^{M_{total}}_{m=1} \left\{2i k_z^{(m)} \right.\nonumber\\
	&\left. \times\int \tilde{E}_m^{inc}(z,y)G\left(z,y,z_0,y_0\right)dy \right\}|_{\left\{(z,y)\& (z_0,y_0)\right\}\in\Gamma_L}
	\label{eq:BIvisualfinalforFisherLee}
\end{align}

Substituting the eigenmode basis expansion of a given incident wave $\tilde{E}^{inc}(z_0,y_0)$ and reflected wave $\tilde{E}^{refl}(z_0,y_0)$, as given in Eq.~(\ref{eq:incwavegenmain}) and Eq.~(\ref{eq:reflwavegenmain}) respectively, into Eq.~(\ref{eq:BIvisualfinalforFisherLee}), and using the property of the Kronecker delta function, the following is obtained : 

\begin{align}
&\sum_{n=1}^{N_{total}}\sum_{m=1}^{M_{total}} \left\{ c_n^+\phi^+_n(z_0)\chi_n(y_0)\delta_{kron}(n,m)|_{(z_0,y_0)\in\Gamma_L}+ \right. \nonumber\\ 
&\left. c_m^+\mathbf{S}_{11}^{n m}\chi_n(y_0)\phi^-_n(z_0)|_{(z_0,y_0)\in\Gamma_L}\right\}=\sum^{M_{total}}_{m=1} \left\{2i k_z^{(m)}\times \right. \nonumber\\
&\left. \int c_m^+ \phi^+_m(z)\chi_m(y)G\left(z,y,z_0,y_0\right)dy \right\}|_{\left\{(z,y)\& (z_0,y_0)\right\}\in\Gamma_L}
\label{eq:S11beginning}
\end{align}
Since Eq.~(\ref{eq:S11beginning}) holds for any incident wave represented by any set of $c^+_m$ values, one can collect the terms multiplied by $c^+_m$ and spanned by $\sum_{m=1}^{M_{total}}$ on both the L.H.S and R.H.S, and solve for $\mathbf{S}^{n m}_{11}$. Eq.~(\ref{eq:S11beginning}) then implies the following : 
\begin{align}
	&\sum_{n=1}^{N_{total}} \phi^+_n(z_0) \chi_n(y_0)\delta_{\text{kron}}(n,m)|_{\left(z_0,y_0\right)\in \Gamma_L}+  \nonumber\\
	& \sum_{n=1}^{N_{total}} \mathbf{S}_{11}^{n m} \chi_n(y_0) \phi^- _n(z_0)|_{\left(z_0,y_0\right)\in \Gamma_L}=\nonumber\\		
	&2i k_z^{(m)}  \phi^+_m(z)\int\chi_m(y)G\left(z,y,z_0,y_0\right)dy|_{(z,y) \&\left(z_0,y_0\right)\in \Gamma_L}
	\label{eq:S11intermed}
\end{align}

To obtain an expression for $\mathbf{S}^{n m}_{11}$ using Eq.~(\ref{eq:S11intermed}), the orthogonality relation involving $\chi_{n'}(y)$ is invoked as the next step. Upon multiplying both sides of Eq.~(\ref{eq:S11intermed}) by $\chi_{n'}(y_0)$ and integrating with respect to $y_0$, yields the following :
\begin{align}
&\phi^+_{n'}(z_0)\delta_{\text{kron}}\left(n',m\right)+\mathbf{S}_{11}^{n' m}  \phi^-_{n'}(z_0)=2i k_z^{(m)} \phi^+_m(z) \times\nonumber \\
& \int \int \chi_m(y)G\left(z,y,z_0,y_0\right)\chi_{n'}(y_0) dy_0dy,
\label{eq:Tosimplify}
\end{align}
where $(z,y)~\&~\left(z_0,y_0\right)\in \Gamma_L$. For further simplification of Eq.~(\ref{eq:Tosimplify}), define the integral that appears on the R.H.S as follows : 
\begin{equation}
	\begin{split}
		&G^{\mathbf{S}_{11}}_{n m}=\\
		&\int\int\chi_m(y) G\left(z,y,z_0,y_0\right)\chi_n(y_0)dy dy_0|_{(z,y)\&\left(z_0,y_0\right)\in \Gamma_L}
	\end{split}
\end{equation}
Here, $G^{\mathbf{S}_{11}}_{n m}$ can be interpreted as a Fourier-like decomposition in the eigenspace, of the Green's function component used for estimating $\mathbf{S}^{n m}_{11}$. By renaming $n'$ as $n$, setting the positions of $z$ and $z_0$ on $\Gamma_{L}$ to 0, and using $G^{\mathbf{S}_{11}}_{n m}$,  Eq.~(\ref{eq:Tosimplify}) can be simplified as : 
\begin{align}
	&\phi^+_{n}(z_0=0)\delta_{\text{kron}}\left(n,m\right)+\mathbf{S}_{11}^{n m}  \phi^-_{n}(z_0=0)=\nonumber\\
	&2i k_z^{(m)} \phi^+_m(z=0) G^{\mathbf{S}_{11}}_{n m}.
	\label{eq:simplified}
\end{align}

There are four different cases to further simplify Eq.~(\ref{eq:simplified}) based on the nature of the incident modes (indexed by $m$) and reflected modes (indexed by $n$) as propagating or evanescent. 

Case 1 is when both the incident and reflecting modes are propagating, such that  $m\leq M_{pr}$ and $n\leq N_{pr}$. Substituting the expressions for the propagating $\phi^+_{n}(z_0=0)$, $\phi^-_{n}(z_0=0)$, and $\phi^+_m(z=0)$ in Eq.~(\ref{eq:simplified}), the following simplification is obtained : 
\begin{align}
\frac{1}{\sqrt{k_z^{(n)}}}\delta_{\text{kron}}(n,m)+\frac{\mathbf{S}_{11}^{n m}}{\sqrt{k_z^{(n)}}}=	2ik_z^{(m)}\frac{1}{\sqrt{k_z^{(m)}}} G^{\mathbf{S}_{11}}_{n m},
\end{align}
which is used to derive the expressions for $\mathbf{S}_{11}^{n m}$ provided in the first row of TABLE \ref{tab:S11}. In the TABLE \ref{tab:S11}, the terms under $Inc(m)$ indicates the nature of the $m^{th}$ incident wave mode, and $Refl(n)$ specifies the nature of the $n^{th}$ reflected wave mode, whether propagating $(pr)$ or evanescent $(ev)$. 

Case 2 corresponds to the scenario where both the incident and reflected modes are evanescent, such that  $m>M_{pr}$ and $n>N_{pr}$. In this case, Eq.~(\ref{eq:simplified}) simplifies as follows : 
\begin{align}
\frac{\delta_{\text{kron}}(n,m)}{\sqrt{\left|k_z^{(n)}\right|}}+\frac{\mathbf{S}_{11}^{n m}}{\sqrt{\left|k_z^{(n)}\right|}}=2i\left(i\left|k_z^{(m)}\right|\right)\frac{1}{\sqrt{\left|k_z^{(m)}\right|}}G^{\mathbf{S}_{11}}_{n m},
\end{align}
which is used to obtain the expression for $\mathbf{S}_{11}^{n m}$ given in the second row of TABLE \ref{tab:S11}. 

Case 3 corresponds to the scenario where the incident modes are propagating and the reflected modes are evanescent, such that $m\leq M_{pr}$ and $n>N_{pr}$. In this situation, $\delta_{kron}(n,m)$ in 	Eq.~(\ref{eq:simplified}) becomes zero, where $M_{pr}=N_{pr}$. Simplifying Eq.~(\ref{eq:simplified})  for  Case 3 leads to the following :
\begin{align}
\frac{\mathbf{S}_{11}^{n m}}{\sqrt{\left|k_z^{(n)}\right|}}=	2i k_z^{(m)}\frac{1}{\sqrt{k_z^{(m)}}}G^{\mathbf{S}_{11}}_{n m},
\end{align}
which yields the expression for $\mathbf{S}_{11}^{n m}$ given in the third row of TABLE \ref{tab:S11}. 

Finally, case 4 considers the scenario where the incident modes are evanescent and the reflected modes are propagating, such that $m>M_{pr}$ and $n\leq N_{pr}$. In this case, Eq.~(\ref{eq:simplified}) simplifies as the follows : 
\begin{align}
\mathbf{S}_{11}^{m n}\frac{1}{\sqrt{k_z^{(n)}}}=2i \left( i |k_z^{(m)}|\right)\frac{1}{\sqrt{\left|k_z^{(m)}\right|}}G^{\mathbf{S}_{11}}_{n m},
\end{align}
which provides the expression for $\mathbf{S}_{11}^{m n}$ given in the fourth row of TABLE \ref{tab:S11}.
\begin{table}[!htp]
	\caption{Elements of $\mathbf{S}_{11}$\label{tab:S11}}
	\begin{flushleft}
		\begin{tabular}{|l|l|l|l|l|l|}
			\hline
			$Inc(m)$&$Refl(n)$& \qquad \qquad \qquad $\mathbf{S}^{nm}_{11}$ \\ 
			\hline
			$pr$&$pr$  & $-\delta_{kron}(n,m)+2i\sqrt{k_z^{(m)}}\sqrt{k_z^{(n)}} G^{\mathbf{S}_{11}}_{n m}$   \\ 
			\hline
			$ev$&$ev$  & $  -\delta_{kron}(n,m)+-2\sqrt{\left|k_z^{(m)}\right|}\sqrt{\left|k_z^{(n)}\right|} G^{\mathbf{S}_{11}}_{n m}$\\ 
			\hline
			$pr$&$ev$  & $2i\sqrt{k_z^{(m)}}\sqrt{\left|k_z^{(n)}\right|} G^{\mathbf{S}_{11}}_{n m}$   \\ 
			\hline
			$ev$&$pr$  & $-2\sqrt{\left|k_z^{(m)}\right|}\sqrt{k_z^{(n)}} G^{\mathbf{S}_{11}}_{n m}$   \\ 
			\hline
		\end{tabular}
	\end{flushleft}	
\end{table}
The same method is applied to estimate the elements of $\mathbf{S}_{21}$. In this case, the boundary integral equation,	Eq.~(\ref{eq:BIvisualfinalforFisherLee}), takes the following form for $\mathbf{S}_{21}$ : 
\begin{align}
&\tilde{E}^{trans}(z_0,y_0)|_{(z_0,y_0)\in \Gamma_{R}}=	\sum_{m=1}^{M_{total}} \left\{2i k_z^{(m)} \times \right.\nonumber\\
&\left. \int \tilde{E}_m^{inc}(z,y)G\left(z,y,z_0,y_0\right)dy \right\} |_{(z,y)\in \Gamma_L\&\left(z_0,y_0\right)\in \Gamma_R},
\label{eq:transboundaryintegral}
\end{align}
where the transmitted wave field is evaluated at the right boundary, denoted by $(z_0,y_0)\in \Gamma_{R}$, and the wave incidence occurs on the left boundary, denoted by $(z,y)\in \Gamma_L$. As done previously, the eigenmode expansion for the incident and the transmitted waves, given in Eq.~(\ref{eq:incwavegenmain}) and Eq.~(\ref{eq:transtotalwave}), respectively, can be substituted into Eq.~(\ref{eq:transboundaryintegral}). Following the same procedure outlined for deriving Eq.~(\ref{eq:S11beginning}) and   Eq.~(\ref{eq:S11intermed}), the following equation is obtained for estimating $\mathbf{S}^{nm}_{21}$ :
\begin{align}
	&S_{21}^{n m}  \phi^+_{n}(z_0=z_R)=2i k_z^{(m)} \phi^+_m(z=0) G^{\mathbf{S}_{21}}_{n m},
	\label{eq:simplifiedtranseqgen}
\end{align}
where 
\begin{align}
G^{\mathbf{S}_{21}}_{n m}=\int\int\chi_m(y) G\left(z,y,z_0,y_0\right) \chi_{n}(y_0) dydy_0,
\end{align}
such that ${(z=0,y)\in \Gamma _L~\text{and}~\left(z_0=z_R,y_0\right)\in \Gamma_R}$.  Similarly, $G^{\mathbf{S}_{22}}_{n m}$, and $G^{\mathbf{S}_{12}}_{n m}$ need to be defined. The integral forms of  $G^{\mathbf{S}_{11}}_{n m}$, $G^{\mathbf{S}_{21}}_{n m}$, $G^{\mathbf{S}_{22}}_{n m}$ and $G^{\mathbf{S}_{12}}_{n m}$ are identical, given as  $\int\int\chi_m(y) G\left(z,y,z_0,y_0\right)\chi_n(y_0)dy dy_0$, except for the differences in their domains of integration, as specified in TABLE \ref{tab:Gnm}.

\begin{table}[!htp]
	\caption{Integral forms of $G^{\mathbf{S}_{11}}_{n m}$, $G^{\mathbf{S}_{21}}_{n m}$, $G^{\mathbf{S}_{22}}_{n m}$ and $G^{\mathbf{S}_{12}}_{n m}$ \label{tab:Gnm}}
	\begin{flushleft}
		\begin{tabular}{|l|l|l|l|l|l|}
			\hline
			\vtop{ \hbox{\strut S matrix}  \hbox{\strut estimation} {\hbox{\strut components}}} &  \vtop{\hbox{\strut $G^{\mathbf{S}_{11}}_{n m}$, $G^{\mathbf{S}_{21}}_{n m}$, $G^{\mathbf{S}_{22}}_{n m}$ and $G^{\mathbf{S}_{12}}_{n m}$}\hbox{\strut $=
					\int\int\chi_m(y) G\left(z,y,z_0,y_0\right)\chi_n(y_0)dy dy_0$}\hbox{except for the change in the domains}\hbox{of integration as the following : }}\\ 
			\hline
			$G^{\mathbf{S}_{11}}_{n m}$ & ${(z=0,y)\in \Gamma_L\quad\&\quad\left(z_0=0,y_0\right)\in \Gamma_L}$ \\ 
			\hline
			$G^{\mathbf{S}_{21}}_{n m}$ & ${(z=0,y)\in \Gamma_L \quad\& \quad\left(z_0=z_R,y_0\right)\in \Gamma_R}$ \\ 
			\hline
			$G^{\mathbf{S}_{22}}_{n m}$ & ${(z=z_R,y)\in \Gamma_R \quad\& \quad\left(z_0=z_R,y_0\right)\in \Gamma_R}$ \\ 
			\hline
			$G^{\mathbf{S}_{12}}_{n m}$ & ${(z=z_R,y)\in \Gamma_R \quad\& \quad\left(z_0=0,y_0\right)\in \Gamma_L}$ \\ 
			\hline
		\end{tabular}
	\end{flushleft}	
\end{table}

Eq.~({\ref{eq:simplifiedtranseqgen}}) can then be used to estimate $S^{nm}_{21}$ for the four cases described earlier, where the nature of the incident and transmitted modes varies between propagating and evanescent. The estimated results are provided in the TABLE \ref{tab:S21}.
\begin{table}[!htp]
	\caption{Elements of $\mathbf{S}_{21}$ \label{tab:S21}}
	\begin{flushleft}
		\begin{tabular}{|l|l|l|l|l|l|}
			\hline
			$Inc(m)$&$Trans(n)$& \qquad \qquad \qquad $\mathbf{S}^{nm}_{21}$ \\ 
			\hline
			$pr$&$pr$  & $2i \sqrt{k_z^{(m)}}\sqrt{k_z^{(n)}}e^{-i z_R k_z^{(n)}} G^{\mathbf{S}_{21}}_{n m}$   \\ 
			\hline
			$ev$&$ev$  & $-2\sqrt{\left|k_z^{(m)}\right|}\sqrt{\left|k_z^{(n)}\right|} G^{\mathbf{S}_{21}}_{n m}$\\ 
			\hline
			$pr$&$ev$  & $2i \sqrt{k_z^{(m)}}\sqrt{\left|k_z^{(n)}\right|} G^{\mathbf{S}_{21}}_{n m}$   \\ 
			\hline
			$ev$&$pr$  & $-2 \sqrt{\left|k_z^{(m)}\right|}\sqrt{k_z^{(n)}}e^{-i z_R k_z^{(n)}} G^{\mathbf{S}_{21}}_{n m}$   \\ 
			\hline
		\end{tabular}
	\end{flushleft}
\end{table}

On comparing TABLE \ref{tab:S11} with TABLE \ref{tab:S21}, it is observed that for $\mathbf{S}^{nm}_{21}$ elements involving transmitted propagating modes, there is an additional exponential phase term, $e^{-i z_R k_z^{(n)}}$,  arising from the $\phi^+_{n}(z_0=z_R)$ term in Eq.~(\ref{eq:simplifiedtranseqgen}). 
The same methodology is applied to estimate the other $S$ matrix components, $\mathbf{S}_{22}$ (given in TABLE \ref{tab:S22}) and $\mathbf{S}_{12}$ (given in TABLE \ref{tab:S12}). Detailed derivations are provided in Supplementary Sec.~\ref{Supp:Fisher} for the sake of conciseness in this paper.  For evaluating $\mathbf{S}^{nm}_{22}$ and $\mathbf{S}^{nm}_{12}$, incident wave propagates from the right side onto $\Gamma_{R}$. Consequently, the total field on $\Gamma_{R}$ consists of both the incident wave and the reflected wave. On the other hand, the total field on $\Gamma_{L}$ corresponds to the transmitted wave propagating to the left side of the disorder.

\begin{table}[!htp]
	\caption{Elements of $\mathbf{S}_{22}$ \label{tab:S22}}
	\begin{flushleft}
		\begin{tabular}{|l|l|l|l|l|l|}
			\hline
			$Inc(m)$&$Refl(n)$& \qquad \qquad \qquad $\mathbf{S}^{nm}_{22}$ \\ 
			\hline
			$pr$&$pr$  & $-\delta_{\text{kron}}(n,m)e^{-i z_R \left(k_z^{(m)}+k_z^{(n)}\right)} $ + \\ &&$2i\sqrt{k_z^{(n)}}\sqrt{k_z^{(m)}}e^{-i z_R \left(k_z^{(m)}+k_z^{(n)}\right)} G^{\mathbf{S}_{22}}_{n m}$   \\ 
			\hline
			$ev$&$ev$  & $ -\delta _{kron}(n,m)+-2\sqrt{\left|k_z^{(m)}\right|}\sqrt{\left|k_z^{(n)}\right|}G^{\mathbf{S}_{22}}_{n m}$\\ 
			\hline
			$pr$&$ev$  & $+2i\sqrt{k_z^{(m)}}\sqrt{\left|k_z^{(n)}\right|}e^{-i z_R k_z^{(m)}} G^{\mathbf{S}_{22}}_{n m}$   \\ 
			\hline
			$ev$&$pr$  & $-2\sqrt{\left|k_z^{(m)}\right|}\sqrt{k_z^{(n)}}e^{-i z_R k_z^{(n)}} G^{\mathbf{S}_{22}}_{n m}$   \\ 
			\hline
		\end{tabular}
	\end{flushleft}
\end{table}

\begin{figure}
	\includegraphics[width=0.35\textwidth]{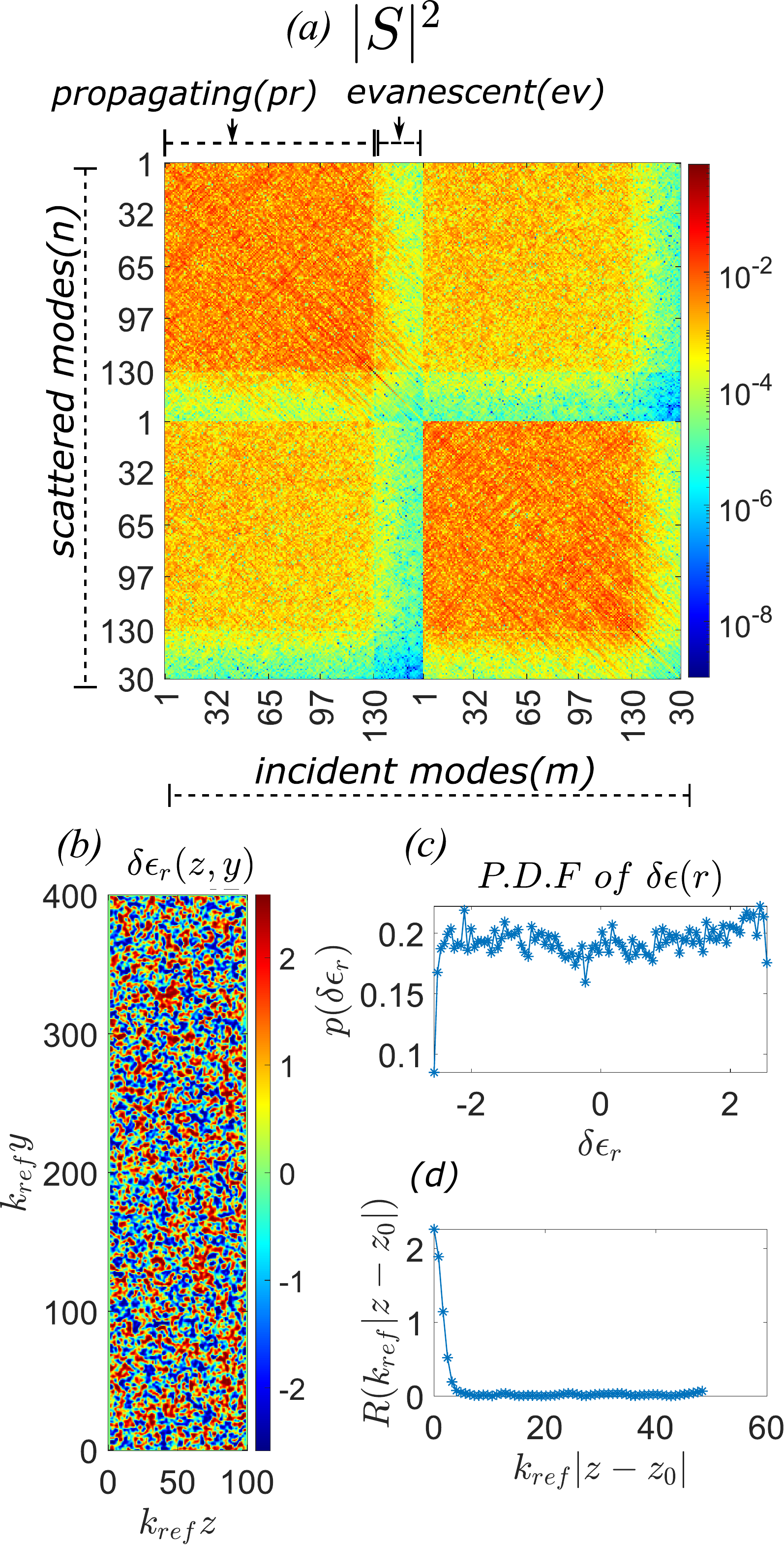}
	\caption{\label{fig:Smatrices} \textbf{$S$ matrix of a diffusive spatially correlated disorder with average transmission $\langle \tau^{pr,pr}_{21} \rangle = 0.198$}. \textbf{(a)} Squared magnitude of the $S$ matrix given in Eq.~(\ref{eq:Smatrix}) (plotted in $log$ scale) estimated using the generalized Fisher-Lee relations given in Tables \ref{tab:S11}, \ref{tab:S21}, \ref{tab:S22}, and \ref{tab:S12}. The number of propagating modes $(M_{pr})$ involved for wave incidence and scattering is 130 and the number of evanescent modes $(M_{ev})$ is 30 for the quasi-$1D$ geometry. As an example, the $\mathbf{S}_{11}$ portion of the $S$ matrix, contains both propagating and evanescent wave modes where the propagating modes appear first in the matrix (with respect to row and column indices) followed by the evanescent modes. \textbf{(b)} Spatially correlated disorder used for the estimation of the $S$ matrix. The disorder is represented by the dielectric-constant perturbation $\delta \epsilon_{r}(z,y)$, modeled as per the Supplementary Sec.~\ref{supp:corrdisorder}. \textbf{(c)} Probability density function of the disorder $\delta \epsilon_{r}(z,y)$ matrix elements such that the $var(\delta \epsilon_{r})=2.26$. \textbf{(d)} Spatial correlation function $R(k_{ref}|z-z_0|)$, defined in Eq.~(\ref{eq:spatcorrfun}) of Supplementary Sec.~\ref{supp:corrdisorder}, is plotted where the dimensionless spatial correlation length $k_{ref}l_c$ along $z$ axis, is 2.01.}
\end{figure}

\begin{table}[!htp]	
	\caption{Elements of $\mathbf{S}_{12}$ \label{tab:S12}}
	\begin{flushleft}
		\begin{tabular}{|l|l|l|l|l|l|}
			\hline
			$Inc(m)$&$Trans(n)$& \qquad \qquad \qquad $\mathbf{S}^{nm}_{12}$ \\ 
			\hline
			$pr$&$pr$  & $ 2i\sqrt{k_z^{(m)}}\sqrt{k_z^{(n)}}e^{-i z_R k_z^{(m)}} G^{\mathbf{S}_{12}}_{n m}$   \\ 
			\hline
			$ev$&$ev$  & $-2\sqrt{\left|k_z^{(m)}\right|}\sqrt{\left|k_z^{(n)}\right|}  G^{\mathbf{S}_{12}}_{n m}$\\ 
			\hline
			$pr$&$ev$  & $2i\sqrt{k_z^{(m)}}\sqrt{\left|k_z^{(n)}\right|}e^{-i z_R k_z^{(m)}} G^{\mathbf{S}_{12}}_{n m}$   \\ 
			\hline
			$ev$&$pr$  & $-2\sqrt{\left|k_z^{(m)}\right|}\sqrt{k_z^{(n)}} G^{\mathbf{S}_{12}}_{n m}$ \\ 
			\hline
		\end{tabular}
	\end{flushleft}
\end{table}
Tables \ref{tab:S11},\ref{tab:S21},\ref{tab:S22}, and \ref{tab:S12} summarize the main results of this section. The $S$ matrix, estimated numerically using the analytic expressions provided in these tables, is shown in FIG.~\ref{fig:Smatrices}(a). 

\section{Verification of the generalized reciprocity  and unitarity properties}
\label{sec:uniandrecivalidation}

In this section, the generalized reciprocity relations \cite{carminati2000reciprocity,byrnes2021symmetry} are summarized for the Landauer-waveguide geometry by comparing the expressions  for $\mathbf{S}_{11},\mathbf{S}_{21},\mathbf{S}_{22},~\text{and}~\mathbf{S}_{12}$ between Tables \ref{tab:S11},\ref{tab:S21},\ref{tab:S22}, and \ref{tab:S12}, respectively. By comparing the expressions given in the first two rows of all the tables, where $m$ and $n$ are either both propagating or both evanescent, we obtain $\mathbf{S}^{nm}_{11}=\mathbf{S}^{mn}_{11}$, $\mathbf{S}^{nm}_{22}=\mathbf{S}^{mn}_{22}$, and $\mathbf{S}^{nm}_{21}=\mathbf{S}^{mn}_{12}$. This arises from the reciprocity symmetry of the perturbed Green's function, such that $G(r, r_0) = G(r_0, r)$, implying $G_{nm} = G_{mn}$. On the other hand, when $m\leq M_{pr}$ and $n>N_{pr}$, the relations $\mathbf{S}^{mn}_{11}=i \mathbf{S}^{nm}_{11}$, $\mathbf{S}^{mn}_{22}=i \mathbf{S}^{nm}_{22}$, 
$\mathbf{S}^{mn}_{12}=i\mathbf{S}^{nm}_{21}$, and $\mathbf{S}^{mn}_{21}=i\mathbf{S}^{nm}_{12}$ hold, requiring the inclusion of a factor of $i$. Similarly,  when $m> M_{pr}$ and $n\leq N_{pr}$, we have $\mathbf{S}^{mn}_{11}=-i \mathbf{S}^{nm}_{11}$, $\mathbf{S}^{mn}_{22}=-i \mathbf{S}^{nm}_{22}$, 
$\mathbf{S}^{mn}_{12}=-i\mathbf{S}^{nm}_{21}$, and $\mathbf{S}^{mn}_{21}=-i\mathbf{S}^{nm}_{12}$, necessitating the inclusion of a factor of $-i$. In order to numerically verify these relations, the matrix representation of the same relations are used which is explained as the following. For the matrix representation of the reciprocity and unitarity relations, the partitioned $S$ matrix ($\mathbf{S}_{parti}$) has been defined \cite{byrnes2021symmetry} for convenience. $\mathbf{S}_{parti}$ rearranges the $S$ matrix, defined in Eq.~(\ref{eq:Smatrix}), into the following block-matrix form :
\begin{figure}[!htp]
	\includegraphics[width=0.4\textwidth]{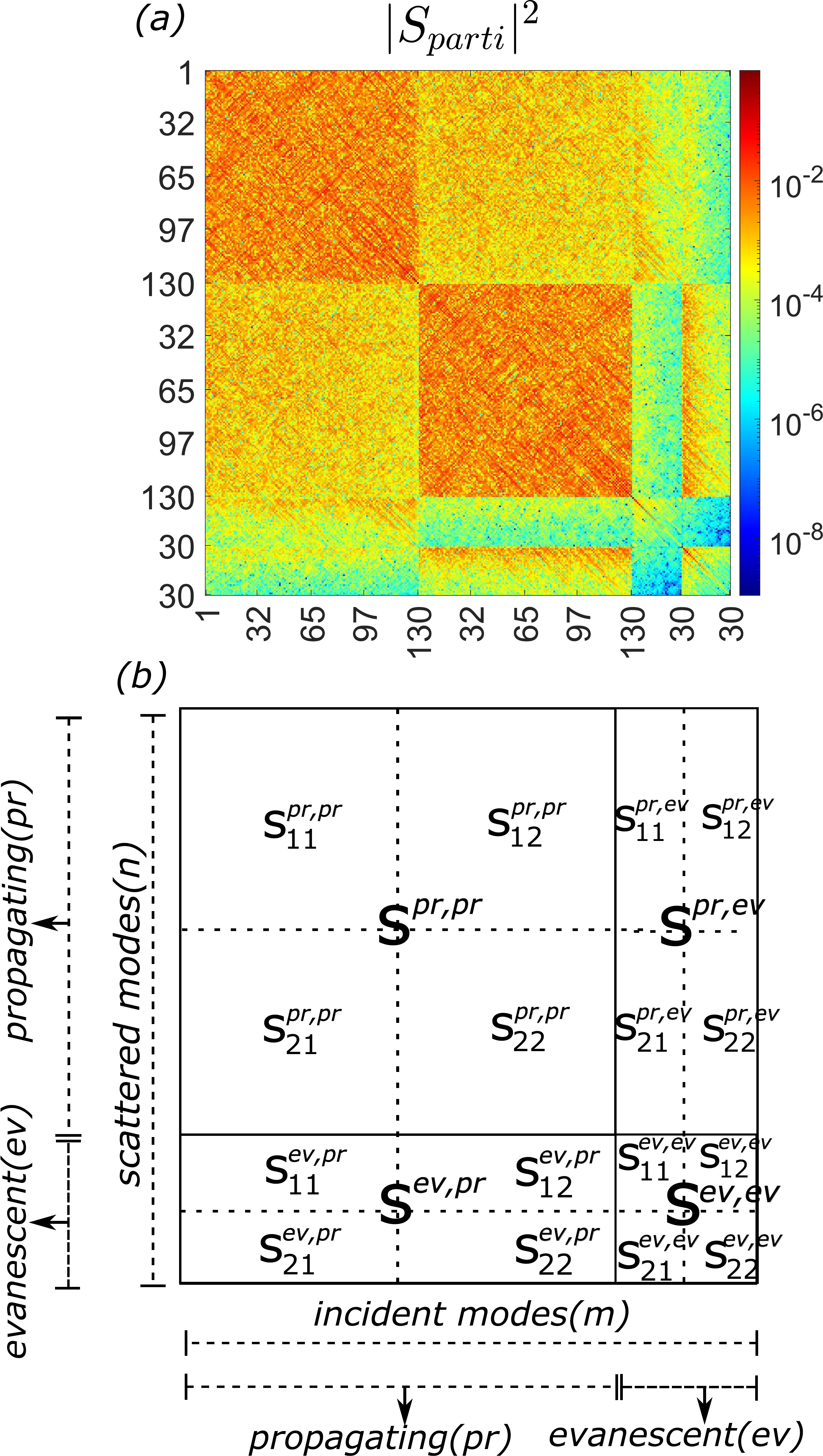}
	\caption{\label{fig:Spartitioned} \textbf{Partitioned $S$ matrix}. \textbf{(a)} Partitioned $S$ matrix, referred to as $S_{parti}$, which rearranges the numerically estimated $S$ matrix given in FIG.~\ref{fig:Smatrices}(a). The method by which the rearrangement is done is explained in Appendix \ref{sec:selectionmask}. \textbf{(b)} Schematic of various submatrices contained within the corresponding partitioned $S$ matrix. The superscripts ($pr : \text{propagating}$, $ev : \text{evanescent}$) on the various $S$ matrix components denote the nature of the scattered and the incident modes. The first superscript (on the left) denotes the nature of the scattered modes and the second superscript (on the right) refers to the nature of the incident modes.} 
\end{figure}

\begin{align}
	\left(
	\begin{array}{c}
		\mathbf{c}_{out}^{pr} \\
		\mathbf{c}_{out}^{ev} \\
	\end{array}
	\right)=\underbrace{\left[
		\begin{array}{cc}
			\mathbf{S}^{pr,pr} & \mathbf{S}^{pr,ev} \\
			\mathbf{S}^{ev,pr} & \mathbf{S}^{ev,ev}\\
		\end{array}
		\right]}_{\mathbf{S}_{parti}}\left(
	\begin{array}{c}
		\mathbf{c}_{inc}^{pr} \\
		\mathbf{c}_{inc}^{ev} \\
	\end{array}
	\right)\label{eq:Smatrixrelation}
\end{align}

where the terms in the superscripts of the submatrices, $pr$ and $ev$, represent propagating and evanescent modes, respectively. Such a partitioned $S$ matrix is shown in FIG.~\ref{fig:Spartitioned}(a), formed by rearranging the $S$ matrix given in FIG.~\ref{fig:Smatrices}(a). The first index in the superscript denotes the nature of the outgoing scattered modes, while the second index denotes the nature of the incident modes as shown in the FIG.~\ref{fig:Spartitioned}(b).  For example, $\mathbf{S}^{pr,ev}$ contains those elements of the partitioned $S$ matrix, corresponding to incident evanescent modes and outgoing propagating modes, grouped together as shown in the FIG.~\ref{fig:Spartitioned}(b). Each of the submatrices $\mathbf{S}^{pr,pr}$, $\mathbf{S}^{pr,ev}$, $\mathbf{S}^{ev,pr}$ and $\mathbf{S}^{ev,ev}$ can be further subdivided into their corresponding $\mathbf{S}_{11}$, $\mathbf{S}_{12}$, $\mathbf{S}_{21}$ and $\mathbf{S}_{22}$ counterparts. For example, $\mathbf{S}^{pr,ev}$ represents the block matrix form 
\begin{align}
	\mathbf{S}^{pr,ev}=\left[
	\begin{array}{cc}
		\mathbf{S}^{pr,ev}_{11} & \mathbf{S}^{pr,ev}_{12} \\
		\mathbf{S}^{pr,ev}_{21} & \mathbf{S}^{pr,ev}_{22} \\
	\end{array}
	\right]
	\label{eq:Smatrixindivparts}
\end{align} 

which is schematically shown in FIG.~\ref{fig:Spartitioned}(b). The four selection masks used to crop the $\mathbf{S}$ matrix and rearrange it to obtain the partitioned $\mathbf{S}_{parti}$ matrix components are shown and explained in FIG.~\ref{fig:Mask} of Appendix~\ref{sec:selectionmask}. Additionally, the coefficients $\mathbf{c}$ given in Eq.~(\ref{eq:Smatrixrelation}) associated with the incident (subscript $inc$) and outgoing (subscript $out$) scattered eigenmodes are also separated into propagating and evanescent modes.

Using the defined $\mathbf{S}^{pr,pr}$, $\mathbf{S}^{pr,ev}$, $\mathbf{S}^{ev,pr}$ and $\mathbf{S}^{ev,ev}$ matrices, 
the analytical forms of the generalized reciprocity relations explained in the beginning of this section, are numerically verified as shown in FIG.~\ref{fig:Reciprocity} and FIG.~\ref{fig:Reciprocity-error} in the Appendix \ref{sec:numericalreciproandunita}. Among these, the important reciprocity relations relevant to this paper are $\mathbf{S}_{12}^{pr,ev}=i \left(\mathbf{S}_{21}^{ev,pr}\right)^T$ and $\mathbf{S}_{12}^{pr,pr}=\left(\mathbf{S}_{21}^{pr,pr}\right){}^T$ where $T$ stands for matrix transpose.

Next, we verify the unitarity property of the $S$ matrix, which implies flux conservation. The compact form \cite{byrnes2021symmetry,gulyaev2005optical} of the generalized unitarity relation, $\mathbf{S}^{\dagger } _{parti}\mathbf{I}_{pr} \mathbf{S}_{parti} = \mathbf{I}_{pr} + i\left\{ \mathbf{S}^{\dagger }_{parti}\mathbf{I}_{ev} -\mathbf{I}_{ev} \mathbf{S}_{parti}\right\}$, is used for the numerical validation of the relation, as shown in FIG.~\ref{fig:Unitarity}(a) of the Appendix \ref{sec:numericalvalunitarity}. Here, $\mathbf{I}_{pr}$  is a matrix with unity only on the diagonal corresponding to the purely propagating portion of $\mathbf{S}_{parti}$ and zeros elsewhere, as shown in the FIG.~\ref{fig:Unitarity}(b). Similarly, $\mathbf{I}_{ev}$ contains unity elements only on the diagonal corresponding to the evanescent portion of $\mathbf{S}_{parti}$, as shown in FIG.~\ref{fig:Unitarity}(c). Additional unitarity properties involving modal coefficients $\mathbf{c}_{out}$ and $\mathbf{c}_{inc}$ is presented in Appendix \ref{sec:additionalunitarity}.

\section{Evanescent mode truncation, disorder correlation and numerical resolution}
\label{sec:truncno}
For numerically representing the analytical forms of the eigenmodes and the Green's functions within the computational code, we use numerical grids based on the finite difference method. These numerical grids have a finite grid cell size, $\Delta_z$ and $\Delta_y$. For the analytical expressions given in the paper to be exact on the grid, $\Delta_z$ and $\Delta_y$ would need to tend to zero, which is practically not feasible due to the memory and computational processing limitations. Therefore, $\Delta_z$ and $\Delta_y$ are chosen as a trade-off such that they are small enough to resolve wave states on the grids (especially evanescent eigenmodes), but not so small as to exhaust available memory and the processing capacity. In the limit, lim  $\Delta_z, \Delta_y \rightarrow 0$, the analytical forms of the wave states and the dispersion relation given in the paper are recovered. For ease of implementation in numeric computations, we set $\Delta_z =\Delta_y = \Delta$ and $\Delta$ is chosen to satisfy two conditions. The discretization step size $\Delta$ and the total number of transverse modes $M_{total}$ are chosen so as to satisfy both the numerical and physical resolution requirements. The first requirement follows from numerical sampling considerations. For an arbitrarily chosen $M_{ev}$, $M_{total}=M_{ev} + M_{pr}$ determines the largest transverse wavevector which is resolvable being $k_y^{\max} = {M_{total}\pi}/{W}$. Here, $M_{pr}\approx2\eta_{ref}W/\lambda_0$ is usually fixed for a given $W/\lambda_0$ ratio. To ensure adequate numerical resolution of the highest spatial frequency, $\Delta$ is chosen such that $k_y^{\max}\Delta \leq 1$ which constitutes a conservative sampling criterion. The second requirement is imposed by the scattering property of the disorder which sets $M_{ev}$ in the particular case when the disorder spatial correlation length $\ell_c$ goes to be a fraction of the wavelength. The spatial frequency of this subwavelength variation is of the order $k_y \sim 2\pi\eta_{ref}/\ell_c$. In order to resolve these fine spatial variations, $k_y^{\max} \geq {2\pi \eta_{ref}}/{\ell_c}$. In that case, the $M_{total}$ should be at least $2 \eta_{ref} W/\ell_c$ and $M_{ev}=2 \eta_{ref} W/\ell_c - M_{pr}$. In this scenario, $\Delta$ is upgraded such that $k_y^{\max} \Delta\leq1$. This also implies that smaller the $\ell_c$, finer discretization is needed which makes $\Delta$ smaller, increasing the $M_{total}$ and $M_{ev}$. This increases the computational memory and processing requirements, making simulations with shorter $\ell_c$ more computationally demanding. Therefore, users of the code can use these guidelines to manually estimate the required number of evanescent modes and choose an appropriate discretization based on their available memory and computational resources.

Once the suitable $M_{total}$ and $\Delta$ are chosen, the finite-difference-based numerical forms of the eigenmodes and Green's functions are constructed (refer to Supplementary Sec.~\ref{supp:discretization}). This is to ensure exact conservation of wave flux in the discrete sense. Specifically, the total flux injected by every eigenmode in the numerical form is exactly unity when evaluated as a discrete summation involving the current density, implying
$\sum_{y_j} J_z^{(m)}(z_i,y_j)\Delta_y = 1$. This ensures that conservation laws such as unitarity and reciprocity are satisfied exactly to the limit of numerical precision when the various numerical integrations described in this paper are evaluated as discrete summations. Such exact discrete flux conservation comes with the cost that the analytical dispersion relation, $\left(k^{(m)}_{y}\right)^2+\left(k^{(m)}_z\right)^2=k^2_{ref}$, is modified and results in a numerical dispersion relation (Eq.~(\ref{eq:disc_disp_rel1_ch1}) of Supplementary Sec.~\ref{supp:discretization}), altering the true form of the dispersion relation. This results in a small numerical dispersion that vanishes as $\Delta\rightarrow0$, which is typical of finite-difference methods.

\section{Focusing of an evanescent mode and universal background transmission}
\label{sec:focussing}
\begin{figure}
\includegraphics[width=0.45\textwidth]{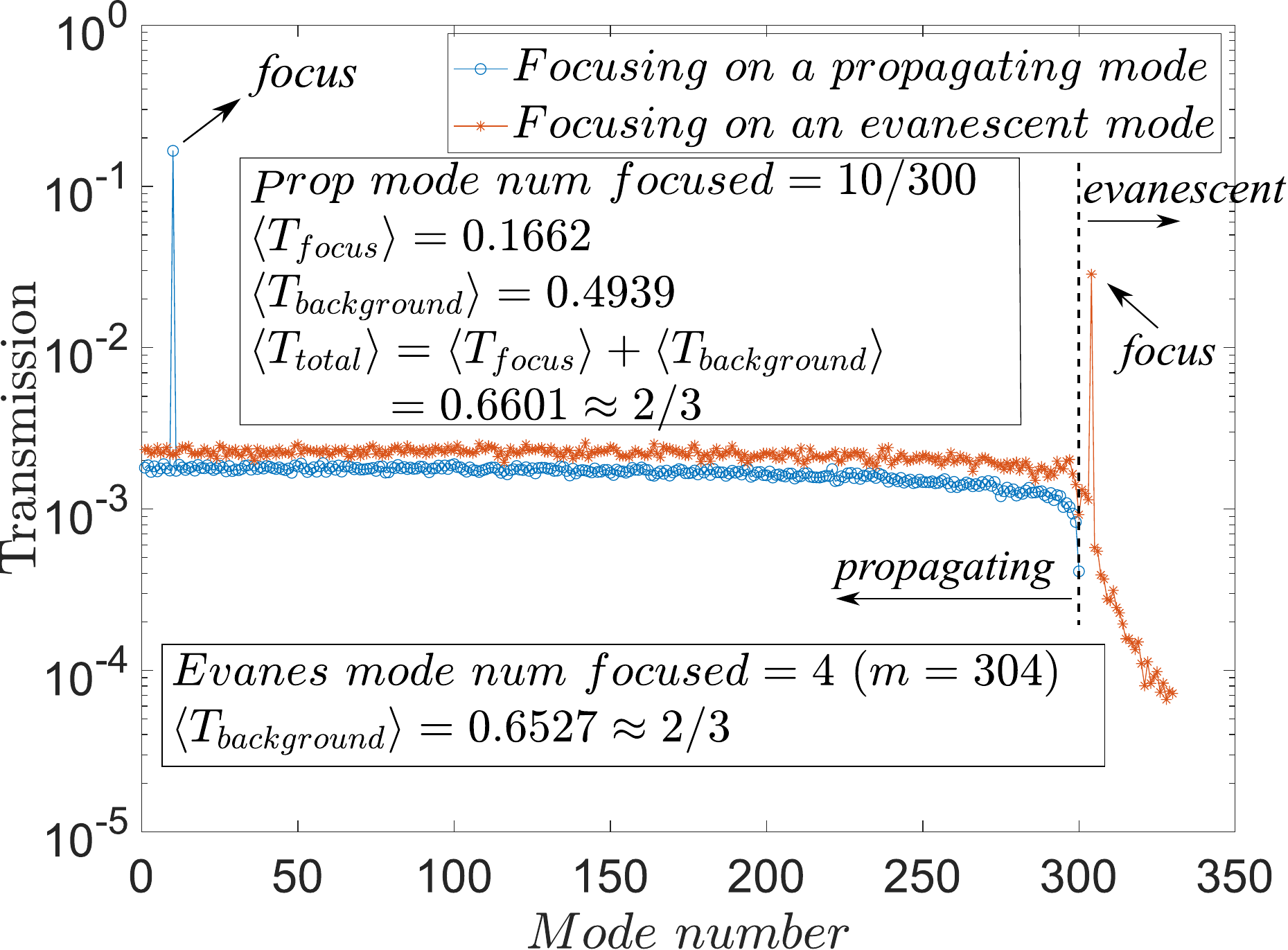} 
		\caption{\label{fig:optimal_transmission} \textbf{Ensemble averaged transmission while focusing via phase conjugation for a collection of cascaded disorders with $\langle \tau^{pr,pr}_{21}\rangle =0.1489$}. Ensemble averaged (involving 1000 disordered slabs) distribution of transmission values while focusing onto a propagating mode and an evanescent mode in two separate focusing scenarios. The optimal transmission $\langle T_{opt} \rangle$ for the focusing of propagating mode is $\langle T_{total} \rangle=\langle T_{focus} \rangle + \langle T_{background} \rangle=0.6601\approx 2/3$, which is a universal constant \cite{vellekoop2008universal} as expected. Here, $\langle \cdot\cdot\cdot \rangle$ denotes averaging over ensemble. $T_{focus}$ is the transmission only due to the focused mode and $\langle T_{background} \rangle$ the total transmission summed over all the modes other than the focus. For the focusing of the evanescent mode, $\langle T_{opt} \rangle $ is the background transmission $\langle T_{background} \rangle$ which is a universal constant such that $\langle T_{background} \rangle=0.6527 \approx 2/3$. The ensemble of generalized $S$ matrices is generated by the $S$ matrix cascading method explained in the supplementary Sec.~\ref{supple:cascading} and in the user manual for the computational codes.} 
\end{figure}

This section covers single-mode focusing onto an evanescent mode through the phase conjugation of the transmitted propagating mode generated by the same evanescent mode in a diffusive disorder. Let there be a single evanescent mode (indexed by $m'$) incident only from the right side onto the disorder. The incident evanescent mode is transmitted as a propagating wave traveling to the left side of the disorder. The transmitted propagating wave on the left side can be represented as $\mathbf{c}_{tr}^{L,pr}=\mathbf{S}_{12}^{pr,ev} \mathbbold{1}^{ev}_{m'}$ where  $\mathbbold{1}^{ev}_{m'}$ is a column vector selecting the $(m')^{th}$ evanescent mode incidence. For example, $\mathbbold{1}^{ev}_{m'=1}=(1,0,0...,0)^T$ selects the first evanescent mode. The flux-normalized phase conjugate of $\mathbf{c}_{tr}^{L,pr}$ is then given as $c_1 (\mathbf{c}_{tr}^{L,pr})^*$, which is incident on the left side of the disorder. Here, $c_1$ is used for the flux normalization, and its value depends on the $\left({m'}\right)^{th}$ evanescent mode used. The propagating phase conjugate wave incident from the left undergoes transmission on the right side, resulting in : 
\begin{subequations}
	\label{eq:allfocusing} 
	\begin{eqnarray}
	\mathbf{c}_{tr}^{R,ev}&=& \mathbf{S}_{21}^{ev,pr}   \left\{c_1\left(\mathbf{S}_{12}^{pr,ev}\right)^* \mathbbold{1}^{ev}_{m'}\right\},\label{eq:focusev}\\
	\mathbf{c}_{tr}^{R,pr}&=&\mathbf{S}_{21}^{pr,pr}   \left\{c_1\left(\mathbf{S}_{12}^{pr,ev}\right)^* \mathbbold{1}^{ev}_{m'}\right\}.\label{eq:focuspr} 
	\end{eqnarray}
\end{subequations}
Here $\mathbf{c}_{tr}^{R,ev}$ contains the complex coefficients for the evanescent component of the transmitted wave on the right side. Similarly,  $\mathbf{c}_{tr}^{R,pr}$ contains the propagating component, forming the transmitting background. Substituting the reciprocity property (discussed in Sec.~\ref{sec:uniandrecivalidation}) that $\mathbf{S}_{12}^{pr,ev}=i \left(\mathbf{S}_{21}^{ev,pr}\right)^T$ into Eq.~(\ref{eq:focusev}), we obtain : 
\begin{align}
&\mathbf{c}_{tr}^{R,ev}=-i c_1 \mathbf{S}_{21}^{ev,pr} {\left(\mathbf{S}_{21}^{ev,pr}\right)}^{\dagger}\mathbbold{1}^{ev}_{m'} \label{eq:focusevinten}
\end{align}
In Eq.~(\ref{eq:focusevinten}), the transmitted wave field peaks at the $(m')^{th}$ evanescent mode resulting in constructive inference leading to focusing as expected and known. Such a focussing scenario is shown in FIG.~\ref{fig:optimal_transmission}, where the focussing onto a propagating and evanescent mode are shown separately. 

For the remainder of the paper, we discuss the optimal transmission resulting from the background propagating wave formed during evanescent wave focusing. It is important to emphasize that the optimization is performed only over the propagating input modes. Although the target of the focusing is an evanescent mode on the right side of the disorder, the incident wave used for optimization is obtained by phase conjugation of the propagating field generated by that evanescent mode on the opposite side. Therefore, the optimization acts within the propagating subspace, while the evanescent focusing arises indirectly through the coupling between propagating and evanescent modes described by the generalized reciprocity relation. The optimal background transmission associated with the focusing of an evanescent mode is defined as $T_{opt}=\left({\mathbf{c}_{tr}^{R,pr}}\right)^{\dagger}{\mathbf{c}_{tr}^{R,pr}}$, where $\mathbf{c}^{R,pr}_{tr}$ is given in Eq.~(\ref{eq:focuspr}). Here, $T_{opt}$ corresponds to the total transmitted flux carried by the propagating modes on the right side, while focusing on an evanescent mode optimally, even though the focused evanescent mode by itself does not transmit flux. For diffusive disorders, the ensemble-averaged background transmission $\langle T_{opt} \rangle$ is observed to be 2/3, similar to the case of the focusing \cite{vellekoop2008universal} of propagating modes. The phenomenon can be justified as follows for diffusive disorders. 

The transmitted propagating mode $\mathbf{c}_{tr}^{L,pr}$ on the left, defined previously due to an incident evanescent mode on the right, can also be generated by an equivalent incident propagating mode as follows : 
\begin{align}
	\mathbf{c}_{tr}^{L,pr}&=\mathbf{S}_{12}^{pr,ev} \mathbbold{1}^{ev}_{m'}
	                     =c_0 \mathbf{S}_{12}^{pr,pr} \mathbf{b}_{12,m'}^{pr},
	\label{eq:equivalence}
\end{align}
where the incident propagating column vector $\mathbf{b}^{pr}_{12,m'}$ is flux normalized as $ \left(\mathbf{b}_{12,m'}^{pr}\right)^{\dagger }\mathbf{b}_{12,m'}^{pr}=1$, and $c_0$ is a coefficient satisfying Eq.~(\ref{eq:equivalence}) whose value depends on $m'$. In other words, the transmitted propagating wave after the disorder can be considered indistinguishable, whether it is generated by a single evanescent mode or by an equivalent propagating wave represented by $c_0 \mathbf{b}^{pr}_{12,m'}$ . This equivalent propagating wave, which yields the same transmitted propagating wave due to the given evanescent mode, is a linear combination of the propagating modes addressed by the propagating transmission matrix. Substituting Eq.~(\ref{eq:equivalence}) into Eq.~(\ref{eq:focuspr}) and using the reciprocity relation  $(\mathbf{S}_{12}^{pr,pr})^*={\mathbf{S}_{21}^{pr,pr}}^\dagger$, we obtain :

\begin{align}
&T_{opt}=\mathbf{c}_{tr}^{R,pr}{}^{\dagger }\mathbf{c}_{tr}^{R,pr}=\nonumber\\
&c_0^2 c_1^2  \left(\mathbf{b}_{12,m'}^{pr}\right){}^T \mathbf{S}_{21}^{pr,pr} \left(\mathbf{S}_{21}^{pr,pr}\right){}^{\dagger } \mathbf{S}_{21}^{pr,pr}  \left(\mathbf{S}_{21}^{pr,pr}\right){}^{\dagger } \left(\mathbf{b}_{12,m'}^{pr}\right)^* \nonumber \\
&=c_0^2  c_1^2  \left(\mathbf{b}_{12,m'}^{pr}\right)^T \mathbf{U}_{21}^{pr,pr} \left(\pmb{\tau}_{21}^{pr,pr}\right){}^2  \mathbf{U}_{21}^{pr,pr\dagger} \left(\mathbf{b}_{12,m'}^{pr}\right)^*,
\label{eq:evanesfoctrans}
\end{align}
where the Singular Value Decomposition (S.V.D) of $\mathbf{S}^{pr,pr}_{21}=\mathbf{U}^{pr,pr}_{21}\sqrt{\pmb{\tau}^{pr,pr}_{21}}{\mathbf{V}^{pr,pr}_{21}}^\dagger$ is utilized. Here, $\pmb{\tau}^{pr,pr}_{21}$ is the diagonal matrix containing the eigenchannel transmission coefficients. Next, the ensemble average of $T_{opt}$, denoted as $\langle T_{opt} \rangle$, is evaluated. For ensemble averaging, the isotropy \cite{mello1990averages,mello1991maximum} assumption is applied. This assumption considers $\left(\pmb{\tau}^{pr,pr}_{21}\right)^2$ and the remaining terms in Eq.~(\ref{eq:evanesfoctrans}) to be statistically uncorrelated during ensemble averaging. Since $\pmb{\tau}^{pr,pr}_{21}$ obeys the bimodal distribution for diffusive media, $\langle T_{opt} \rangle$ has the universal value of 2/3 similar to the case of focusing \cite{vellekoop2008universal} of propagating modes. The associated proof, involving the averaging over unitary groups, is provided in Appendix~\ref{sec:proof2by3}. The relation $\langle T_{opt} \rangle = 2/3$ for evanescent mode focusing is also numerically validated using the developed code, as shown in FIG.~\ref{fig:optimal_transmission}. Here, the ensemble averaging is performed by generating an ensemble of $S$ matrices using the $S$ matrix cascading method \cite{jin2014new} (see the supplementary Sec.~\ref{supple:cascading}) by randomly shuffling the cascading order. For further details on the numerical validation, refer to  the user manual.    

\section{Conclusion}
\label{sec:conclusion}
The analytical description and the associated numerical modeling for estimating the generalized $S$ matrix are presented from the perspective of classical wave scattering in the Landauer-waveguide geometry. Although the Landauer-waveguide geometry is predominantly used in the context of quantum transport, classical wave scattering methods are intentionally adopted in this paper to make the mesoscopic wave transport theory more accessible to researchers from engineering domains. The authors believe that the presented method offers analytical tractability and generality through the use of Green's functions, despite the fact that the Green's function perturbation method is not the most memory-efficient computational approach among the available algorithms for $S$ matrix estimation. For a numerically efficient strategy for $S$ matrix estimation, readers may refer to the recent work by Lin et al. \cite{lin2022fast}. The reduced memory efficiency associated with the Green's function method is mitigated in this paper through the incorporation of the $S$ matrix cascading method. In this approach, the generalized $S$ matrices of thinner layers are cascaded to obtain the effective $S$ matrix of a thicker layer formed by the cascading of these thinner layers. This allows the analysis of a thicker sample to be broken down into a cascade of thinner samples, thereby overcoming the computational burden of solving a large problem in one step. Furthermore, shuffling the order of $S$ matrices during cascading enables the generation of disorder ensembles, facilitating the estimation of various averages. Since the Green's function perturbation method is employed, starting from the analytical expression for the free-space Green's function with a truncated number of evanescent modes, precise numerical control over the evanescent modes under consideration is achieved. Finally, and most importantly, the usefulness of the model in wavefront-shaping applications, specifically for the focusing of an evanescent mode, is demonstrated. The universal background transmission associated with the focusing of an evanescent mode is also discussed. The authors believe that these results and the accompanying code may be of interest to wavefront-shaping researchers studying transport phenomena in general.

\section*{Code repository}
The open-source (MIT license) MATLAB\textsuperscript{\textregistered} codes \cite{Raju_Matlab_code_packages_2024} used for modeling the results are hosted in Github  \small{\texttt{https://github.com/michaelraju/Generalized-S-Matrix}}. The same repository is also hosted in Zenodo at \small{\texttt{https://doi.org/10.5281/zenodo.10963078}}. The saved run data associated with the code packages are  also hosted at \texttt{https://doi.org/10.5281/zenodo.10960970}. Users can download the saved run data from Zenodo and place it appropriately within the local Github repository to regenerate all the presented results without performing a new computational run from scratch. Alternatively, users can perform a new computational run initializing a new disorder,  setting the appropriate flag described in the user manual. The user manual, included in the repository, provides a description of the file structure of the code packages and instructions for their use. The computational simulations were performed on a desktop PC with a configuration of 64 GB of RAM and an eight-core processor (AMD Ryzen 7 2700X). This represents a modest computational setup without the use of specialized high-performance computing resources. Using this configuration, the total runtime required to generate the results presented in this paper was approximately 3.5 hours.
\section*{Author contributions}
M.R. derived the analytical and numerical descriptions of the methodologies presented, and developed the associated simulation code packages. B.J. and S.A.E. supervised M.R's PhD project. All authors were involved in the discussion, validation and interpretation of the results. M.R wrote the manuscript and the supplementary documents with support from the other authors.
\begin{acknowledgments}
We would like to thank Taighde \'Eireann~--~Research Ireland for funding this research through S.A.E.'s Professorship Grant ``Novel applications and techniques for in-vivo optical imaging and spectroscopy'' under grant numbers SFI/15/RP/2828 and SFI/22/RP-2TF/10293. 
\end{acknowledgments}

\appendix
\begin{figure}[t]
	\includegraphics[width=0.5\textwidth]{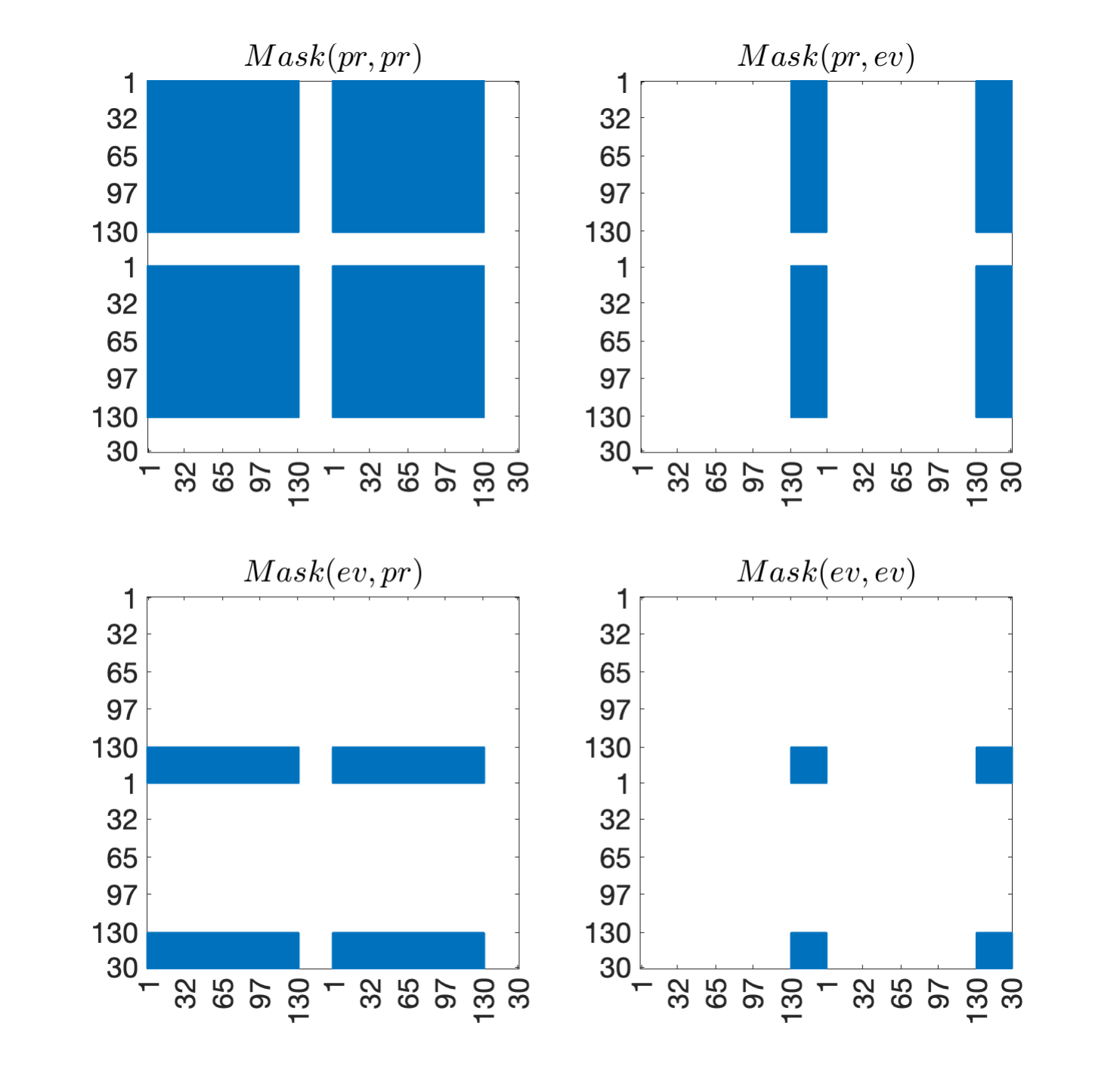}
	\caption{\label{fig:Mask} \textbf{Plotting the sparsity of the selection mask matrix $\mathbf{Mask}$ used for generating the partitioned $S$ matrix, $\mathbf{S}_{parti}$}. Blue colour stands for unity and white for zeros. This mask is used to re-arrange the $S$ matrix given in FIG.~\ref{fig:Smatrices}(a) to form the partitioned $S$ matrix ($\mathbf{S}_{parti}$) based on \cite{byrnes2021symmetry}. As an example, $\mathbf{Mask}(ev,pr)$ selects that part of the $S$ matrix which contains only modes such that the incident modes are propagating and the scattered being evanescent as explained in Sec.~\ref{sec:uniandrecivalidation}.} 
\end{figure}

\section{Selection masks for estimating partitioned S matrix}
\label{sec:selectionmask}
The selection masks used to rearrange the $S$ matrix to form the partitioned $S$ matrix, $S_{parti}$ (shown in FIG.~\ref{fig:Spartitioned}) are given in FIG.~\ref{fig:Mask}. These selection masks, $\mathbf{Mask}(pr,pr)$, $\mathbf{Mask}(ev,pr)$, $\mathbf{Mask}(ev,ev)$ and $\mathbf{Mask}(pr,ev)$, have the same dimensions as the $S$ matrix and contains unity (shown in blue) and zeros (shown in white) as their elements. For instance, $\mathbf{Mask}(pr,pr)$, shown in FIG.~\ref{fig:Mask}, has unity elements when both the incident and scattered modes are propagating. Taking the  Hadamard (element-wise) product of the S matrix with $\mathbf{Mask}(pr,pr)$ yields $\mathbf{S}\odot \mathbf{Mask}(pr,pr)$, which isolates the purely propagating contribution of the S matrix while setting to zero the remaining contribution involving evanescent modes. By discarding the evanescent portion of $\mathbf{S}\odot \mathbf{Mask}(pr,pr)$ (which was set to zero) and reshaping the matrix to retain only the propagating contribution, we obtain the $\mathbf{S}^{pr,pr}_{parti}$ component of the partitioned S matrix, as given in Eq.~\ref{eq:Smatrixrelation}. Similarly, $\mathbf{Mask}(ev,pr)$ extracts the portion of the $S$ matrix where the incident modes are propagating and the scattered modes are evanescent, yielding $\mathbf{S}^{ev,pr}_{parti}$. Analogously, $\mathbf{Mask}(ev,ev)$ extracts the purely evanescent contribution of the $S$ matrix, while $\mathbf{Mask}(pr,ev)$ extracts the contribution where the incident modes are evanescent and the scattered modes are propagating, yielding $\mathbf{S}^{pr,ev}_{parti}$. Using these four masks, $\mathbf{S}_{parti}$ is constructed. 

\section{Numerical validation of the reciprocity relations}
\label{sec:numericalreciproandunita}
\begin{figure}[!htp]
	\includegraphics[width=0.48\textwidth]{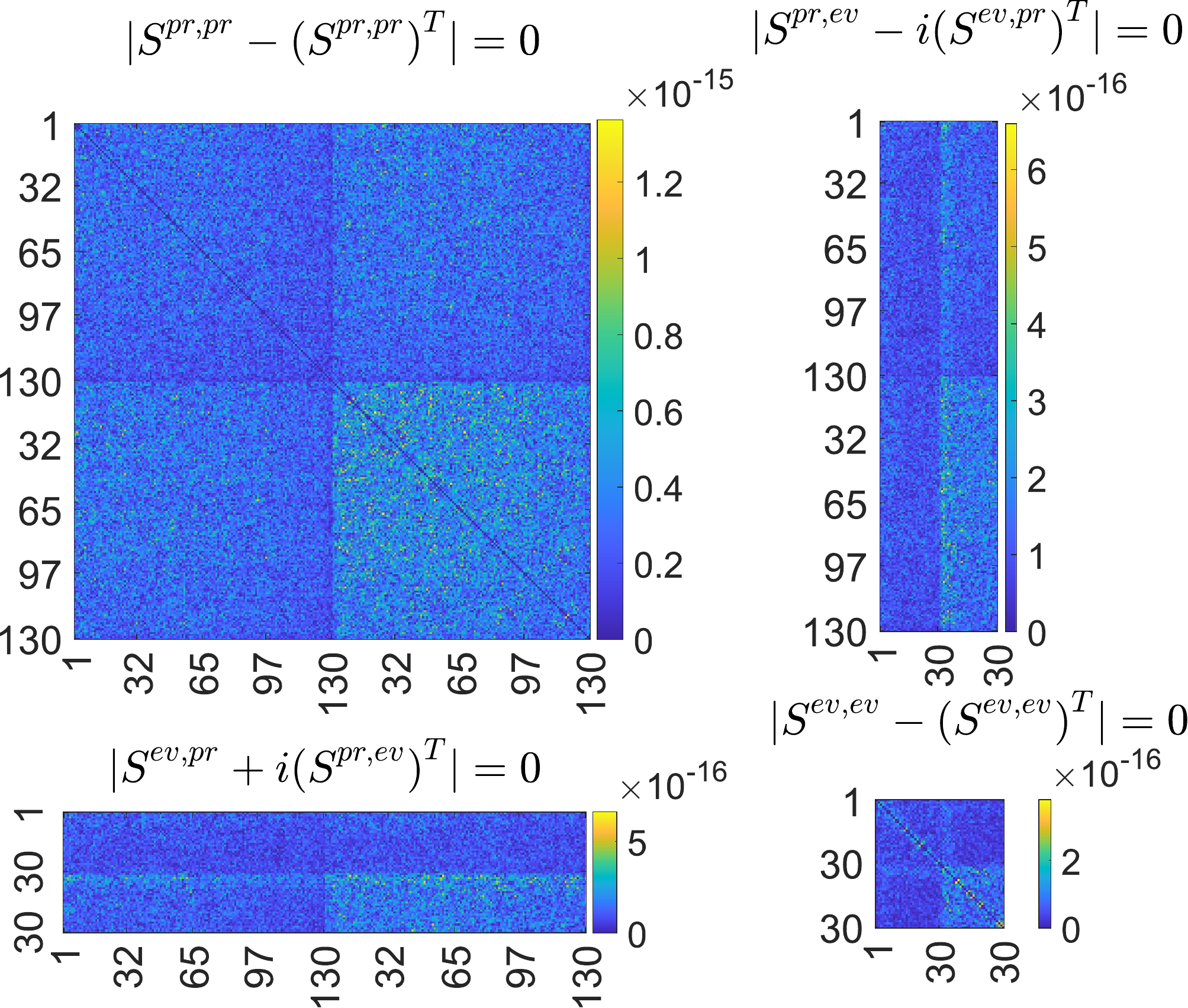}
	\caption{\label{fig:Reciprocity} \textbf{Numerical validation of the generalized reciprocity relations.} Plotting the magnitude of the numerically estimated generalized reciprocity relations in terms of $\mathbf{S}_{parti}$. The modulus terms on the L.H.S of the reciprocity relations are plotted whose magnitudes are of the order of $10^{-15}$, approximately equal to zero. $T$ stands for matrix transpose.} 
\end{figure}

\begin{figure}[!htp]
	\includegraphics[width=0.5\textwidth]{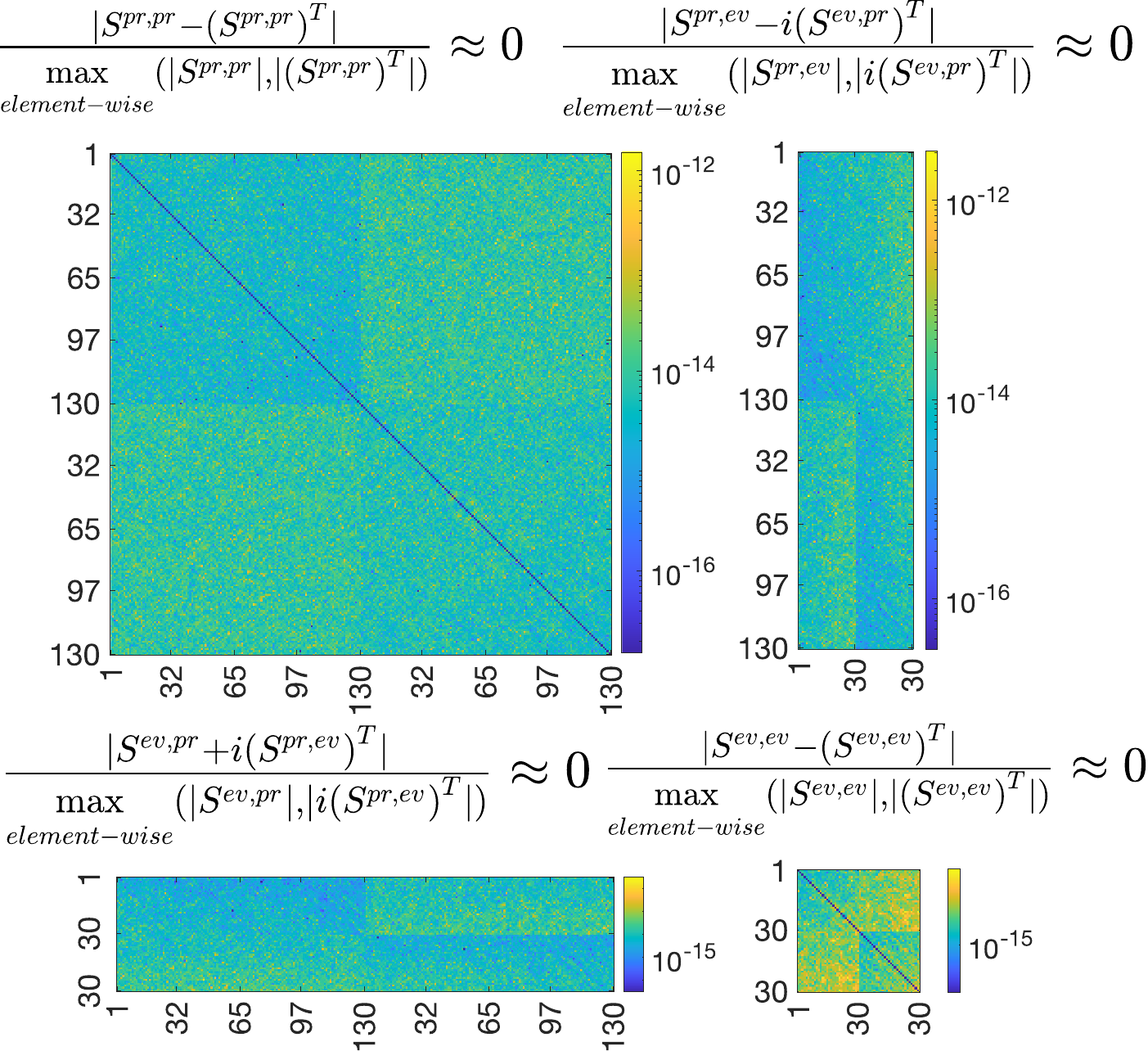}
	\caption{\label{fig:Reciprocity-error} \textbf{Relative error of the numerically estimated generalized reciprocity relations}. Relative errors are plotted as matrices and the $\max\limits_{element-wise}$ in the denominator denotes the element-wise maximum and all operations are understood in an entry-wise sense. Maximum relative error is of the order of $10^{-12}$.} 
\end{figure}

The generalized reciprocity relations are first validated in an absolute sense, as shown in Fig.~\ref{fig:Reciprocity}, where the absolute differences between various submatrices of the partitioned scattering matrix $S$ are displayed. These differences are found to be on the order of $10^{-15}$, which corresponds to the numerical precision limit of double-precision floating-point arithmetic. To further quantify the accuracy of the reciprocity relations, we evaluate the corresponding relative error, which is presented in Fig.~\ref{fig:Reciprocity-error}. For example, the relative error matrix associated with the reciprocity condition of $\mathbf{S}^{pr,pr}$ is defined as
\begin{align}
	\frac{\big|\mathbf{S}^{pr,pr}-\left(\mathbf{S}^{pr,pr}\right)^T\big|}
	{\max\limits_{element-wise}\!\left(\big|\mathbf{S}^{pr,pr}\big|,\big|\left(\mathbf{S}^{pr,pr}\right)^T\big|\right)},
	\label{eq:relative_error_prpr}
\end{align}
where $\max\limits_{element-wise}$ denotes the element-wise maximum and all operations are understood in an entry-wise sense. The resulting relative error is found to be on the order of $10^{-12}$, as shown in Fig.~\ref{fig:Reciprocity-error}, confirming that the generalized reciprocity relations are satisfied to within the limits of numerical precision.

\section{Numerical validation of the unitarity relations}
\label{sec:numericalvalunitarity}
\begin{figure}[t]
	\includegraphics[width=0.4\textwidth]{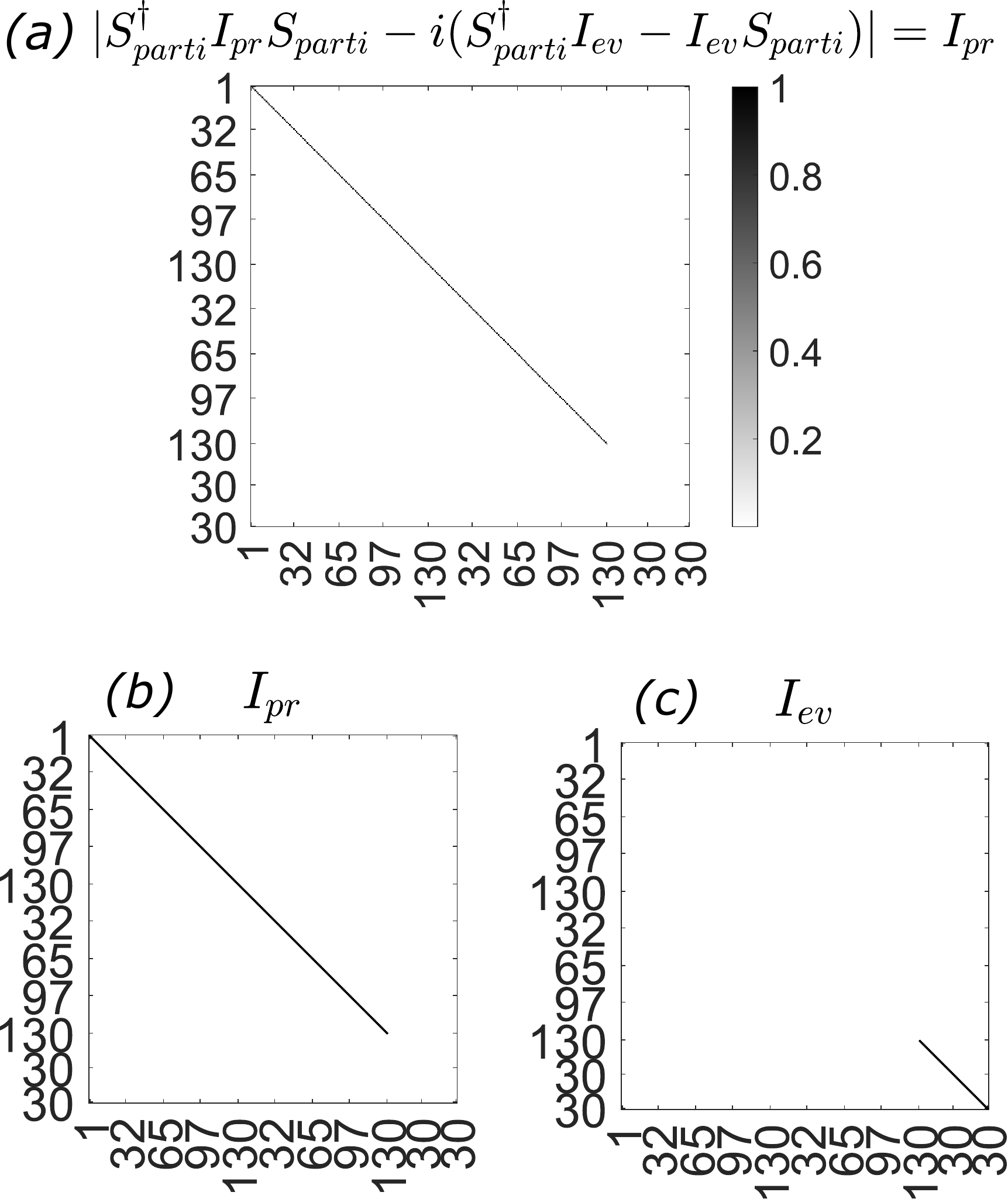}
	\caption{\label{fig:Unitarity} \textbf{Numerical validation of the generalized unitarity relations.} \textbf{(a)} Plotting the magnitude of the numerically estimated generalized unitarity relation based on the analytical result given in \cite{byrnes2021symmetry,gulyaev2005optical}. The modulus term on the L.H.S of the unitarity relation is plotted. The non-zero diagonal elements inside the modulus term on the L.H.S (shown in black colour) are all unities and the magnitude of all the other elements are of the order of $10^{-13}$, approximated to zero.  $\dagger$ stands for the conjugate transpose.   \textbf{(b)} Plotting the sparsity of the matrices $\mathbf{I}_{pr}$ and \textbf{(c)} $\mathbf{I}_{ev}$ defined in relation to the $\mathbf{S}_{parti}$ used in the unitarity relation (black colour stands for unity and white for zeros).} 
\end{figure}

The compact form \cite{byrnes2021symmetry,gulyaev2005optical} of the generalized unitary relation, $\mathbf{S}^{\dagger } _{parti}\mathbf{I}_{pr} \mathbf{S}_{parti} = \mathbf{I}_{pr} + i\left\{ \mathbf{S}^{\dagger }_{parti}\mathbf{I}_{ev} -\mathbf{I}_{ev} \mathbf{S}_{parti}\right\}$,  is the combination of the following relations :  $\left(\mathbf{S}^{pr,pr}\right)^{\dagger }\mathbf{S}^{pr,pr}=\mathbf{I}$,  $\left(\mathbf{S}^{pr,pr}\right)^{\dagger } \mathbf{S}^{pr,ev} =i \left(\mathbf{S}^{ev,pr}\right)^{\dagger }$, and 
$\left(\mathbf{S}^{pr,ev}\right)^{\dagger }\mathbf{S}^{pr,ev}=i \left\{\left(\mathbf{S}^{ev,ev}\right)^{\dagger }-\mathbf{S}^{ev,ev}\right\}$. Such a compact form is used to verify the unitary property for the numerically estimated partitioned $S$ matrix as given in FIG. ~\ref{fig:Unitarity}(a). The non-zero diagonal elements inside the modulus term on the L.H.S (shown in black colour) are all unities and the magnitude of all the other elements are of the order of $10^{-13}$, close to zero. $\mathbf{I}_{pr}$ and $\mathbf{I}_{ev}$ are given in FIG.~\ref{fig:Unitarity}(b) and FIG.~\ref{fig:Unitarity}(c). For additional numerical verification involving the expanded forms of the compact unitarity relation, refer to supplementary FIG.~\ref{fig:additionalunitarity} in supplementary Sec.~\ref{supple:additionalunitarity}.

\section{Additional unitarity properties}
\label{sec:additionalunitarity}
Flux conservation in terms of the modal coefficients given in Eq.~(\ref{eq:Smatrixrelation}) is explained next. The complex vector $\mathbf{c}_{inc}$ and  $\mathbf{c}_{out}$, with their elements being the complex modal coefficients defined before, are defined as follows : 
\begin{align}
	\mathbf{c}_{inc}=\left(
	\begin{array}{c}
		\mathbf{c}_{inc}^{+,L} \\
		\mathbf{c}_{inc}^{-,R} \\
	\end{array}
	\right)~\text{and}~
	\mathbf{c}_{out}=\left(
	\begin{array}{c}
		\mathbf{r}_{out}^{-,L} \\
		\mathbf{t}_{out}^{+,R} \\
	\end{array}
	\right)
\end{align}
where $\mathbf{c}_{inc}^{+,L}$ is the column vector containing the modal coefficients of the right going $(+)$ incident wave incident on $\Gamma_{L}$. Similarly, the column vector $\mathbf{c}_{inc}^{-,R}$ contains the modal coefficients of the left going $(-)$ incident wave incident on $\Gamma_{R}$. Furthermore, $\mathbf{r}_{out}^{-,L}$ is the column vector containing the modal coefficients of the outgoing wave emerging from $\Gamma_L$, due to the wave incidence from both the left and right sides of the disorder. Likewise, $\mathbf{t}_{out}^{+,R}$ is the column vector containing the modal coefficients of the outgoing wave emerging from $\Gamma_R$ due to the wave incidence from both the left and right sides of the disorder. The propagating and evanescent components of $\mathbf{c}_{inc}$ can be separated into $\mathbf{c}^{pr}_{inc}$ and $\mathbf{c}^{ev}_{inc}$, respectively, (which are used in the Eq.~(\ref{eq:Smatrixrelation})), as follows : 
\begin{align}
	\mathbf{c}_{inc}^{pr}=\left(
	\begin{array}{c}
		\mathbf{c}_{inc}^{+,L,pr} \\
		\mathbf{c}_{inc}^{-,R,pr} \\
	\end{array}
	\right)~\text{and}~
	\mathbf{c}_{inc}^{ev}=\left(
	\begin{array}{c}
		\mathbf{c}_{inc}^{+,L,ev} \\
		\mathbf{c}_{inc}^{ -,R,ev} \\
	\end{array}
	\right)
	\label{eq:split1}
\end{align}
Similarly, the propagating and evanescent components of $\mathbf{c}_{out}$ can be separated into $\mathbf{c}^{pr}_{out}$ and $\mathbf{c}^{ev}_{out}$. 

The flux conservation method presented in \cite{carminati2000reciprocity,byrnes2021symmetry} can be applied onto the modal coefficients. It states that the incident flux, $F_{inc}=\int_0^W \mathfrak{Im}\left\{\tilde{E}^{*}_{inc} \frac{\partial}{\partial z} \tilde{E}_{inc}\right\} dy$, must equal the outgoing flux, $F_{out}=\int_0^W \mathfrak{Im}\left\{\tilde{E}^{*}_{out} \frac{\partial}{\partial z} \tilde{E}_{out}\right\} dy$, such that $F_{inc}=F_{out}$. This leads to the following generalized flux conservation relation for the Landauer-waveguide geometry, 
\begin{align}
	\left(\mathbf{c}_{inc}^{pr}\right)^{\dagger }\mathbf{c}_{inc}^{pr}-\left(\mathbf{c}_{out}^{pr}\right)^{\dagger }\mathbf{c}_{out}^{pr} =i\left\{ \left(\mathbf{c}_{inc}^{ev}\right)^{\dagger}\mathbf{c}_{out}^{ev} - \left(\mathbf{c}_{out}^{ev}\right)^{\dagger }\mathbf{c}_{inc}^{ev} \right\}
	\label{eq:genFluxconserv}
\end{align}
This obtained relation is the discrete version (for the Landauer-waveguide geometry) of the relation given in Eq. (11) of \cite{carminati2000reciprocity}  and Eq.~(17) of \cite{byrnes2021symmetry}, which were defined for a different geometry involving a continuous angular spectrum representation. The obtained relation generalizes the well-known flux conservation involving only propagating modes, given as $ \left(\mathbf{c}_{inc}^{pr}\right)^{\dagger }\mathbf{c}_{inc}^{pr}= \left(\mathbf{c}_{out}^{pr}\right)^{\dagger }\mathbf{c}_{out}^{pr}$. For the physical interpretation of the relation given in Eq.~(\ref{eq:genFluxconserv}) in terms of counter-propagating interference of evanescent modes (on the R.H.S.), one may refer to the original proposition in \cite{carminati2000reciprocity} and the discussions in \cite{barabanenkov2007energy,byrnes2021symmetry}.

In the case of purely-propagating incident and scattered modes, where the incident wave modes exist only on the left side of the disorder, a portion of the relation $\left(\mathbf{S}_{parti}^{pr,pr}\right)^{\dagger }\mathbf{S}_{parti}^{pr,pr}=\mathbf{I}$ leads to the well-known trace formula \cite{fisher1981relation} for transmission and reflection averaged over different modes, $T_{pr}=\left({1}/{M_{pr}} \right)\mathbf{Tr}\left\{ \left(\mathbf{S}_{21}^{pr,pr}\right)^{\dagger }\mathbf{S}_{21}^{pr,pr}\right\}$ and  $R_{pr}=\left({1}/{M_{pr}}\right)  \mathbf{Tr}\left\{ \left(\mathbf{S}_{11}^{pr,pr}\right)^{\dagger }\mathbf{S}_{11}^{pr,pr}\right\}$ such that $T_{pr}+R_{pr}=1$. Here, $\mathbf{Tr}$ denotes the trace of the matrix. For a purely evanescent incident wave, the scenario differs, as described below. In the case of incident evanescent wave modes (existing only on the left side of the disorder) and using the generalized unitarity relation $\left(\mathbf{S}^{pr,ev}_{parti}\right)^{\dagger }\mathbf{S}^{pr,ev}_{parti}=i \left\{\left(\mathbf{S}^{ev,ev}_{parti}\right)^{\dagger }-\mathbf{S}^{ev,ev}_{parti}\right\}$, the following holds : 
\begin{align}
	&{ \frac{1}{M_{ev}}\mathbf{Tr}\left\{{\mathbf{S}_{21}^{pr,ev}}^{\dagger }\mathbf{S}_{21}^{pr,ev}\right\}} + 
	\frac{1}{M_{ev}} \mathbf{Tr} \left\{{\mathbf{S}_{11}^{pr,ev}}^{\dagger }\mathbf{S}_{11}^{pr,ev}\right\} \nonumber\\
	&=\frac{2}{M_{ev}}  \mathfrak{Im} \pmb{\left\{\vphantom{\frac{1}{2}}\right.} \mathbf{Tr} \{\mathbf{S}_{11}^{ev,ev}\} \pmb{\left.\vphantom{\frac{1}{2}}\right\}}
	\label{eq:evanesincidencetransmission1}
\end{align}
This can be considered as the optical theorem for the Landauer-waveguide geometry involving evanescent wave incidence, where the imaginary part of the sum of back-scattered evanescent reflection coefficients, along the same direction of the incident evanescent modes, determine the outgoing propagating wave intensity. 

\section{Proof that $\langle T_{opt} \rangle = 2/3$ for evanescent wave focusing}
\label{sec:proof2by3}
The goal of this section is to prove that for diffusive disorders, the ensemble average of $T_{opt}$ given in Eq.~(\ref{eq:evanesfoctrans}),
\begin{align}
\langle T_{opt} \rangle&=\langle c_0^2  c_1^2  \left(\mathbf{b}_{12,m'}^{pr}\right)^T \mathbf{U}_{21}^{pr,pr} \left(\pmb{\tau}_{21}^{pr,pr}\right){}^2  \mathbf{U}_{21}^{pr,pr\dagger} \left(\mathbf{b}_{12,m'}^{pr}\right)^* \rangle \nonumber \\
&=2/3.
\label{eq:evanesfocderive}
\end{align}
Using summation indices, the above equation can also be represented as 
\begin{align}
	&\langle T_{opt} \rangle=\langle c_1^2 c_0^2 \sum _{m_1} \sum _{m_2} \sum _{m_3} \left\{\mathbf{b}_{12,m'}^{pr}\right\}_{m_1,1} \left\{\mathbf{b}_{12,m'}^{pr}\right\}^*_{m_3,1} \times \nonumber\\
	&  \left\{\mathbf{U}_{21}^{pr,pr}\right\}_{m_1,m_2} \left\{\left(\mathbf{U}_{21}^{pr,pr}\right)^*\right\}_{m_3,m_2} \left\{\left(\pmb{\tau}_{21}^{pr,pr}\right)^2\right\}_{m_2,m_2} \rangle
\label{eq:evanesfocsum}
\end{align}
Taking the isotropy assumption
\begin{align}
 &\langle T_{opt} \rangle=\langle c_1^2 c_0^2 \rangle \sum _{m_1} \sum _{m_2} \sum _{m_3} \langle  \left\{\mathbf{b}_{12,m'}^{pr}\right\}_{m_1,1} \left\{\mathbf{b}_{12,m'}^{pr}\right\}^*_{m_3,1} \rangle \times \nonumber\\
	&  \langle \left\{\mathbf{U}_{21}^{pr,pr}\right\}_{m_1,m_2} \left\{\left(\mathbf{U}_{21}^{pr,pr}\right)^*\right\}_{m_3,m_2} \rangle \langle \left\{\left(\pmb{\tau}_{21}^{pr,pr}\right)^2\right\}_{m_2,m_2} \rangle
	\label{eq:evanesfocisotropy1}
\end{align}
As per Sec. 3 of \cite{mello1990averages}, $Q_{b \beta }^{a \alpha }= \langle \mathbf{U}_{b \beta } \left(\mathbf{U}_{a \alpha }\right)^*  \rangle = {\delta_{a b} \delta _{\alpha  \beta }}/{M_{pr}}$ where $\delta$ is the Kronecker delta function and denoting $\sum _{m_2}\left\{\left(\pmb{\tau}_{21}^{pr,pr}\right)^2\right\}_{m_2,m_2} = T_2$, $\langle T_{opt} \rangle$ can be simplified as,

\begin{align}
	&\langle T_{opt} \rangle=\langle c_1^2 c_0^2 \rangle \sum _{m_1}  \sum _{m_3} \langle  \left\{\mathbf{b}_{12,m'}^{pr}\right\}_{m_1,1} \left\{\mathbf{b}_{12,m'}^{pr}\right\}^*_{m_3,1} \rangle \times \nonumber\\
	&  \left( \delta_{m_1,m_3}/M_{pr} \right) \langle T_2 \rangle
	\label{eq:evanesfocisotropy2}
\end{align}

As $ \left(\mathbf{b}_{12,m'}^{pr}\right)^{\dagger }\mathbf{b}_{12,m'}^{pr}=1$, $\langle T_{opt} \rangle$ is simplified as
\begin{align}
\langle T_{opt} \rangle =\langle c_1^2 c_0^2 \rangle \langle T_2 \rangle /M_{pr}
\label{eq:evanesfocisotropy3}
\end{align}
It can be shown (given towards the end of this section) that $\langle c_1^2 c_0^2 \rangle = 1/\langle \tau_{avg}\rangle$ where $ \tau_{avg}$ is the numerical average of the eigenchannel transmission coefficients contained in the diagonal of the matrix $\pmb{\tau}_{21}^{pr,pr}$. For the bimodal eigenchannel transmission coefficient distribution, $\langle T_2 \rangle = 2g/3$ where $g=M_{pr} \langle\tau_{avg}\rangle$ is the optical conductance.  Substituting these relations in Eq.~(\ref{eq:evanesfocisotropy3}), we get $\langle T_{opt} \rangle=2/3$ which is a universal constant. 

For the completeness of the proof, $\langle c_1^2 c_0^2 \rangle = 1/\langle \tau_{avg}\rangle$ is derived as the following. As given in the first paragraph of Sec.~\ref{sec:focussing}, the flux normalized propagating phase conjugate wave incident on the left side of the  disorder to create an evanescent focus on the right is represented as $c_1 (c_{tr}^{L,pr})^*=c_1 \left(\mathbf{S}_{12}^{pr,ev}\right)^* \mathbbold{1}_{m'}^{ev}$. But, it was seen that $\left(\mathbf{S}_{12}^{pr,ev}\right)^* \mathbbold{1}_{m'}^{ev}=c_0 \left(\mathbf{S}_{12}^{pr,pr}\right)^* \left(\mathbf{b}_{12,m'}^{pr} \right)^*$. As flux conservation implies $\left(c_1 (\mathbf{c}_{tr}^{L,pr})^*\right)^\dagger\left(c_1 (\mathbf{c}_{tr}^{L,pr})^*\right)=1$, it leads to
\begin{align}
c_1^2 c_0^2 \left(\mathbf{b}_{12,m'}^{pr}\right)^T \left(\mathbf{S}_{12}^{pr,pr}\right)^T  \left(\mathbf{S}_{12}^{pr,pr}\right)^*\left( \mathbf{b}_{12,m'}^{pr}\right)^*=1
\end{align}
Using the reciprocity relation that $ \left(\mathbf{S}_{12}^{pr,pr}\right)^T  \left(\mathbf{S}_{12}^{pr,pr}\right)^*=\mathbf{S}_{21}^{pr,pr}{\mathbf{S}_{21}^{pr,pr}}^\dagger$ and implementing the $S.V.D$ of $\mathbf{S}_{21}^{pr,pr}$,
\begin{align}
c_1^2 c_0^2 \left(\mathbf{b}_{12,m'}^{pr}\right)^T \mathbf{U}_{21}^{pr,pr} \pmb{\tau}_{21}^{pr,pr}  \left(\mathbf{U}_{21}^{pr,pr}\right){}^{\dagger } \left( \mathbf{b}_{12,m'}^{pr}\right)^*=1
\end{align}
Taking the ensemble averaging with the isotropy assumption and using the summation indices,
\begin{align}
&\langle c_1^2 c_0^2 \rangle \sum_{m_3}\sum _{m_2}\sum _{m_1} 
\langle \left\{\mathbf{b}_{12,m'}^{pr}\right\}_{m_1,1} \left\{\mathbf{b}_{12,m'}^{pr}\right\}_{m_3,1}^*\rangle \times\nonumber \\
& \langle \left\{\mathbf{U}_{21}^{pr,pr}\right\}_{m_1,m_2} \left\{\mathbf{U}_{21}^{pr,pr}\right\}^*_{m_3,m_2}\rangle \langle \left\{\pmb{\tau}_{21}^{pr,pr}\right\}_{m_2,m_2} \rangle = 1
\end{align}

Further simplifying, using the averaging over unitary groups, 
\begin{align}
	&\langle c_1^2 c_0^2 \rangle \sum_{m_3}\sum _{m_1} 
	\langle \left\{\mathbf{b}_{12,m'}^{pr}\right\}_{m_1,1} \left\{\mathbf{b}_{12,m'}^{pr}\right\}_{m_3,1}^*\rangle \times \nonumber \\
	& \left(\delta_{m_1,m_3}/M_{pr}\right) \langle \sum _{m_2}\left\{\pmb{\tau}_{21}^{pr,pr}\right\}_{m_2,m_2} \rangle = 1.
\end{align}
As $ \left(\mathbf{b}_{12,m'}^{pr}\right)^{\dagger }\mathbf{b}_{12,m'}^{pr}=1$,
\begin{align}
\langle c_1^2 c_0^2 \rangle 
	\left(1/M_{pr}\right) \left( M_{pr} \langle \tau_{avg} \rangle \right)= 1.
\end{align}
This results in the relation $\langle c_1^2 c_0^2 \rangle$= $1/\langle \tau_{avg}\rangle$, which is used to simplify Eq.~(\ref{eq:evanesfocisotropy3}). 
\bibliography{bibliography.bib}

\end{document}


	
	\title{Supplementary Material : Modeling scattering matrix containing evanescent modes for wavefront shaping applications in disordered media}
	
	\author{Michael Raju}
	\email[]{michaelraju@umail.ucc.ie}
	\affiliation{Tyndall National Institute, Lee Maltings Complex, Dyke Parade, Cork, Ireland, T12 R5CP}
	\affiliation{School of Physics, University College Cork, College Road, Cork, Ireland, T12 K8AF}
	\author{Baptiste Jayet}
	\affiliation{Tyndall National Institute, Lee Maltings Complex, Dyke Parade, Cork, Ireland, T12 R5CP}
	\author{Stefan Andersson-Engels}
	\affiliation{Tyndall National Institute, Lee Maltings Complex, Dyke Parade, Cork, Ireland, T12 R5CP}
	\affiliation{School of Physics, University College Cork, College Road, Cork, Ireland, T12 K8AF}
	
	\maketitle
	
	\section{Spatially correlated disorders}
	\label{supp:spatialcorrdisordergeneration}
If the random disorder ensemble (characterized by its spatial refractive index fluctuations) drawn from a particular probability distribution is considered as a spatial random process, then according to the Wiener–Khinchin theorem, the randomness power spectrum of the disorder can be estimated by the spatial Fourier transform of its correlation function. Therefore, to create a random disorder with a particular correlation length, either the spatial correlation function (in real space) or the randomness power spectrum (in the spatial Fourier space) could be used.  Here, we use the Fourier space method (as described in the Supplementary Sec.~\ref{supp:corrdisorder}), for generating a uniformly-distributed Gaussian-correlated disorder. The disorder is represented in terms of the spatial dielectric-constant perturbation 
	$\delta\tilde{\epsilon_r}(z,y)$ defined such that it is uniformly distributed with zero-mean and has a Gaussian spatial correlation. To characterize the disorder, variance is defined as
	$var_{disorder} = \langle \delta\tilde{\epsilon_r}(r)^2\rangle- \langle\delta\tilde{\epsilon_r}(r)\rangle^2~=~ \langle\delta\tilde{\epsilon_r}(r)^2\rangle$. The spatial correlation function can be defined as  $R(k_{ref}r,k_{ref}r_0)~=~\langle\delta\tilde{\epsilon_r}(k_{ref}z,k_{ref}y)\delta\tilde{\epsilon_r}(k_{ref}z_0,k_{ref}y_0)\rangle$, where~$R(k_{ref}r,k_{ref}r_0)$ is the un-normalized spatial correlation function. For a statistically homogeneous disorder, $R(k_{ref}r,k_{ref}r_0)=R(k_{ref}|r-r_0|)$. It can be noted from the definition of the disorder variance that $R(k_{ref}r_0,k_{ref}r_0)=R(0)=var_{disorder}$. 
	Therefore, to represent the un-normalized spatial correlation function of the disorder, $R(0)$ value can be combined with a normalized  correlation function $C_{disorder}(k_{ref}|r-r_0|)$ as given below
	\begin{equation}
		R(k_{ref}|r-r_0|)=R(0) C_{disorder}(k_{ref}|r-r_0|), 
		\label{eq:spatcorrfun}
	\end{equation}
	where the normalized two-point Gaussian spatial correlation function, $C_{disorder}(k_{ref}|r-r_0|)$ for $\delta\tilde{\epsilon_r}(r)$ as the following
	\begin{eqnarray}
		C_{disorder}(k_{ref}|r-r_0|)=e^{-(k_{ref}|r-r_0|)^2/(k^2_{ref}l^2_c)}
		\label{eq:correlation_spatial}
	\end{eqnarray} 
	The power spectral density of the disorder ($P_{PSD}(p_z,p_y)$) is defined as the Fourier transform ($\mathcal{F}$) of the spatial correlation function.
	\begin{eqnarray}
		P_{PSD}(p_z,p_y)&&=\mathcal{F} \left\{R(k_{ref}|r-r_0|) \right\}=var_{disorder} \mathcal{F} \left\{ C_{disorder}(r,l_c)\right\}
		\label{eq:PSD}
	\end{eqnarray}
	where $p_z$ and $p_y$ are angular spatial frequencies. Having reviewed the essential concepts, using the numerical algorithm given in supplementary Sec.~\ref{supp:corrdisorder}, a uniformly distributed-spatially correlated disorder is modeled. 
	
	\subsection{Method for generating uniformly-distributed spatially-correlated disorder}\label{supp:corrdisorder}
	The steps for estimating the uniformly-distributed Gaussian-correlated disorder are given as the following.
	\begin{enumerate}
		\item Estimate $\delta\epsilon^{white}_{\texttt{fft}}=\texttt{fft2}(\delta\epsilon^{white}_{r}(z,y))$ where $\texttt{fft2}$ is the two dimensional Fourier transform and $\delta\epsilon^{white}_{r}(z,y)$ is the white-noise like uncorrelated disorder generated from a uniform probability distribution.
		
		\item Take an element-wise product in the Fourier space to obtain  $\delta\epsilon^{corr}_{\texttt{fft}}=\delta\epsilon^{white}_{\texttt{fft}} \mathcal{F} \left\{ C_{disorder}(r,l_c)\right\}$
		
		\item Apply inverse Fourier transform to estimate the correlated disorder in the real space with spatial correlation length $l_c$, given by
		$\delta\epsilon^{corr}_{r}(z,y)=\texttt{ifft2}\left(\delta\epsilon^{corr}_{\texttt{fft}}\right)$ where \texttt{ifft2} is the two dimensional inverse Fourier transform.
		
		\item  Normalize $\delta\epsilon^{corr}_{r}(z,y)$ by dividing it with the standard deviation of the matrix elements. This results in a Gaussian distributed (with unit variance) and Gaussian correlated (with correlation length $l_c$) disorder. 
		\item To transform the Gaussian distributed-Gaussian correlated disorder to a uniform distributed-Gaussian correlated disorder, a Gaussian cumulative probability distribution function $(C.D.F)$ can be used. A zero mean Gaussian $C.D.F$ takes a normally distributed random variable and converts it into a transformed random variable between $(0,1)$ giving a uniform distribution. Hence, every matrix element of Gaussian distributed-Gaussian correlated disorder can be transformed between $(0,1)$ using the Gaussian $C.D.F$. Such a uniformly distributed disorder between $(0,1)$ can then be scaled between $min\left\{\delta\epsilon^{white}\right\}$ and $max\left\{\delta\epsilon^{white}\right\}$ to have similar spread in values with that of the uncorrelated white-noise like disorder. 
		
		\item Real space spatial correlation of the obtained disorder can be numerically estimated and validated against the $l_c$ value used for defining the correlation function by the following method. It involves the Fourier space method for evaluating spatial autocorrelation of a $2D$ matrix. For an $M\times N$ correlated disorder	$\delta\epsilon^{corr}_{r}(z,y)$, 
		\begin{equation}
				R_{disorder}(k_{ref}|r-r_0|) =
				|\texttt{ifft2}\left[\texttt{fft2}(\delta\epsilon^{corr}_{r})
				(\texttt{fft2}(\delta\epsilon^{corr}_{r}))^*\right]|/(\texttt{MN}) 
		\end{equation}
		where $*$ denotes complex conjugation. Numerically estimated correlation length can be defined as the length at which the correlation function decays from $R(0)$ to $R(0)/e$.
	\end{enumerate}

\section{Numerical discretization scheme implemented}
\label{supp:discretization}
To represent the analytical functions presented in this paper as functions of $z$ and $y$ in numerical computations, we use numerical grids from the finite difference methods. These numerical grids have a finite grid cell size, $\Delta_z$ and $\Delta_y$. For the analytical expressions given in the paper to be exact on the grid, $\Delta_z$ and $\Delta_y$ would need to tend to zero, which is practically infeasible due to the memory and computational processing limitations. Therefore, $\Delta_z$ and $\Delta_y$ are chosen as a trade-off (to be explained later) such that they are small enough to resolve wave states on the grids (especially evanescent eigenmodes), but not so small as to exhaust available memory and the processing capacity. Let there be $N_{z}$ grid points/nodes along the $z$ direction and $N_{y}$ along the $y$ direction so that the total number of grid points is $N_z N_y$, with the grid spacing $\Delta_z$ and $\Delta_y$. The longitudinal size of the region is then $(N_z-1)\Delta_z$, and transverse size is $(N_y-1)\Delta_y$. Let the indices be $z_i=[1,2,...,N_z]$ and $y_j=[1,2,...,N_y]$. Then, the discrete values of $z$ and $y$ are given by  $z=(z_i-1)\Delta_z$ and $y=(y_j-1)\Delta_y$, respectively, which form an array of values for $z$ and $y$. Additionally, we define $W=(N_y-1)\Delta_y$. The analytical form of the propagating eigenmode, $	\psi^{\pm}_{m,pr}(z,y)=\frac{1}{\sqrt{k^{(m)}_z}} e^{\pm i k^{(m)}_z z}\chi_m(y)$, needs to be discretized as follows. Let the discrete version of $\psi^{+}_{m,pr}(z_i,y_j)$ on the finite difference grid be written as 
\begin{align}
	\psi^{+}_{m,pr}(z_i,y_j)= \underbrace{c^+_m e^{+i k^{(m)}_z \Delta_z(z_i-1)}}_{\Phi^+_{m,pr}(z_i)}\chi_m(y_j)\label{eq:rt_eig_num}
\end{align}
where the coefficient $c^+_m$  is determined next, using the fact that the wave current injected by all normalized eigenmodes are unity. Here,
\begin{align}
	\chi_{m}(y_j)=\sqrt{\frac{2}{W}}\sin\{k^{(m)}_y \Delta_y (y_j-1)\}
\end{align}
with the discrete form of orthogonality relationship given by
\begin{align}
	\sum_{y_j} 	\chi_m(y_j) \chi_n(y_j) \Delta_y=\delta_{kron}(m,n) \label{eq:dis_ortho}
\end{align} 
The magnitude of the Z component of the flux density of the $m^{th}$ eigenmode is defined as \\
$J^{(m)}_z(z,y) =\mathfrak{Im}\left\{\psi^{*}_m(z,y) \frac{\partial}{ \partial z} \psi_m (z,y)\right\}$, where $\mathfrak{Im}$ refers to the imaginary part. Using the centered first-order finite difference formula,
\begin{align}
	\frac{d}{dz} \Phi^{+}_{m,pr}(z_i) = \frac{\Phi^{+}_{m,pr}(z_i+1)-\Phi^{+}_{m,pr}(z_i-1)}{2\Delta_z},
\end{align}
the discrete version of the flux density along Z can be evaluated as 
\begin{align}
	&{J}^{(m)}_z(z_i,y_j)=\frac{\chi_m(y_j)^2}{2i} \{ { {\Phi^{+}_{m,pr}(z_i)}^{*}\frac{d}{dz} \Phi^{+}_{m,pr}(z_i)}_{} - {\Phi^{+}_{m,pr}(z_i) \frac{d}{dz} {\Phi^{+}_{m,pr}(z_i)}^*}_{} \}
\end{align}
which, upon simplification yields,
\begin{align}
	{J}^{(m)}_z(z_i,y_j)=\frac{\chi_m(y_j)^2}{2i} \left\{   2i~{c^+_m}^2 \left(\frac{sin(k^{(m)}_z\Delta_z)}{\Delta_z}\right)\right\} \label{eq:dis_curr}
\end{align}
Next, we numerically enforce the wave current (or flux) injected by the propagating eigenmode to be unity. Integrating the ${J}^{(m)}_z(z_i,y_j)$ along Y (using discrete summation)  to get the total flux and equating to unity,

\begin{align}
	&\sum_{y_j}{J}_z^{(m)}(z_i,y_j)\cdot  \Delta_y =1 \implies 
	\sum_{y_j}\chi_m(y_j)^2 \Delta_y{c^+_m}^2 \left(\frac{sin(k^m_z\Delta_z)}{\Delta_z}\right)=1\implies\nonumber\\
	&{c^+_m}^2 \left(\frac{sin(k^{(m)}_z\Delta_z)}{\Delta_z}\right)\underbrace{\sum_{y_j}\chi_m(y_j)^2 \Delta_y}_{=1}=1\implies
	c^+_m=\frac{1}{\sqrt{\frac{sin(k^{(m)}_z\Delta_z)}{\Delta_z}}}
\end{align}

Substituting for $c^+_m$ into $\psi^{+}_m(z_i,y_j)$, the discrete form of eigenmode is obtained as

\begin{align}
	\psi^{\pm}_{m,pr}(z_i,y_j)= \overbrace{\frac{1}{\sqrt{\frac{\sin\left(k^{(m)}_z\Delta_z\right)}{\Delta_z}}} e^{\pm i k^{(m)}_z \Delta_z(z_i-1)}}^{\Phi^\pm_m(z_i)}\chi_m(y_j)\label{eq:disc_eigmode_main}
\end{align}
This discrete form of eigenmodes ensures flux conservation upon discretization and approaches the continuous analytic form $\psi^{\pm}_{m,pr}(z,y)=\frac{1}{\sqrt{k^{(m)}_z}} e^{\pm i k^{(m)}_z z}\chi_m(y)$, in the limit $\Delta_z \rightarrow 0$ and $\Delta_y \rightarrow 0$, since 
$lim_{\Delta_z \rightarrow 0} \frac{\sin\left(k^{(m)}_z \Delta_z\right)}{\Delta_z}=k^{(m)}_z$. In numerical implementation, the product $k^{(m)}_y \Delta_y$ is treated as a single term (dimensionless), to avoid precision errors when evaluating $\chi_m(y)$, since $k^{(m)}_y$ is significantly larger than $\Delta_y$. Similarly, $k^{(m)}_z \Delta_z$ is also treated as a single variable.

For the evanescent modes where $k^{(m)}_z = i k^{(m)}_{z,ev}$, the analytical expression\\ $\Phi^{\pm}_{m,ev}(z,z_{L/R})=\frac{1}{\sqrt{|k^{(m)}_{z}|}} e^{- k^{(m)}_{z,ev} |z-z_{L/R}|}$ gets replaced by its discretized form given as 
\begin{align}
	\Phi^{\pm}_{m,ev}(z_i,z_{i_{L/R}})=\frac{1}{\sqrt{\frac{sin\left(|k^{(m)}_z|\Delta_z\right)}{\Delta_z}}} e^{- k^{(m)}_{z,ev}\Delta_z |z_i-z_{i_{L/R}}|}
\end{align}
Here, the subscript $L/R$ denotes the left ($L$) or the right ($R$) boundary. Next, we discretize the continuous analytical dispersion relation $\left( \left(k^{(m)}_z\right)^2 + \left(k^{(m)}_y\right)^2=k^2_{ref}\right)$.  By substituting the discrete form of eigenmodes in the discrete Helmholtz wave equation, the discrete dispersion relation is obtained as    

\begin{align}
	\frac{4}{\Delta_z^2}  sin^2\left(\frac{k^{(m)}_z \Delta_z}{2}\right)+\frac{4}{\Delta_y^2}  sin^2\left(\frac{k^{(m)}_y \Delta_y}{2}\right) = k^2_{ref} \label{eq:disc_disp_rel1_ch1}
\end{align}

Taking the limits, $lim_{\Delta_z \rightarrow 0}$ and $lim \Delta_y \rightarrow 0$, the continuous analytic dispersion relation is recovered. 

For ease of implementation in numeric computations, we set $\Delta_z =\Delta_y = \Delta$ and $\Delta$ is chosen such that it adequately resolves the evanescent eigenmode with the highest transverse spatial frequency, $k^{(m)}_{y, max}= M_{total} \pi /W $ where $M_{total}=M_{pr}+M_{ev}$ is the total number of eigenmodes considered.  The discretization resolution is chosen such that $k^{(m)}_{y, max} \Delta=1$. 

Similar to that of eigenmodes, the unperturbed Green's function is also discretized for numerical implementation as follows : 

\begin{align}
	&G_0(z_i,y_j,{z_0}_i,{y_0}_j)=\nonumber\\
	&\sum_{m=1}^{\infty} d_m e^{i k^{(m)}_z |\Delta_z (z_i-z_{0_i})|}\sqrt{\frac{2}{W}} sin\left(k^{(m)}_y \Delta_y(y_j-1)\right)\sqrt{\frac{2}{W}} sin\left(k^{(m)}_y (y_{0_j}-1)\right) 
	\label{eq:Greensfundefdiscrete}
\end{align}
where, 
\begin{align}
	d_m=\begin{cases}
		\frac{1}{2i \left(\frac{\sin(k^m_z\Delta_z)}{\Delta_z}\right)}, & \text{for propagating modes, $m\leq M_{pr}$}\\
		\frac{1}{-2\left(\frac{\sin(|k^m_z|\Delta_z)}{\Delta_z}\right)}, & \text{for evanescent modes, $m > M_{pr}$}
	\end{cases}
\end{align}

\section{Estimating the perturbed Green's function $G(r,r_0)$}\label{supp:GreensfunPertb}
The discretized Dyson's equation in $2D$ is $G_{i j}=G^{(0)}_{i j}+ \sum_{k=1}^{k=N^{P}_z N^{P}_y} G^{(0)}_{i k} V_k \Delta_{area} G_{k j}$ where $k$ is used as a index.  The entire computational domain which includes the disordered perturbation region and the surrounding region for field visualization, has the size of $[N_y\times N_z]$. The disorder perturbation region has the size of $[N^P_y\times N^P_z]$, which is a subset of the computational domain $[N_y\times N_z]$. Therefore, there are $N^P_z N^P_y$ scatterer perturbation points.
Here, the index $i$  denotes the field evaluation location accessed linearly as a $1D$ array. Similarly, the index $j$ denotes the source location accessed linearly as a $1D$ array, placed on the boundaries $\Gamma_{L}$ or $\Gamma_{R}$. The index $k$ accessed linearly as a $1D$ array, is associated with the locations at which the scatterer perturbation exists. Two methods are presented as the following where the first method is the direct matrix inversion method. Such a method is easier to understand, but it is memory intensive. The second method is the iterative method which requires iteration steps, but it is memory efficient.

Both the methods start by estimating the free space Green's function $G^{(0)}_{i k}$ and $G^{(0)}_{ij}$, where $i$ spans all the field evaluation points of size $N_y N_z$ for visualisation purposes, $k$ spans all the $N^{P}_z N^{P}_y$ scatterer position points where the perturbation is about to be brought in, and $j$ the $N_y$ points associated with the left or the right boundary. $G^{(0)}_{i k}$ is of the dimension $\left[N_y N_z\times N^P_y N^P_z\right]$. Other Green's function terms to be appearing in the methods can be derived as a subset from $G^{(0)}_{i k}$ and $G^{(0)}_{ij}$.
\subsection{Direct method}
\paragraph*{Step 1 : Using the Dyson's equation, solve for $G_{k j}$ for different locations of $k=k_{1}, k_{2}, k_{3}, ..., k_{N^{P}_z N^{P}_y}$ and a given $j$ location}

\begin{align}
	&G_{k_1j }= G_{k_1 j}^{(0)} + \left[G_{k_1 k_1}^{(0)}V_{k_1}\quad  G_{k_1 k_2}^{(0)}V_{k_2}\quad\text{...}\quad G_{k_1 k_{N^P_{y}N^P_{z}}}^{(0)}V_{k_{N^P_{z}N^P_{y}}}\right] \left(
	\begin{array}{c}
		\Delta _{area} G_{k_1 j }  \\
		\Delta _{area} G_{k_2 j } \\
		\text{...} \\
		\Delta_{area} G_{k_{(N^P_{y}N^P_{z})} j } \\
	\end{array} \right) \\
	&G_{k_2j }= G_{k_2 j}^{(0)} + \left[G_{k_2 k_1}^{(0)}V_{k_1}\quad  G_{k_2 k_2}^{(0)}V_{k_2}\quad\text{...}\quad G_{k_2 k_{N^P_{y}N^P_{z}}}^{(0)}V_{k_{N^P_{z}N^P_{y}}}\right] \left(
	\begin{array}{c}
		\Delta_{area} G_{k_1 j }  \\
		\Delta _{area} G_{k_2 j } \\
		\text{...} \\
		\Delta_{area} G_{k_{(N^P_{y}N^P_{z})} j } \\
	\end{array} \right)\\
	&\hspace{5cm}\textbf{\huge\vdots}\nonumber 
\end{align}
\begin{align}
		G_{k_{(N^P_z N^P_y)}j }= G_{k_{(N^P_z N^P_y)} j}^{(0)} + \left[G_{k_{(N^P_z N^P_y)} k_1}^{(0)}V_{k_1}  ~G_{k_{(N^P_z N^P_y)} k_2}^{(0)}V_{k_2} \textbf{...}\quad G_{k_{(N^P_z N^P_y)} k_{N^P_{y}N^P_{z}}}^{(0)}V_{k_{N^P_{z}N^P_{y}}}\right]\left(
		\begin{array}{c}
			\Delta_{area} G_{k_1 j }  \\
			\Delta_{area} G_{k_2 j } \\
			\text{...} \\
			\Delta_{area} G_{k_{(N^P_{y}N^P_{z})} j } 
		\end{array}\right)
\end{align}
These set of equations can be written in a matrix form as
\begin{align}
	&\left(\begin{array}{c}
			G_{k_1j } \\
			G_{k_2j } \\
			G_{k_3j }\\
			\textbf{\vdots} \\
			G_{k_{(N^P_{z}N^P_y)}j }\\
		\end{array}\right)
		\text{ = }
		\left(\begin{array}{c}
			G_{k_1 j}^{(0)}  \\
			G_{k_2 j}^{(0)}  \\
			G_{k_3 j}^{(0)}  \\
			\textbf{\vdots} \\
			G_{k_{(N^P_z N^P_y)} j}^{(0)}  \\
		\end{array}\right)
		\text{  + }\nonumber\\
		&\underbrace{\left(
			\begin{array}{cccc}
				G_{k_1 k_1}^{(0)}V_{k_1}  &  G_{k_1 k_2}^{(0)}V_{k_2} & \textbf{...} & G_{k_1 k_{(N^P_z N^P_y)}}^{(0)}V_{k_{(N^P_z N^P_y)} } \\
				G_{k_2 k_1}^{(0)}V_{k_1}\text{   } & G_{k_2 k_2}^{(0)}V_{k_2} & \textbf{...} & G_{k_2 k_{(N^P_z N^P_y)} }^{(0)}V_{k_{(N^P_z N^P_y)} } \\
				G_{k_3 k_1}^{(0)}V_{k_1} & G_{k_3 k_2}^{(0)}V_{k_2}\text{  } & \textbf{...} & G_{k_3 k_{(N^P_z N^P_y)} }^{(0)}V_{k_{(N^P_z N^P_y)} }\\
				\textbf{\vdots} & \textbf{\vdots} & \textbf{\vdots} & \textbf{\vdots} \\
				G_{k_{(N^P_z N^P_y)} k_1}^{(0)}V_{k_1} & G_{k_{(N^P_z N^P_y)}  k_2}^{(0)}V_{k_2} & \textbf{...} & G_{k_{(N^P_z N^P_y)}  k_{(N^P_z N^P_y)} }^{(0)}V_{k_{(N^P_z N^P_y)} } \\
			\end{array}
			\right)}_{M}\left(
		\begin{array}{c}
			\Delta _{area}G_{k_1 j }\\
			\Delta _{area}G_{k_2 j }\\
			\Delta _{area}G_{k_3 j }\\
			\textbf{\vdots} \\
			\Delta _{area}G_{k_{(N^P_z N^P_y)} j }\\
		\end{array}\right)
	\label{eq:mainGreensSolving}	
\end{align}	 

Representing the same system of equations as a matrix equation, $G_{k j}$ is solved involving an $M$ matrix ($M$ matrix defined in the above given equation), as given below
\begin{align}
	&\left[G_{k j}\right] = \left[G_{k j}^{(0)}\right]+ \left[M\right] \left[G_{k j}\right] \implies\nonumber\\
	&\left[G_{k j}\right]=\left[I - \left[M\right] \right]^{-1} \left[G_{k j}^{(0)}\right]
	\label{eq:directinversion}
\end{align}
\paragraph*{Step 2 : Once $[G_{k j}]$ is obtained, evaluate $[G_{i j}]$ for a given $j$} 

\begin{align}
		&\left(\begin{array}{c}
			G_{i_1j } \\
			G_{i_2j } \\
			G_{i_3j }\\
			\textbf{\vdots} \\
			G_{i_{(N_{z}N_y)}j }\\
		\end{array}\right)
		\text{ = }
		\left(\begin{array}{c}
			G_{i_1 j}^{(0)}  \\
			G_{i_2 j}^{(0)}  \\
			G_{i_3 j}^{(0)}  \\
			\textbf{\vdots} \\
			G_{i_{(N_z N_y)} j}^{(0)}  \\
		\end{array}\right)
		\text{  + }\nonumber\\
		&\left(
		\begin{array}{cccc}
			G_{i_1 k_1}^{(0)}V_{k_1}  &  G_{i_1 k_2}^{(0)}V_{k_2} & \textbf{...} & G_{i_1 k_{(N^P_z N^P_y)}}^{(0)}V_{k_{(N^P_z N^P_y)} } \\
			G_{i_2 k_1}^{(0)}V_{k_1}\text{   } & G_{i_2 k_2}^{(0)}V_{k_2} & \textbf{...} & G_{i_2 k_{(N^P_z N^P_y)} }^{(0)}V_{k_{(N^P_z N^P_y)} } \\
			G_{i_3 k_1}^{(0)}V_{k_1} & G_{i_3 k_2}^{(0)}V_{k_2}\text{  } & \textbf{...} & G_{i_3 k_{(N^P_z N^P_y)} }^{(0)}V_{k_{(N^P_z N^P_y)} } \\
			\textbf{\vdots} & \textbf{\vdots} & \textbf{\vdots} & \textbf{\vdots} \\
			G_{i_{(N_z N_y)} k_1}^{(0)}V_{k_1} & G_{i_{(N_z N_y)}  k_2}^{(0)}V_{k_2} & \textbf{...} & G_{i_{(N_z N_y)}  k_{(N^P_z N^P_y)} }^{(0)}V_{k_{(N^P_z N^P_y)} } \\
		\end{array}
		\right) \left(
		\begin{array}{c}
			\Delta _{area}G_{k_1 j }\\
			\Delta _{area}G_{k_2 j }\\
			\Delta _{area}G_{k_3 j }\\
			\textbf{\vdots} \\
			\Delta _{area}G_{k_{(N^P_z N^P_y)} j }\\
		\end{array}\right)
\end{align}	
where there is $N_z N_y$ field evaluation points due to $N^P_z N^P_y$ scattering particles. With step 1 and step 2 evaluated for different source locations $j$, $G_{ij}$ can be fully solved. Although the direct method presented in this subsection is short and easy to understand, it is memory intensive because of the formation of the large non-sparse $M$ matrix (of size $N^P_zN^P_y \times N^P_zN^P_y$) and the associated inversion process in the Eq.~\ref{eq:directinversion}. Also, the free space Green's function $G^{(0)}_{ik}$ needs to be evaluated for the $i$ corresponding to the entire $[N_z\times N_y]$ grid for the purpose of visualization. The location $k$ needs to be spanned through all the scattering particle locations (to evaluate the $M$ matrix), consuming memory. To save memory the following iterative method is used instead.
\subsection{Iterative method involving slices of scatterers}
In the iterative method, the total scattering block is divided into slices and the Green's function is updated accounting the scattering effect of the slices taken one at a time. As an example, let's say the entire scattering region is divided into 10 slices. $G^{(0)}_{ik}$ needs to be evaluated like before for the source $k$ location spanning all the particle positions. $G^{(0)}_{ij}$ is also evaluated where $j$ is on $\Gamma_{L}$ or $\Gamma_{R}$. Accounting the scattering perturbation of slice number 1 yields the new updated Green's function $G^{(1)}_{ik}$. While evaluating $G^{(1)}_{ik}$, the source $k$ locations needs to be only those scatterer positions where the remaining 9 slices are to be brought in and also on the boundaries. One could overwrite the $G^{(0)}_{ik}$ array data (only for the source locations of the particles to be brought in, corresponding to the remaining 9 slices) for saving memory. Now the second slice is brought in and $G^{(1)}_{ik}$ is used to evaluate $G^{(2)}_{ik}$ accounting the scattering effect of the second slice. For evaluating $G^{(2)}_{ik}$, the source $k$ locations needs to be only those scatterer locations where the remaining 8 slices are to be brought in and also on the boundaries. The method is repeated by bringing in the remaining slices one by one and updating the Green's function by overwriting the previous version, for the remaining particle positions. Therefore, the process becomes faster upon progressing through the slices.

First, let the first transverse slice of the scatterers be chosen, numbered $1:N^P_y$ denoted as $1\rightarrow N^P_y$. Solve the Eq.~\ref{eq:mainGreensSolving} just for the first slice.	

\paragraph*{Main step 1 for the iteration method involving the first slice of scatterers}
\paragraph*{$\bullet$ Sub-step 1:}

\begin{align}
		\left(\begin{array}{c}
			G^{(1)}_{k_1j } \\
			G^{(1)}_{k_2j } \\
			G^{(1)}_{k_3j }\\
			\textbf{\vdots} \\
			G^{(1)}_{k_{(N^P_y)}j }\\
		\end{array}\right)
		\text{ = }
		\left(\begin{array}{c}
			G_{k_1 j}^{(0)}  \\
			G_{k_2 j}^{(0)}  \\
			G_{k_3 j}^{(0)}  \\
			\textbf{\vdots} \\
			G_{k_{(N^P_y)} j}^{(0)}  \\
		\end{array}\right)
		\text{  + }
		\underbrace{\left(
			\begin{array}{cccc}
				G_{k_1 k_1}^{(0)}V_{k_1}  &  G_{k_1 k_2}^{(0)}V_{k_2} & \textbf{...} & G_{k_1 k_{( N^P_y)}}^{(0)}V_{k_{(N^P_y)} } \\
				G_{k_2 k_1}^{(0)}V_{k_1}\text{   } & G_{k_2 k_2}^{(0)}V_{k_2} & \textbf{...} & G_{k_2 k_{(N^P_y)} }^{(0)}V_{k_{(N^P_y)} } \\
				G_{k_3 k_1}^{(0)}V_{k_1} & G_{k_3 k_2}^{(0)}V_{k_2}\text{  } & \textbf{...} & G_{k_3 k_{(N^P_y)} }^{(0)}V_{k_{(N^P_y)} }\\
				\textbf{\vdots} & \textbf{\vdots} & \textbf{\vdots} & \textbf{\vdots} \\
				G_{k_{(N^P_y)} k_1}^{(0)}V_{k_1} & G_{k_{(N^P_y)}  k_2}^{(0)}V_{k_2} & \textbf{...} & G_{k_{(N^P_y)}  k_{(N^P_y)} }^{(0)}V_{k_{(N^P_y)} } \\
			\end{array}
			\right)}_{M^{(0)}_{(1\rightarrow N^P_y)}}\left(
		\begin{array}{c}
			\Delta _{area}G^{(1)}_{k_1 j }\\
			\Delta _{area}G^{(1)}_{k_2 j }\\
			\Delta _{area}G^{(1)}_{k_3 j }\\
			\textbf{\vdots} \\
			\Delta _{area}G^{(1)}_{k_{(N^P_y)} j }\\
		\end{array}\right)
	\label{eq:mainGreensSolvingstep1}	
\end{align}

\paragraph*{$\bullet$ Sub-step 2:}
Representing the above given system of equations as a matrix equation, $G^{(1)}_{k_{(1\rightarrow N^{p}_y)} j}$ is solved involving the $M^{(0)}_{(1\rightarrow N^P_y)}$ matrix (defined in the previous equation), as given below
\begin{align}
	&\left[G^{(1)}_{k_{(1\rightarrow N^{p}_y)} j}\right] = \left[G_{k_{(1\rightarrow N^{p}_y)} j}^{(0)}\right]+ \left[M^{(0)}_{(1\rightarrow N^P_y)}\right] \left[G^{(1)}_{k_{(1\rightarrow N^{p}_y)} j}\right] \implies\nonumber\\
	&\left[G^{(1)}_{k_{(1\rightarrow N^{p}_y)}  j}\right]=\left[I - \left[M^{(0)}_{(1\rightarrow N^P_y)}\right] \right]^{-1} \left[G_{k_{(1\rightarrow N^{p}_y)}  j}^{(0)}\right]
	\label{eq:iterationstep1}
\end{align}

\paragraph*{$\bullet$ Sub-step 3:}
Once $[G^{(1)}_{k_{(1\rightarrow N^{p}_y)} j}]$ is obtained, evaluate $[G^{(1)}_{i j}]$ for a given $j$

\begin{align}
		\left(\begin{array}{c}
			G^{(1)}_{i_1j } \\
			G^{(1)}_{i_2j } \\
			G^{(1)}_{i_3j }\\
			\textbf{\vdots} \\
			G^{(1)}_{i_{(N_{z}N_y)}j }\\
		\end{array}\right)
		\text{ = }
		\left(\begin{array}{c}
			G_{i_1 j}^{(0)}  \\
			G_{i_2 j}^{(0)}  \\
			G_{i_3 j}^{(0)}  \\
			\textbf{\vdots} \\
			G_{i_{(N_z N_y)} j}^{(0)}  \\
		\end{array}\right)
		\text{  + }
		\left(
		\begin{array}{cccc}
			G_{i_1 k_1}^{(0)}V_{k_1}  &  G_{i_1 k_2}^{(0)}V_{k_2} & \textbf{...} & G_{i_1 k_{( N^P_y)}}^{(0)}V_{k_{(N^P_y)} } \\
			G_{i_2 k_1}^{(0)}V_{k_1}\text{   } & G_{i_2 k_2}^{(0)}V_{k_2} & \textbf{...} & G_{i_2 k_{(N^P_y)} }^{(0)}V_{k_{(N^P_y)} } \\
			G_{i_3 k_1}^{(0)}V_{k_1} & G_{i_3 k_2}^{(0)}V_{k_2}\text{  } & \textbf{...} & G_{i_3 k_{(N^P_y)} }^{(0)}V_{k_{(N^P_y)} } \\
			\textbf{\vdots} & \textbf{\vdots} & \textbf{\vdots} & \textbf{\vdots} \\
			G_{i_{(N_z N_y)} k_1}^{(0)}V_{k_1} & G_{i_{(N_z N_y)}  k_2}^{(0)}V_{k_2} & \textbf{...} & G_{i_{(N_z N_y)}  k_{(N^P_y)} }^{(0)}V_{k_{(N^P_y)} } \\
		\end{array}
		\right) \left(
		\begin{array}{c}
			\Delta _{area}G^{(1)}_{k_1 j }\\
			\Delta _{area}G^{(1)}_{k_2 j }\\
			\Delta _{area}G^{(1)}_{k_3 j }\\
			\textbf{\vdots} \\
			\Delta _{area}G^{(1)}_{k_{(N^P_y)} j }\\
		\end{array}\right)
		\label{eq:substep3}
\end{align}

\paragraph*{$\bullet$ Sub-step 4:}

Before moving onto the next main iteration step (main iteration step number 2), the following sub-step evaluations has to be performed to estimate $[G^{(1)}_{k_{(1\rightarrow N^{p}_y)} k_{(N^{p}_y+1\rightarrow N^{p}_y N^{p}_z)}}]$ and $[G^{(1)}_{i k_{(N^{p}_y+1\rightarrow N^{p}_y N^{p}_z)}}]$ to be used in the next main iteration step. Here $(N^{p}_y+1\rightarrow N^{p}_y N^{p}_z)$ corresponds to the remaining scatter position where the perturbation is brought-in in the future iteration steps by introducing slice number 2,3,4 etc..

Using Eq.~\ref{eq:iterationstep1},
\begin{align}
	\left[G^{(1)}_{k_{(1\rightarrow N^{p}_y)}  k_{(N^{p}_y+1\rightarrow N^{p}_y N^{p}_z)}}\right]=\left[I - \left[M^{(0)}_{(1\rightarrow N^P_y)}\right] \right]^{-1} \left[G_{k_{(1\rightarrow N^{p}_y)}  k_{(N^{p}_y+1\rightarrow N^{p}_y N^{p}_z)}}^{(0)}\right]
\end{align}

Taking Eq.~\ref{eq:substep3} and substituting for $j$ as $k_{(N^{p}_y+1\rightarrow N^{p}_y N^{p}_z)}$, together with the above given equation, \\$[G^{(1)}_{i k_{(N^{p}_y+1\rightarrow N^{p}_y N^{p}_z)}}]$ can be estimated. The estimated $[G^{(1)}_{i k_{(N^{p}_y+1\rightarrow N^{p}_y N^{p}_z)}}]$ can be overwritten upon the existing memory taken by $G^{(0)}_{ik}$.
\paragraph*{Main step 2 involving the second slice of scatterers denoted by $N^P_y+1\rightarrow 2N^P_y$,}
\paragraph*{$\bullet$ Sub-step 1:}
\begin{align}
		&\left(\begin{array}{c}
			G^{(2)}_{k_{(N^P_y+1)}j } \\
			G^{(2)}_{k_{(N^P_y+2)}j } \\
			G^{(2)}_{k_{(N^P_y+3)}j }\\
			\textbf{\vdots} \\
			G^{(2)}_{k_{(2N^P_y)}j }\\
		\end{array}\right)
		\text{ = }
		\left(\begin{array}{c}
			G_{k_{(N^P_y+1)} j}^{(1)}  \\
			G_{k_{(N^P_y+2)} j}^{(1)}  \\
			G_{k_{(N^P_y+3)} j}^{(1)}  \\
			\textbf{\vdots} \\
			G_{k_{(2N^P_y)} j}^{(1)}  \\
		\end{array}\right)
		\text{  + }\nonumber\\
		&\underbrace{\left(
			\begin{array}{cccc}
				G_{k_{(N^P_y+1)} k_{(N^P_y+1)}}^{(1)}V_{k_{(N^P_y+1)}}  &  G_{k_{(N^P_y+1)} k_{(N^P_y+2)}}^{(1)}V_{k_{(N^P_y+2)}} & \textbf{...} & G_{k_{(N^P_y+1)} k_{(2N^P_y)}}^{(1)}V_{k_{(2N^P_y)} } \\
				G_{k_{(N^P_y+2)} k_{(N^P_y+1)}}^{(1)}V_{k_{(N^P_y+1)}}\text{   } & G_{k_{(N^P_y+2)} k_{(N^P_y+2)}}^{(1)}V_{k_{(N^P_y+2)}} & \textbf{...} & G_{k_{(N^P_y+2)} k_{(2N^P_y)} }^{(1)}V_{k_{(2N^P_y)} } \\
				G_{k_{(N^P_y+3)} k_{(N^P_y+1)}}^{(1)}V_{k_{(N^P_y+1)}} & G_{k_{(N^P_y+3)} k_{(N^P_y+2)}}^{(1)}V_{k_{(N^P_y+2)}}\text{  } & \textbf{...} & G_{k_{(N^P_y+3)} k_{(2N^P_y)} }^{(1)}V_{k_{(2N^P_y)} }\\
				\textbf{\vdots} & \textbf{\vdots} & \textbf{\vdots} & \textbf{\vdots} \\
				G_{k_{(2N^P_y)} k_{(N^P_y+1)}}^{(1)}V_{k_{(N^P_y+1)}} & G_{k_{(2N^P_y)}  k_{(N^P_y+2)}}^{(1)}V_{k_{(N^P_y+2)}} & \textbf{...} & G_{k_{(2N^P_y)}  k_{(2N^P_y)} }^{(1)}V_{k_{(2N^P_y)} } \\
			\end{array}
			\right)}_{M^{(1)}_{(N^P_y+1\rightarrow 2N^P_y)}}\left(
		\begin{array}{c}
			\Delta _{area}G^{(2)}_{k_{(N^P_y+1)} j }\\
			\Delta _{area}G^{(2)}_{k_{(N^P_y+2)} j }\\
			\Delta _{area}G^{(2)}_{k_{(N^P_y+3)} j }\\
			\textbf{\vdots} \\
			\Delta _{area}G^{(2)}_{k_{(2N^P_y)} j }\\
		\end{array}\right)\label{eq:mainGreensSolvingstep2}	
\end{align}

\paragraph*{$\bullet$ Sub-step 2:}
Representing the above given system of equations as a matrix equation, $G^{(2)}_{k_{(N^{p}_y+1\rightarrow 2N^{p}_y)} j}$ is solved involving the $M^{(1)}_{(N^{p}_y+1\rightarrow 2N^P_y)}$ matrix (defined in the previous equation), as given below
\begin{align}
	&\left[G^{(2)}_{k_{(N^{p}_y+1\rightarrow 2N^{p}_y)} j}\right] = \left[G_{k_{(N^{p}_y+1\rightarrow 2N^{p}_y)} j}^{(1)}\right]+ \left[M^{(1)}_{(N^{p}_y+1\rightarrow 2N^P_y)}\right] \left[G^{(2)}_{k_{(N^{p}_y+1\rightarrow 2N^{p}_y)} j}\right] \implies\nonumber\\
	&\left[G^{(2)}_{k_{(N^{p}_y+1\rightarrow 2N^{p}_y)}  j}\right]=\left[I - \left[M^{(1)}_{(N^{p}_y+1\rightarrow 2N^P_y)}\right] \right]^{-1} \left[G_{k_{(N^{p}_y+1\rightarrow 2N^{p}_y)}  j}^{(1)}\right]
	\label{eq:iterationstep2}
\end{align}

\paragraph*{$\bullet$ Sub-step 3:}
Once $[G^{(2)}_{k_{(N^{p}_y+1\rightarrow 2N^{p}_y)} j}]$ is obtained, evaluate $[G^{(2)}_{i j}]$ for a given $j$.

\begin{align}
		&\left( \begin{array}{c}
			G^{(2)}_{i_1j } \\
			G^{(2)}_{i_2j } \\
			G^{(2)}_{i_3j }\\
			\textbf{\vdots} \\
			G^{(2)}_{i_{(N_{z}N_y)}j }\\
		\end{array}\right)
		\text{ = }
		\left(\begin{array}{c}
			G_{i_1 j}^{(1)}  \\
			G_{i_2 j}^{(1)}  \\
			G_{i_3 j}^{(1)}  \\
			\textbf{\vdots} \\
			G_{i_{(N_z N_y)} j}^{(1)}  \\
		\end{array}\right)
		\text{  + }\nonumber\\
		&\left(
		\begin{array}{cccc}
			G_{i_1 k_{(N^{p}_y+1)}}^{(1)}V_{k_{(N^{p}_y+1)}}  &  G_{i_1 k_{(N^{p}_y+2)}}^{(1)}V_{k_{(N^{p}_y+2)}} & \textbf{...} & G_{i_1 k_{( 2N^P_y)}}^{(1)}V_{k_{(2N^P_y)} } \\
			G_{i_2 k_{(N^{p}_y+1)}}^{(1)}V_{k_{(N^{p}_y+1)}}\text{   } & G_{i_2 k_{(N^{p}_y+2)}}^{(1)}V_{k_{(N^{p}_y+2)}} & \textbf{...} & G_{i_2 k_{(2N^P_y)} }^{(1)}V_{k_{(2N^P_y)} } \\
			G_{i_3 k_{(N^{p}_y+1)}}^{(1)}V_{k_{(N^{p}_y+1)}} & G_{i_3 k_{(N^{p}_y+2)}}^{(1)}V_{k_{(N^{p}_y+2)}}\text{  } & \textbf{...} & G_{i_3 k_{(2N^P_y)} }^{(1)}V_{k_{(2N^P_y)} } \\
			\textbf{\vdots} & \textbf{\vdots} & \textbf{\vdots} & \textbf{\vdots} \\
			G_{i_{(N_z N_y)} k_{(N^{p}_y+1)}}^{(1)}V_{k_{(N^{p}_y+1)}} & G_{i_{(N_z N_y)}  k_{(N^{p}_y+2)}}^{(1)}V_{(k_{N^{p}_y+2)}} & \textbf{...} & G_{i_{(N_z N_y)}  k_{(2N^P_y)} }^{(1)}V_{k_{(2N^P_y)} } \\
		\end{array}
		\right) \left(
		\begin{array}{c}
			\Delta _{area}G^{(2)}_{k_{(N^{p}_y+1)} j }\\
			\Delta _{area}G^{(2)}_{k_{(N^{p}_y+2)} j }\\
			\Delta _{area}G^{(2)}_{k_{(N^{p}_y+3)} j }\\
			\textbf{\vdots} \\
			\Delta _{area}G^{(2)}_{k_{(2N^P_y)} j }\\
		\end{array}\right)
		\label{eq:substep3iteration2}
\end{align}	

\paragraph*{$\bullet$ Sub-step 4:}

Before moving onto the next main iteration step (main iteration step number 3), the following sub-step evaluations has to be performed to estimate $[G^{(2)}_{k_{(N^{p}_y+1\rightarrow 2N^{p}_y)} k_{(2N^{p}_y+1\rightarrow N^{p}_y N^{p}_z)}}]$ and $[G^{(2)}_{i k_{(2N^{p}_y+1\rightarrow N^{p}_y N^{p}_z)}}]$ to be used in the next main iteration step. Here $(2N^{p}_y+1\rightarrow N^{p}_y N^{p}_z)$ corresponds to the remaining scatter position where the perturbation is brought-in in the future iteration steps by introducing slice number 3,4,5 etc..

Using Eq.~\ref{eq:iterationstep2},
\begin{align}
	\left[G^{(2)}_{k_{(N^{p}_y+1\rightarrow 2N^{p}_y)}  k_{(2N^{p}_y+1\rightarrow N^{p}_y N^{p}_z)}}\right]=\left[I - \left[M^{(1)}_{(N^{p}_y+1\rightarrow 2N^P_y)}\right] \right]^{-1} \left[G_{k_{(N^{p}_y+1\rightarrow 2N^{p}_y)}  k_{(2N^{p}_y+1\rightarrow N^{p}_y N^{p}_z)}}^{(1)}\right]
\end{align}

Taking Eq.~\ref{eq:substep3iteration2} and substituting for $j$ as $k_{(2N^{p}_y+1\rightarrow N^{p}_y N^{p}_z)}$, together with the above given equation,\\$[G^{(2)}_{i k_{(2N^{p}_y+1\rightarrow N^{p}_y N^{p}_z)}}]$ can be estimated. The estimated $[G^{(2)}_{i k_{(2N^{p}_y+1\rightarrow N^{p}_y N^{p}_z)}}]$ can be overwritten upon the existing memory taken by $G^{(1)}_{ik}$. Following the same logic, the main step 3 involving the slice of scatterers denoted by $2N^P_y+1\rightarrow 3N^P_y$ can be evaluated. The iteration process continues until the last slice of scatterers are brought in, obtaining the fully perturbed Green's function $G_{ij}$ .

\section{Homogeneous boundary conditions satisfied by the Green's function and the eigenmodes}
\label{supple:homoBC}

\begin{eqnarray}
G(r,r_0)|_{z < z_0,(z,y)\in\Gamma_L}= \sum_{m=1}^{\infty} G_{m}(z,y,z_0,y_0)|_{z < z_0,(z,y)\in\Gamma_L} 
= \sum_{m=1}^{\infty} \underbrace{ g^{-}_{m}(z_0,y_0) \psi^{-}_m(z,y) |_{z < z_0,(z,y)\in\Gamma_L}}_{G_{m}(z,y,z_0,y_0)|_{z < z_0,(z,y)\in\Gamma_L}} 
\label{eq:PertGreenexpL}
\end{eqnarray}

\begin{eqnarray}
G(r,r_0)|_{z \geq z_0,(z,y)\in\Gamma_R}= \sum_{m=1}^{\infty} G_{m}(z,y,z_0,y_0)|_{z \geq z_0,(z,y)\in\Gamma_R} 
= \sum_{m=1}^{\infty} \underbrace{ g^{+}_{m}(z_0,y_0) \psi^{+}_m(z,y) |_{z \geq z_0,(z,y)\in\Gamma_R}}_{G_{m}(z,y,z_0,y_0)|_{z \geq z_0,(z,y)\in\Gamma_R}} 
\label{eq:PertGreenexpR}
\end{eqnarray} 
Both the propagating and the evanescent eigenmodes satisfy the following homogeneous boundary condition at the boundaries $\Gamma_{L}$ and $\Gamma_{R}$,
\begin{equation}
	\frac{\partial }{\partial z}\psi_m^{\pm}(z,y)\mp i k_z^m\psi_m^{\pm}(z,y)=0,
	\label{eq:homBCeigcombined}
\end{equation}
where in particular $k_z^{(m)}=i k_{z,ev}^{(m)}$ for the evanescent eigenmodes. As the disorder perturbation doesn't change the boundary conditions of eigenmodes associated with the outgoing Green's function waves, the following homogeneous boundary condition can be formed in terms of $G_{m}(z,y,z_0,y_0)$ similar to that of  Eq.~\ref{eq:homBCeigcombined}. 

\begin{align}
	&\left\{\frac{\partial}{\partial z}G_{m}(r,r_0) + i k^{(m)}_z G_{m}(r,r_0)\right\}|_{z < z_0,(z,y)\in\Gamma_L}=0\label{eq:homogeneousbc1}\\
	&\left\{\frac{\partial}{\partial z}G_{m}(r,r_0) - i k^{(m)}_z G_{m}(r,r_0)\right\}|_{z \geq z_0,(z,y)\in\Gamma_R}=0\label{eq:homogeneousbc2}
\end{align}

\section{Simplifying Kirchhoff–Helmholtz boundary integral equation}\label{supp:BIsimplification}
As there is no flux flowing out of the transverse boundaries $\Gamma_{U}$ and $\Gamma_{D}$, the boundary integral equation  Eq.~\ref{eq:generalBoundInt} can be simplified as
\begin{widetext}
	\begin{equation}
		\begin{split}
			&\int\left\{\tilde{E}(r)\frac{\partial}{\partial z}G\left(r,r_0\right)\hat{e}_z-G\left(r,r_0\right)\frac{\partial}{\partial z}\tilde{E}(r)\hat{e}_z\right\}|_{(z,y)\in\Gamma_L}\cdot \left(-\hat{e}_z\right) dy+\\
			&\int\left[\tilde{E}(r)\frac{\partial}{\partial z}G\left(r,r_0\right)\hat{e}_z-G\left(r,r_0\right)\frac{\partial}{\partial z}\tilde{E}(z,y)\hat{e}_z\right]|_{(z,y)\in\Gamma_R} \cdot \hat{e}_z dy =\tilde{E}\left(z_0,y_0\right)|_{(z_0,y_0)\in\Omega}
			\label{eq:generalBoundIntsimp}
		\end{split}
	\end{equation}
\end{widetext}
There is a minus sign associated with the $-\hat{e}_z dy$ on the first integral of the L.H.S in Eq.~\ref{eq:generalBoundIntsimp}. This is due to the sign convention that the surface normal $\hat{n}$ points outwards on $\Gamma_L$ which is in opposite direction to $\hat{e}_z\frac{\partial }{\partial z}$ terms. The boundary integral equation  Eq.~\ref{eq:generalBoundIntsimp}, can be further simplified as both $G(z,z,y,y_0)$ and the total field $\tilde{E}(z,y,z_0,y_0)|_{(z,y)\in\Gamma_{L}or{(z,y)\in\Gamma_{R}}}$ can be expanded in terms of eigenmodes $\psi^{\pm}_{m}(z,y)|_{(z,y)\in\Gamma_{L/R}}$ followed by the use of associated boundary conditions to be explained as the following. Let the wave incidence be from the left side of the disorder onto $\Gamma_{L}$. The total wave on $\Gamma_{L}$ is the sum of incident wave modes and reflected wave modes. As the $E(z,y)$ in  Eq.~\ref{eq:generalBoundIntsimp} is the total field, the first step to simplify is to provide explicit expression for  $\tilde{E}(z,y)$ close to the boundary $\Gamma_L$ and $\Gamma_R$ followed by the use boundary conditions involving $z$ derivatives of $\tilde{E}(z,y)$ and $G(z,y,z_0,y_0)$ to be explained in the following section. For ${(z,y)\in\Gamma_L}$,
\begin{eqnarray}
&&\tilde{E}(z,y)=\tilde{E}^{inc}(z,y) + \tilde{E}^{refl}(z,y)\label{eq:maintotalfieldleft},\\
&&\text{where}~\tilde{E}_{inc}(z,y)|_{(z,y)\in\Gamma_L}=\sum_{m=1}^{M_{total}} c_m^+ \chi_m(y)\phi^+_m(z)\label{eq:incwavegen}\\
	&&\text{such that}~\phi^+_m(z)=\begin{cases}
		&\phi^+_{m,pr}(z)~\text{when}~m\leq M_{pr}\nonumber\\
		&\phi^+_{m,ev}(z,z_L)~\text{when}~m > M_{pr}
	\end{cases}~\text{and}\nonumber\\
	&&\tilde{E}_{refl}(z,y)|_{(z,y)\in\Gamma_L}=\sum_{n=1}^{N_{total}}r^-_n\chi_n(y)\phi^-_n(z)\label{eq:reflwavegen}
	\text{such that}~\phi^-_n(z)=\begin{cases}
		&\phi^-_{n,pr}(z)~\text{when}~n\leq N_{pr}\\
		&\phi^-_{n,ev}(z,z_L)~\text{when}~n > N_{pr}
	\end{cases}	\\
	&&\text{and}~r^-_n=\sum_{m=1}^{M_{total}} S^{n m}_{11} c^{+}_m\label{eq:reflcoeff}			
\end{eqnarray}

Here, $r^-_n$ is the reflection coefficient. Substituting equations  Eq.~\ref{eq:incwavegen} and  Eq.~\ref{eq:reflwavegen} in Eq.~\ref{eq:maintotalfieldleft}, and using the property of the Kronecker delta function, the following is obtained.

\begin{eqnarray}
	&&\tilde{E}(z,y)|_{(z,y)\in\Gamma_L}=
	\sum_{n=1}^{N_{total}}\sum_{m=1}^{M_{total}}c_m^+\phi^+_m(z)\chi_m(y)\delta_{kron}(n,m)|_{(z,y)\in\Gamma_L}+
	\sum _{n=1}^{N_{total}} \sum_{m=1}^{M_{total}}c_m^+S_{11}^{n m}\chi_n(y)\phi^-_n(z)|_{(z,y)\in\Gamma_L}
\end{eqnarray}

Due to the Kronecker delta on the first term on the R.H.S of the above given equation, the index $m$ can be replaced by $n$ to obtain 
\begin{eqnarray}
	&&\tilde{E}(z,y)|_{(z,y)\in\Gamma_L}=
	\sum_{n=1}^{N_{total}}\underbrace{\sum_{m=1}^{M_{total}}c_m^+\left\{\phi^+_n(z)\chi_n(y)\delta_{kron}(n,m)+S_{11}^{ n m}\chi_n(y)\phi^-_n(z)\right\}}_{\tilde{E}_n(z,y)|_{(z,y)\in\Gamma_L}} =\sum_{n=1}^{N_{total}} \tilde{E}_n(z,y)|_{(z,y)\in\Gamma_L}\label{eq:TFleftseriessum}
\end{eqnarray}
Eq.~\ref{eq:TFleftseriessum} denotes that the total field on $\Gamma_{L}$ can be written as a series summation involving $\tilde{E}_n(z,y)|_{(z,y)\in\Gamma_{L}}$ waves. Next, the boundary condition associated with $\tilde{E}_n(z,y)|_{(z,y)\in\Gamma_{L}}$ waves for simplifying the boundary integral equation is provided. 
\begin{eqnarray}
	&\frac{\partial}{\partial z} \tilde{E}_n(z,y)|_{(z,y)\in\Gamma_L}+i k^{(n)}_z \tilde{E}_{n}(z,y)|_{(z,y)\in\Gamma_L}
	=\begin{cases}
		2 i k^{(n)}_z \tilde{E}_{n}^{inc}(z,y)|_{(z,y)\in\Gamma_L}, & \text{if}~ 1\leq n\leq M_{total} \\
		0, & \text{if}~ n> M_{total}
	\end{cases} \label{eq:inhomoBCL}
\end{eqnarray}
where $\tilde{E}_n^{inc}(z,y)|_{(z,y)\in\Gamma_L}=c^{+}_n \psi^{+}_n(z,y)|_{(z,y)\in\Gamma_L}$ is the $n^{th}$ eigenmode term associated with the incident wave incident on the disorder from the left and $k_z^{(n)}=i k_{z,ev}^{(n)}$ for evanescent waves. Next is to evaluate the transmitted total field on $\Gamma_R$ due to the wave incidence from the left side upon $\Gamma_{L}$ and the associated boundary condition.

\begin{eqnarray}
	&&\tilde{E}(z,y)|_{(z,y)\in\Gamma_{R}}=\tilde{E}^{trans}(z,y)|_{(z,y)\in\Gamma_{R}}
	=\sum_{n=1}^{N_{total}} t_n^+\phi^+_n(z) \chi_n(y)|_{(z,y)\in\Gamma_{R}} \\
	&&\text{such that}~\phi^+_n(z)=\begin{cases}
		&\phi^+_{n,pr}(z)~\text{when}~n\leq N_{pr}\\
		&\phi^+_{n,ev}(z,z_R)~\text{when}~n > N_{pr}
	\end{cases}\nonumber
\end{eqnarray}
where the field transmission coefficient term $t^+_n$ is related to the transmission matrix $S_{21}$ by
\begin{eqnarray}
	t^+_n&= \sum_{m=1}^{M_{total}} S^{n m}_{21} c^{+}_m 
\end{eqnarray}
Therefore,
\begin{eqnarray}
	&&\tilde{E}(z,y)|_{(z,y)\in\Gamma_{R}}=
	\sum_{n=1}^{N_{total}} \underbrace{\left\{\sum_{m=1}^{M_{total}} c_m^+ S_{21}^{n m}\right\} \phi_n^+(z) \chi_n(y)}_{\tilde{E}_n(z,y)|_{(z,y)\in \Gamma_{R}}}|_{(z,y)\in\Gamma_{R}}\label{eq:BCright1}\\
	&&=\sum_{n=1}^{N_{total}}\tilde{E}_n(z,y)|_{(z,y)\in\Gamma_{R}}\label{eq:BCright2}
\end{eqnarray}
Similar to that of  Eq.~\ref{eq:inhomoBCL}, one could form a boundary condition for $E_n(z,y)|_{(z,y)\in\Gamma_{R}}$ on the right boundary $\Gamma_{R}$ as given below, to be used to simplify the boundary integral equation given in Eq.~\ref{eq:generalBoundIntsimp}. This homogeneous boundary condition can be easily verified using equations Eq.~\ref{eq:BCright1} and Eq.~\ref{eq:BCright2}.

\begin{equation}
	\frac{\partial}{\partial z}\tilde{E}_n(z,y)|_{(z,y)\in\Gamma_{R}}-i k_z^{(n)}\tilde{E}_n(z,y)|_{(z,y)\in\Gamma_{R}}=0
	\label{eq:homoBCR}
\end{equation}

A key aspect of simplifying Eq.~\ref{eq:generalBoundIntsimp} is to substitute the basis expansion for $G(z,z_0,y,y_0)=\sum^{M_{total}}_{m=1}G_m(z,y,z_0,y_0)$ and $\tilde{E}(z,y)|_{(z,y)\in\Gamma_{L/R}}=\sum^{N_{total}}_{n=1}\tilde{E}_n(z,y)|_{(z,y)\in\Gamma_{L/R}}$ and their corresponding boundary conditions while simplifying. 
\begin{widetext}
	\begin{eqnarray}
		&\int\left\{\sum_n\tilde{E}_n(z,y)\frac{\partial}{\partial z}\sum_m G_m\left(z,y,z_0,y_0\right)\hat{e}_z-\sum_m G_m\left(z,y,z_0,y_0\right)\frac{\partial}{\partial z}\sum_n\tilde{E}_n(z,y)\hat{e}_z\right\}|_{(z,y)\in\Gamma_L}\cdot \left(-\hat{e}_z\right) dy+\nonumber\\
		&\int\left\{\sum_n\tilde{E}_n(z,y)\frac{\partial}{\partial z}\sum_m G_m\left(z,y,z_0,y_0\right)\hat{e}_z-\sum_m G_m\left(z,y,z_0,y_0\right)\frac{\partial}{\partial z}\sum_n\tilde{E}_n(z,y)\hat{e}_z\right\}|_{(z,y)\in\Gamma_R} \cdot \hat{e}_z dy \nonumber\\
		&=\tilde{E}\left(z_0,y_0\right)|_{(z_0,y_0)\in\Omega}
		\label{eq:generalBoundIntsimplification}
	\end{eqnarray}
\end{widetext}
As there is a $\chi_{m}(y)$ in $G_m\left(z,y,z_0,y_0\right)$ and a $\chi_{n}(y)$ in $\tilde{E}_n(z,y)$, due to orthogonality relation that $\int \chi_m(y)\chi_n(y)dy=\delta_{kron}(m,n)$, $y$ integral of all the cross product terms where $m\ne n$, is zero. Hence only the terms $m=n$ remains non-zero. Therefore,
\begin{widetext}
	\begin{eqnarray}
		&\int\sum_n\left\{\tilde{E}_n(z,y)\frac{\partial}{\partial z} G_n\left(z,y,z_0,y_0\right)\hat{e}_z- G_n\left(z,y,z_0,y_0\right)\frac{\partial}{\partial z}\tilde{E}_n(z,y)\hat{e}_z\right\}|_{(z,y)\in\Gamma_L}\cdot \left(-\hat{e}_z\right) dy+\nonumber\\
		&\int\sum_n\left\{\tilde{E}_n(z,y)\frac{\partial}{\partial z} G_n\left(z,y,z_0,y_0\right)\hat{e}_z- G_n\left(z,y,z_0,y_0\right)\frac{\partial}{\partial z}\tilde{E}_n(z,y)\hat{e}_z\right\}|_{(z,y)\in\Gamma_R} \cdot \hat{e}_z dy 
		=\tilde{E}\left(z_0,y_0\right)|_{(z_0,y_0)\in\Omega}
		\label{eq:generalBoundIntsimplification2}
	\end{eqnarray}
\end{widetext}
Substituting the previously obtained boundary conditions (equations Eq.~\ref{eq:homogeneousbc2}, Eq.~\ref{eq:homoBCR}, Eq.~\ref{eq:homogeneousbc1} and Eq.~\ref{eq:inhomoBCL}) in the above given form of boundary integral, the following simplified expression is obtained after some simple calculations.

\begin{eqnarray}
	&&\int\sum_{n}2i k_z^{(n)} \tilde{E}_n^{inc}(z,y)G_n\left(z,y,z_0,y_0\right)|_{(z,y)\in\Gamma_L}dy=
	\tilde{E}(z_0,y_0)|_{(z_0,y_0)\in\Omega} 
	\label{eq:BIsimp}
\end{eqnarray}
One could safely add a summation $\sum_{m}$ to the $G_m\left(z,y,z_0,y_0\right)$ term in the above equation due to the orthogonality relationship $\int \chi_m(y)\chi_n(y)dy=\delta_{kron}(m,n)$. Therefore, the meaning of the integral given in Eq.~\ref{eq:BIsimp} doesn't change if it is written in the following form
\begin{eqnarray}
	&&\int\sum_{n}2i k_z^{(n)} \tilde{E}_n^{inc}(z,y)\sum_{m}G_m\left(z,y,z_0,y_0\right)|_{(z,y)\in\Gamma_L}dy=
	\tilde{E}(z_0,y_0)|_{(z_0,y_0)\in\Omega} 
	\label{eq:BIsimp1}
\end{eqnarray}
Replacing the expression for $\sum_m G_m(z,y,z_0,y_0)$ as $G(z,y,z_0,y_0)$ and changing the index $n$ to $m$, the boundary integral takes a simpler form as given below. For
wave incidence only from the left to right of the disorder $(\text{incidence onto}~\Gamma_{L})$,
\begin{equation}
	\begin{split}
		&\sum^{M_{total}}_{m=1}2i k_z^{(m)}\int \tilde{E}_m^{inc}(z,y)G\left(z,y,z_0,y_0\right)|_{(z,y)\in \Gamma_L}dy
		=\tilde{E}\left(z_0,y_0\right)|_{(z_0,y_0)\in\Omega}
	\end{split}
\end{equation}
where $\tilde{E}_m^{inc}(z,y)|_{(z,y)\in\Gamma_L}=c^{+}_m \psi^{+}_m(z,y)|_{(z,y)\in\Gamma_L}$, $k_z^{(m)}=i k_{z,ev}^{(m)}$ when $m>M_{pr}$ and $M_{total}=M_{pr}+M_{ev}$, the total number of eigenmodes considered.

\section{Derivation of generalized Fisher-Lee relationships} 
\label{Supp:Fisher}
Generalized Fisher-Lee relations are obtained by substituting expression for the total field $\tilde{E}\left(z_0,y_0\right)$ in the boundary integral equation and solving for the reflection or transmission coefficients.
\subsection{Derivation for $S_{21}$}
\label{Supp:FisherS21}
Let the wave incidence be from the left side of the disorder onto $\Gamma_{L}$. For estimating the transmission matrix $S_{21}$, need to take $\left(z_0,y_0\right)\in\Gamma_R$. Then the simplified boundary integral equation becomes,
\begin{align}
	\sum_{m=1}^{M_{total}}2i k_z^{(m)}\int \tilde{E}_m^{inc}(z,y)G\left(z,y,z_0,y_0\right)dy|_{(z,y)\in \Gamma_L\&\left(z_0,y_0\right)\in \Gamma_R}&=\tilde{E}(z_0,y_0)|_{(z_0,y_0)\in\Gamma_{R}}
	=\sum_{n=1}^{N_{total}}\tilde{E}_n\left(z_0,y_0\right)|_{\left(z_0,y_0\right)\in \Gamma_R}\nonumber\\
	&=\sum_{n=1}^{N_{total}} t^+_n \chi_n(y_0) \phi^+_n(z_0)|_{\left(z_0,y_0\right)\in \Gamma_R}\nonumber\\
	&=\sum_{n=1}^{N_{total}} \left\{\sum_{m=1}^{M_{total}} c^+_m S_{21}^{n m}\right\} \chi_n(y_0) \phi^+_n(z_0) |_{\left(z_0,y_0\right)\in \Gamma_R}
\end{align}

Since this equation holds for any incident wave represented by any set of $c^+_m$ values, one can collect the terms multiplied by $c^+_m$ and spanned by $\sum_{m=1}^{M_{total}}$ on both the L.H.S and R.H.S, and solve for $\mathbf{S}^{n m}_{21}$.



\begin{align}
	2i k_z^{(m)}\int \phi^+_m(z) \chi_m(y) G\left(z,y,z_0,y_0\right)dy|_{(z,y) \in \Gamma_L\&\left(z_0,y_0\right)\in \Gamma_R}=\sum _{n=1}^{N_{total}} S_{21}^{n m} \chi_n(y_0) \phi^+_n(z_0)|_{\left(z_0,y_0\right)\in \Gamma_R}
\end{align}

Using the orthogonality relationship involving $\chi_{n'}(y)$ and integrating with respect to $y_0$ on both sides of the above given equation,
\begin{align}
	2i k_z^{(m)} \phi^+_m(z)\int_y \int_{y_0} \chi_m(y) G\left(z,y,z_0,y_0\right) \chi_{n'}(y_0) dydy_0|_{(z,y)\in \Gamma_L\&\left(z_0,y_0\right)\in \Gamma _R}= S_{21}^{n'm} \phi^+_{n'}(z_0)|_{z_0\in \Gamma_R}
\end{align}
Simplification : Rename $n'$ as $n$ and let $z$ be positioned at $0$ on the left boundary and $z_0=z_R$ at $\Gamma_{R}$ on the right boundary. Therefore,
\begin{align}
	&2i k_z^{(m)} \phi^+_m(z=0) \int_y \int_{y_0} \chi_m(y) G\left(z=0,y,z_0=z_R,y_0\right) \chi_{n}(y_0) dydy_0|_{(z=0,y)\in \Gamma _L\&\left(z_0=z_R,y_0\right)\in \Gamma_R}=\nonumber\\
	&S_{21}^{n m}  \phi^+_{n}(z_0=z_R)|_{\left(z_0=z_R\right)\in \Gamma_R}
\end{align}

Mathematically, there are 4 different possibilities with respect to the nature of incident and transmitted modes.\\
Case 1 : Both input and output modes are propagating $m\leq M_{prop}$ and $n\leq N_{prop}$\\
\begin{align}
	2i k_z^{(m)}\frac{1}{\sqrt{k_z^{(m)}}}\int_y \int_{y_0}  \chi_m(y) G\left(z,y,z_0,y_0\right) \chi_n(y_0)dy dy_0|_{(z=0,y)\in \Gamma _L\&\left(z_0=z_R,y_0\right)\in \Gamma _R}=S_{21}^{n m} \frac{1}{\sqrt{k_z^{(n)}}} e^{+i k_z^{(n)}z_R}
\end{align}
Case 2 : Input mode and output modes are evanescent $m>M_{prop}$ and $n>N_{prop}$\\
\begin{align}
	2i k_z^{(m)}\frac{1}{\sqrt{\left|k_z^{(m)}\right|}}\int_y \int_{y_0} \chi_m(y) G\left(z,y,z_0,y_0\right) \chi _n(y_0)dydy_0|_{(z,y) \in \Gamma_L\&\left(z_0,y_0\right)\in \Gamma_R}=S_{21}^{n m}\frac{1}{\sqrt{\left|k_z^{(n)}\right|}}
\end{align}
Case 3 : Input modes are propagating and output modes are evanescent $m\leq M_{prop}$ and $n>N_{prop}$\\
\begin{align}
	2ik_z^{(m)}\frac{1}{\sqrt{k_z^{(m)}}}\int_y\int_{y_0} \chi_m(y) G\left(z,y,z_0,y_0\right)\chi _n(y_0 )dydy_0|_{(z,y)\in \Gamma _L\&\left(z_0,y_0\right)\in \Gamma_R}=S_{21}^{n m}\frac{1}{\sqrt{\left|k_z^{(n)}\right|}}
\end{align}
Case 4 : Input modes are evanescent and output modes are propagating $m>M_{prop}$ and $n\leq N_{prop}$\\
\begin{align}
	2i k_z^{(m)}\frac{1}{\sqrt{\left|k_z^{(m)}\right|}}\int_y \int_{y_0} \chi_m(y) G\left(z,y,z_0,y_0\right) \chi_n(y_0) dydy_0|_{(z,y)\in \Gamma_L\&\left(z_0,y_0\right)\in \Gamma_R}=S_{21}^{n m}\frac{ e^{+i z_R k_z^{(n)}}}{\sqrt{k_z^{(n)}}}
\end{align}
Solving for $S_{21}$ for cases $1,2,3$ and $4$ (using the information that for evanescent modes $\frac{k^{(m)}_z}{\sqrt{|k^{(m)}_z}|}=i\sqrt{|k^{(m)}_z|}$, is imaginary) are summarised in the Table \ref{tab:S21}.
\subsection{Derivation for $S_{22}$}
\label{Supp:FisherS22}
Let the wave incidence be from the right side of the disorder onto $\Gamma_{R}$. For estimating the reflection matrix $S_{22}$, need to take $\left(z_0,y_0\right)\in\Gamma_R$. Then the simplified boundary integral equation becomes,

\begin{align}
	&\sum_{m=1}^{M_{total}}2i k_z^{(m)}\int E_m^{\text{inc}}(z,y)G\left(z,y,z_0,y_0\right)dy|_{(z,y)\in \Gamma_R\&\left(z_0,y_0\right)\in \Gamma _R}=\sum _{n=1}^{N_{total}}\tilde{E}_n\left(z_0,y_0\right)|_{\left(z_0,y_0\right)\in \Gamma_R}\nonumber\\
	&=\sum_{m=1}^{M_{total}}\sum_{n=1}^{N_{total}}c^-_m\left[\delta_{\text{kron}}(n,m) \phi^-_n(z_0) \chi_n(y_0)+S_{22}^{n m} \chi_n(y_0) \phi^+_n(z_0)\right]|_{\left(z_0,y_0\right)\in \Gamma_R}
\end{align}
Performing the same set of operations as used for the derivation of $S_{11}$ and setting $z=z_R$ and $z_0=z_R$, the following is obtained
\begin{align}
	&2i k_z^{(m)} \phi^-_m(z=z_R) \int _{y_0} \int_y \chi_m(y) G\left(z=z_R,y,z_0=z_R,y_0\right) \chi_n(y_0)|_{(z=z_R,y)\in \Gamma _R\&\left(z_0=z_R,y_0\right)\in \Gamma_R}dy dy_0=\nonumber\\
	&\left\{\delta_{\text{kron}}(n,m) \phi^-_n(z_0=z_R)+S_{22}^{n m}  \phi^+_n(z_0=z_R)\right\}|_{z_0=z_R\in \Gamma_R}
\end{align}

Mathematically, there are 4 different possibilities with respect to the nature of incident and reflected modes.\\

Case 1 : Both input and output modes are propagating $m\leq M_{prop}$ and $n\leq N_{prop}$\\
\begin{align}
	&2ik_z^{(m)}\frac{ e^{-i z_R k_z^{(m)}}}{\sqrt{k_z^{(m)}}}\int_{y_0}\int_y \chi_m(y) G\left(z,y,z_0,y_0\right) \chi_n(y_0)|_{\left(z=z_R,y\right)\in \Gamma_R\&\left(z_0=z_R,y_0\right)\in \Gamma _R}dy dy_0=\nonumber\\
	&\delta_{\text{kron}}(n,m)\frac{e^{-i z_R k_z^{(n)}}}{\sqrt{k_z^{(n)}}}+S_{22}^{n m}\frac{ e^{+i z_R k_z^{(n)}}}{\sqrt{k_z^{(n)}}}
\end{align}
Case 2 : Input mode and output modes are evanescent $m>M_{prop}$ and $n>N_{prop}$\\
\begin{align}
	&2i k_z^{(m)}\frac{1}{\sqrt{\left|k_z^{(m)}\right|}}\int_{y_0}\int_y \chi_m(y)G\left(z,y,z_0,y_0\right) \chi_n(y_0)|_{\left(z=z_R,y\right)\in\Gamma_R\&\left(z_0=z_R,y_0\right)\in \Gamma_R}dy dy_0=\nonumber\\
	&\delta _{\text{kron}}(n,m)\frac{1}{\sqrt{\left|k_z^{(n)}\right|}}+S_{22}^{n m}\frac{1}{\sqrt{\left|k_z^{(n)}\right|}}
\end{align}

Case 3 : Input modes are propagating and output modes are evanescent $m\leq M_{prop}$ and $n>N_{prop}$\\
\begin{align}
	&2ik_z^{(m)}\frac{ e^{-i z_R k_z^{(m)}}}{\sqrt{k_z^{(m)}}}\int_{y_0}\int_y \chi_m(y) G\left(z,y,z_0,y_0\right) \chi_n(y_0)|_{(z,y)\in \Gamma _R\&\left(z_0,y_0\right)\in \Gamma_R}dy dy_0=\nonumber\\
	&\delta_{\text{kron}}(n,m)\frac{e^{-i z_R k_z^{(n)}}}{\sqrt{k_z^{(n)}}}+S_{22}^{n m}\frac{1}{\sqrt{\left|k_z^{(n)}\right|}}
\end{align}
But $\delta_{\text{kron}}(n,m)=0$, if $m\leq M_{\text{prop}}$ and $n>N_{\text{prop}}$.\\
Case 4 : Input modes are evanescent and output modes are propagating $m>M_{prop}$ and $n\leq N_{prop}$
\begin{align}
	&2i k_z^{(m)}\frac{1}{\sqrt{\left|k_z^{(m)}\right|}}\int_{y_0}\int_y \chi_m(y) G\left(z,y,z_0,y_0\right) \chi_n(y_0)|_{(z,y)\in \Gamma_R\&\left(z_0,y_0\right)\in \Gamma_R}dy dy_0=\nonumber\\
	&\delta_{\text{kron}}(n,m)\frac{1}{\sqrt{\left|k_z^{(n)}\right|}}+S_{22}^{n m}\frac{ e^{+i z_R k_z^{(n)}}}{\sqrt{k_z^{(n)}}}
\end{align}
But $\delta_{\text{kron}}(n,m)=0$, if $m>M_{\text{prop}}$ and $n\leq N_{\text{prop}}$.\\
Solving $S_{22}$ for cases $1,2,3$ and $4$ (using the information that for evanescent modes $\frac{k^{(m)}_z}{\sqrt{|k^{(m)}_z}|}=i\sqrt{|k^{(m)}_z|}$, is imaginary) are summarised in the Table \ref{tab:S22}.
\subsection{Derivation for $S_{12}$}
\label{Supp:FisherS12}
Let the wave incidence be from the right side of the disorder onto $\Gamma_{R}$. For estimating the transmission matrix $S_{12}$, need to take $\left(z_0,y_0\right)\in\Gamma_L$. Then the simplified boundary integral equation becomes,
\begin{align}
	\sum_{m=1}^{M_{total}}2i k_z^{(m)}\int \tilde{E}_m^{inc}(z,y)G\left(z,y,z_0,y_0\right)dy|_{(z,y)\in\Gamma_R\&\left(z_0,y_0\right)\in\Gamma _L}&=\sum _{n=1}^{N_{total}}\tilde{E}_n\left(z_0,y_0\right)|_{\left(z_0,y_0\right)\in \Gamma_L}\nonumber\\
	&=\sum_{n=1}^{N_{total}} \left\{\sum_{m=1}^{M_{total}} c^+_m S_{12}^{ n m}\right\} \chi_n(y_0) \phi^-_n(z_0) |_{\left(z_0,y_0\right)\in \Gamma_L}
\end{align}

Performing the same set of operations as used for the derivation of $S_{21}$ and setting $z=z_R$ and $z_0=z_L$, the following is obtained
\begin{align}
	&2ik_z^{(m)} \phi^-_m(z=z_R)\int_y \int_{y_0} \chi_m(y) G\left(z=z_R,y,z_0=z_L,y_0\right) \chi_n(y_0) dydy_0|_{(z=z_R,y)\in \Gamma_R\&\left(z_0=z_L,y_0\right)\in \Gamma _L}=\nonumber\\
	&S_{12}^{n m} \phi^-_{n'}(z_0)|_{\left(z_0=z_L\right)\in \Gamma_L}
\end{align}

Mathematically, there are 4 different possibilities with respect to the nature of incident and transmitted modes.\\

Case 1 : Both input and output modes are propagating $m\leq M_{prop}$ and $n\leq N_{prop}$\\
\begin{align}
	2i k_z^{(m)} \frac{ e^{-i z_R k_z^{(m)}}}{\sqrt{k_z^{(m)}}}\int_y\int_{y_0} \chi_m(y)G\left(z,y,z_0,y_0\right) \chi_n(y_0)dydy_0|_{(z=z_R,y)\in \Gamma_R\&\left(z_0 = z_L,y_0\right)\in \Gamma_L}=S_{12}^{n m}\frac{1}{\sqrt{k_z^{(n)}}}
\end{align}
Case 2 : Input mode and output modes are evanescent $m>M_{prop}$ and $n>N_{prop}$\\
\begin{align}
	2ik_z^{(m)}\frac{1}{\sqrt{\left|k_z^{(m)}\right|}}\int_y\int_{y_0}  \chi_m(y) G\left(z,y,z_0,y_0\right) \chi_n(y_0)dydy_0|_{(z=z_R,y)\in\Gamma_R\&\left(z_0=z_L,y_0\right)\in \Gamma_L}=S_{12}^{n m}\frac{1}{\sqrt{\left|k_z^{(n)}\right|}}
\end{align}
Case 3 : Input modes are propagating and output modes are evanescent $m\leq M_{prop}$ and $n>N_{prop}$\\
\begin{align}
	2ik_z^{(m)}\frac{e^{-i z_R k_z^{(m)}}}{\sqrt{k_z^{(m)}}}\int_y \int_{y_0} \chi_m(y) G\left(z,y,z_0,y_0\right) \chi_n(y_0)dydy_0|_{(z=z_R,y)\in \Gamma_R\&\left(z_0=z_L,y_0\right)\in \Gamma_L}=S_{12}^{n m}\frac{1}{\sqrt{\left|k_z^{(n)}\right|}}
\end{align}
Case 4 : Input modes are evanescent and output modes are propagating $m>M_{prop}$ and $n\leq N_{prop}$
\begin{align}
	2i k_z^{(m)}\frac{1}{\sqrt{\left|k_z^{(m)}\right|}}\int_y\int_{y_0} \chi_m(y) G\left(z,y,z_0,y_0\right) \chi_n(y_0)dydy_0|_{(z,y)\in\Gamma_L\&\left(z_0,y_0\right)\in \Gamma _R}=\frac{S_{12}^{n m}}{\sqrt{k_z^{(n)}}}
\end{align}
Solving $S_{12}$ for cases $1,2,3$ and $4$ (using the information that for evanescent modes $\frac{k^{(m)}_z}{\sqrt{|k^{(m)}_z}|}=i\sqrt{|k^{(m)}_z|}$, is imaginary) are summarized in the Table \ref{tab:S12}.

\begin{figure*}[!htp]
	\includegraphics[width=0.5\textwidth]{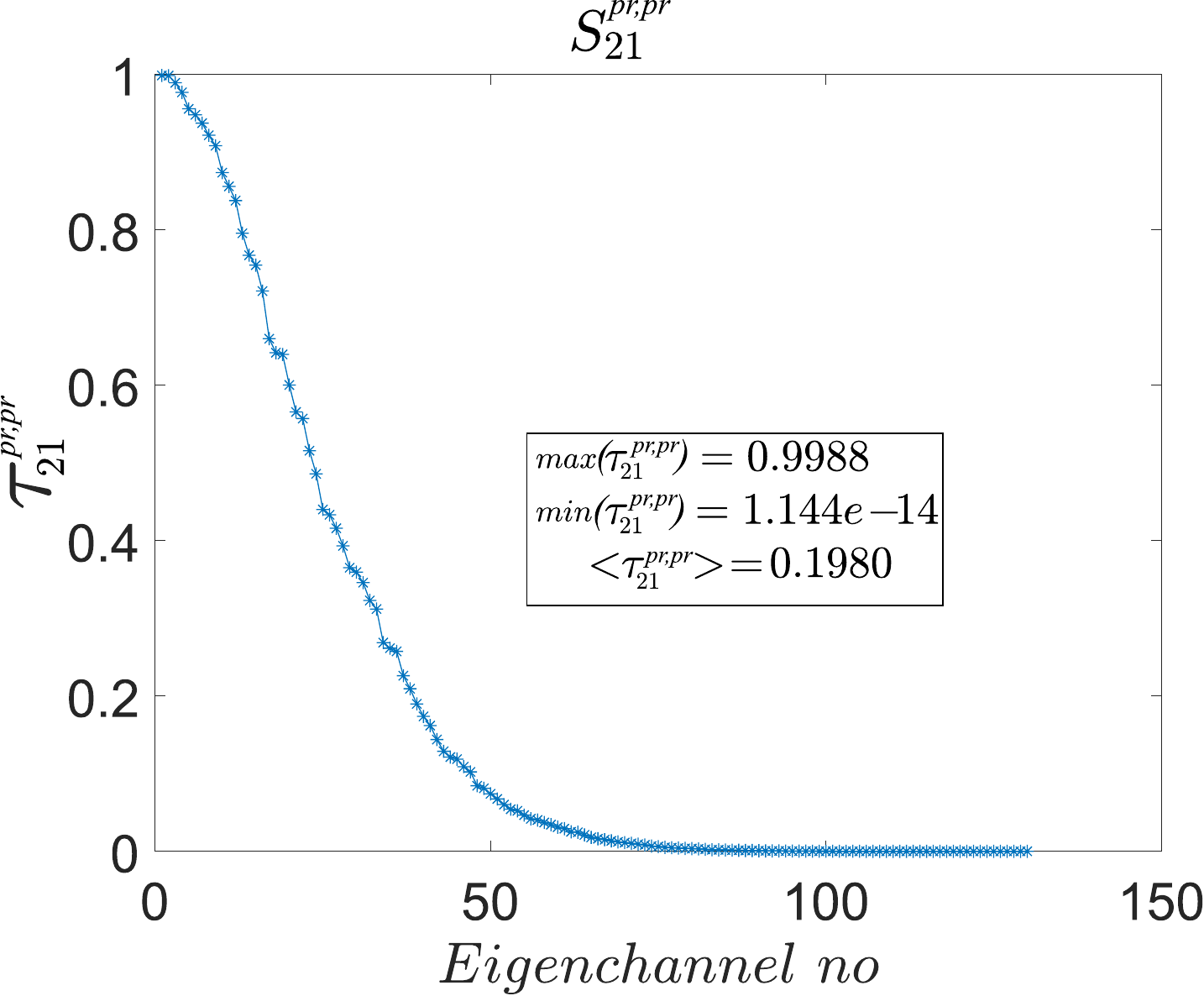}
	\caption{\textbf{Eigenchannel transmission coefficients of $S^{pr,pr}_{21}$ are plotted for a disorder with the average eigenchannel transmission of 0.198}. Eigenchannel transmission coefficients contained in the diagonal of the matrix $\tau^{pr,pr}_{21}$, (obtained from the $S.V.D$ of the $S^{pr,pr}_{21}$ matrix given in FIG.~\ref{fig:Smatrices} of the main paper), is plotted. The maximum, minimum and average values of $\tau^{pr,pr}_{21}$ are also given. The averaging is done with respect to all the eigenchannels of the given disorder. Some of the corresponding eigenchannels are plotted in the supplementary FIG.~\ref{fig:eigenchannels}.}
	\label{fig:eigenchanneltranscoeff}
\end{figure*}
\section{Visualization of transmission eigenchannels}
\label{supp:visualisingeigchannel}
\begin{figure*}[!htp]
	\includegraphics[width=0.88\textwidth]{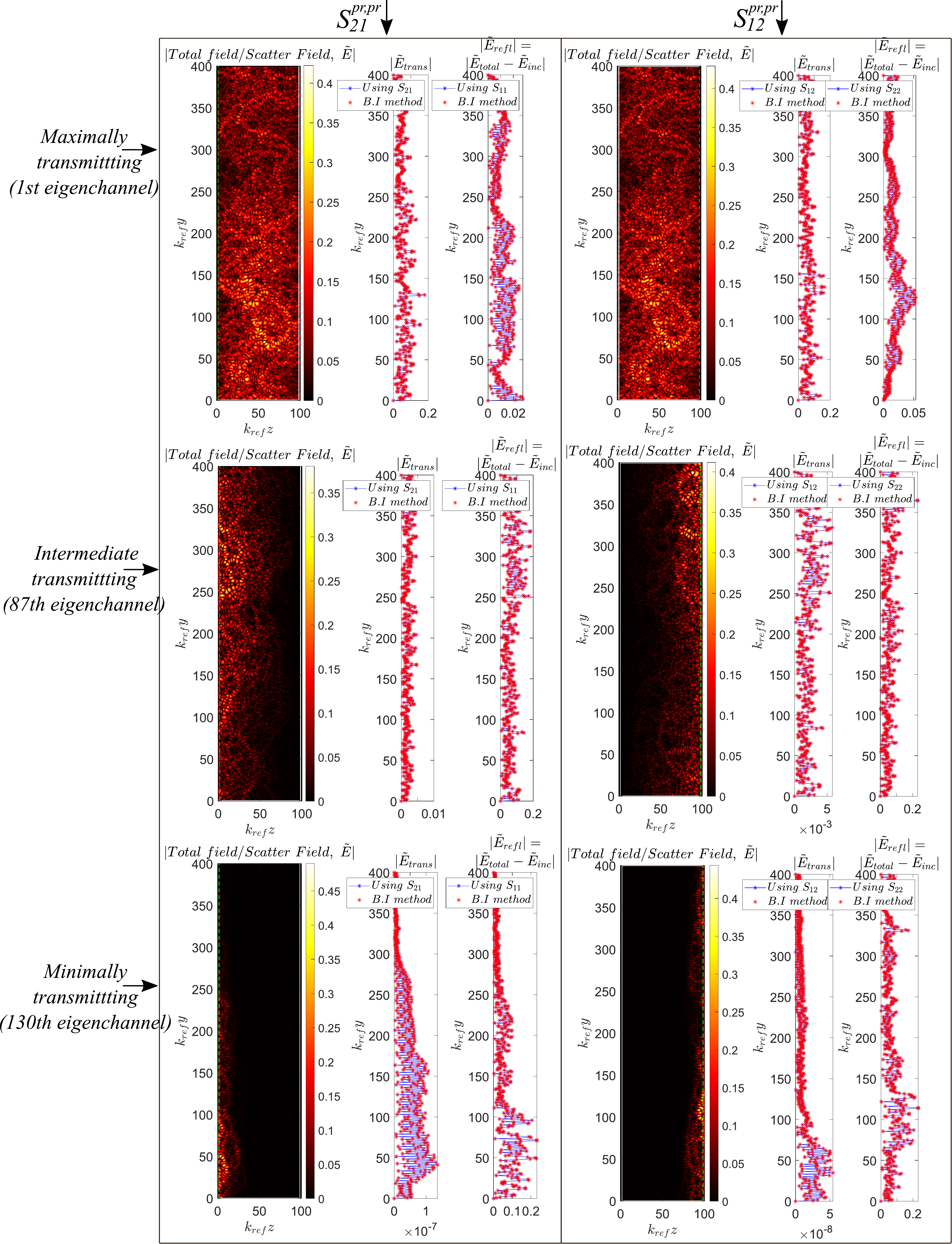}
	\caption{\label{fig:eigenchannels}  \textbf{Visualizing various eigenchannels of $S^{pr,pr}_{21}$ and $S^{pr,pr}_{12}$, containing 130 propagating eigenmodes and 30 evanescent eigenmodes, with $\langle \tau^{pr,pr}_{21}\rangle=0.198$, the transmission averaged over various eigenchannels of the given disorder shown in FIG.~\ref{fig:Smatrices}(b) of the main paper.} \textbf{(a)} Maximally transmitted (first) eigenchannel of $S^{pr,pr}_{21}$, where transmission, $T=0.9988$ and reflection $R=0.0012$ \textbf{(b)} An intermediate eigenchannel numbered 87 out of the 130 eigenchannel modes, of $S^{pr,pr}_{21}$, where $T=0.0013,R=0.9987$. \textbf{(c)} Minimally transmitted (last eigenchannel numbered 130) eigenchannel of $S^{pr,pr}_{21}$, where $T=1.1438e-14,R=1$. Similarly, \textbf{(d)} Maximally transmitted (first) eigenchannel of $S^{pr,pr}_{12}$, where $T=0.9988,R=0.0012$ \textbf{(e)} An intermediate eigenchannel numbered 87 out of the 130 eigenchannel modes, of $S^{pr,pr}_{12}$, where $T=0.0013,R=0.9987$. \textbf{(f)} Minimally transmitted (last eigenchannel, numbered 130) eigenchannel of $S^{pr,pr}_{12}$, where $T=1.1438e-14,R=1$. The transport paths associated with the maximally transmitted eigenchannel of $S^{pr,pr}_{21}$ and $S^{pr,pr}_{12}$ looks identical in the inside of the disorder. It is because the transport path taken from left to right with the maximum transmission close to unity, is the same path taken from right to left for maximal transmission.}
\end{figure*}
\begin{figure*}[!htp]
	\includegraphics[width=0.9\textwidth]{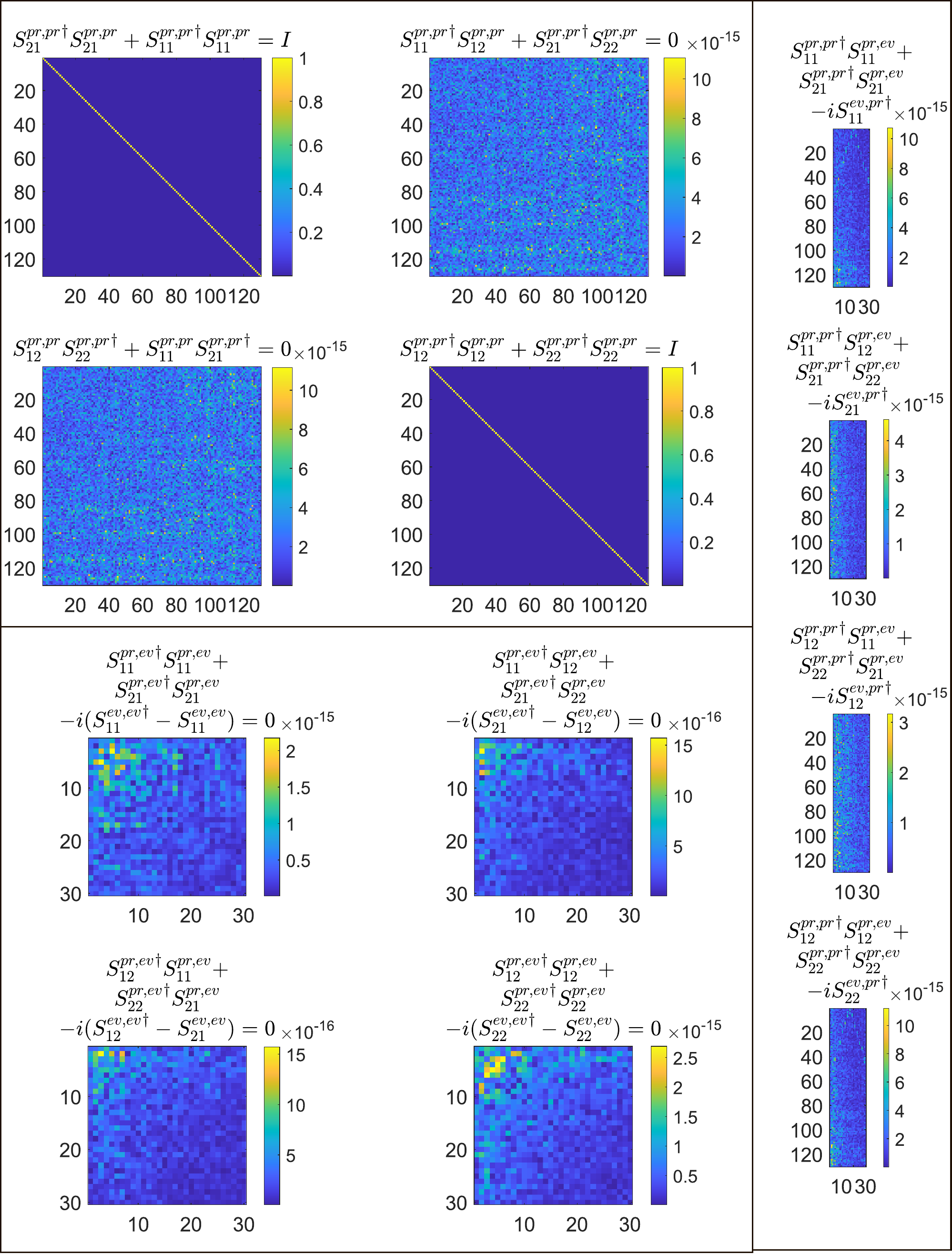}
	\caption{\label{fig:additionalunitarity} \textbf{Additional numerical validation done for verifying the generalized unitarity relations}. For those unitarity relations where the R.H.S is a zero matrix analytically, the simulation code yields a numerical matrix close to be approximated to zero (of the order of $10^{-15}$) as shown in the plots. For those unitarity relations where the R.H.S is a unit matrix analytically, the diagonal elements are all unity.}
\end{figure*}
For accessing the propagating eigenchannels, singular value decomposition $(S.V.D)$ of the components of the $S^{pr,pr}$ matrix is performed. As an example, the propagating transmission eigenchannels for wave incidence only from the left-side of the disorder, is accessed by performing the $S.V.D$ of $S^{pr,pr}_{21}=U_{21} \Sigma_{21}  V_{21}^\dagger$ and using the first singular vector of $V_{21}$ to maximize the transmission. $\tau^{pr,pr}_{21}=\left(\Sigma^{pr,pr}_{21}\right)^2$ is the diagonal matrix containing the eigenchannels' transmission coefficients ranging from 0 to 1 as shown in the supplementary FIG.~\ref{fig:eigenchanneltranscoeff}. For the disordered slab shown in FIG.~\ref{fig:Smatrices}(b) of the main paper, various eigenchannels, for the wave incidence from both the left side and the right side of the disorder, are shown in the supplementary FIG.~\ref{fig:eigenchannels}.

\section{Additional numerical validation of the Generalized unitarity relation}
\label{supple:additionalunitarity}
In addition to the numerical validation of the compact form of the generalized unitary relation given in the FIG.~\ref{fig:Unitarity}(a) of the main paper, additional validation is performed by decomposing the compact form into some of its individual components as given in the supplementary  FIG.~\ref{fig:additionalunitarity}.

\section{$S$ matrix cascading and ensemble averaging}
\label{supple:cascading}
\begin{figure*}[!htp]
	\includegraphics[width=0.75\textwidth]{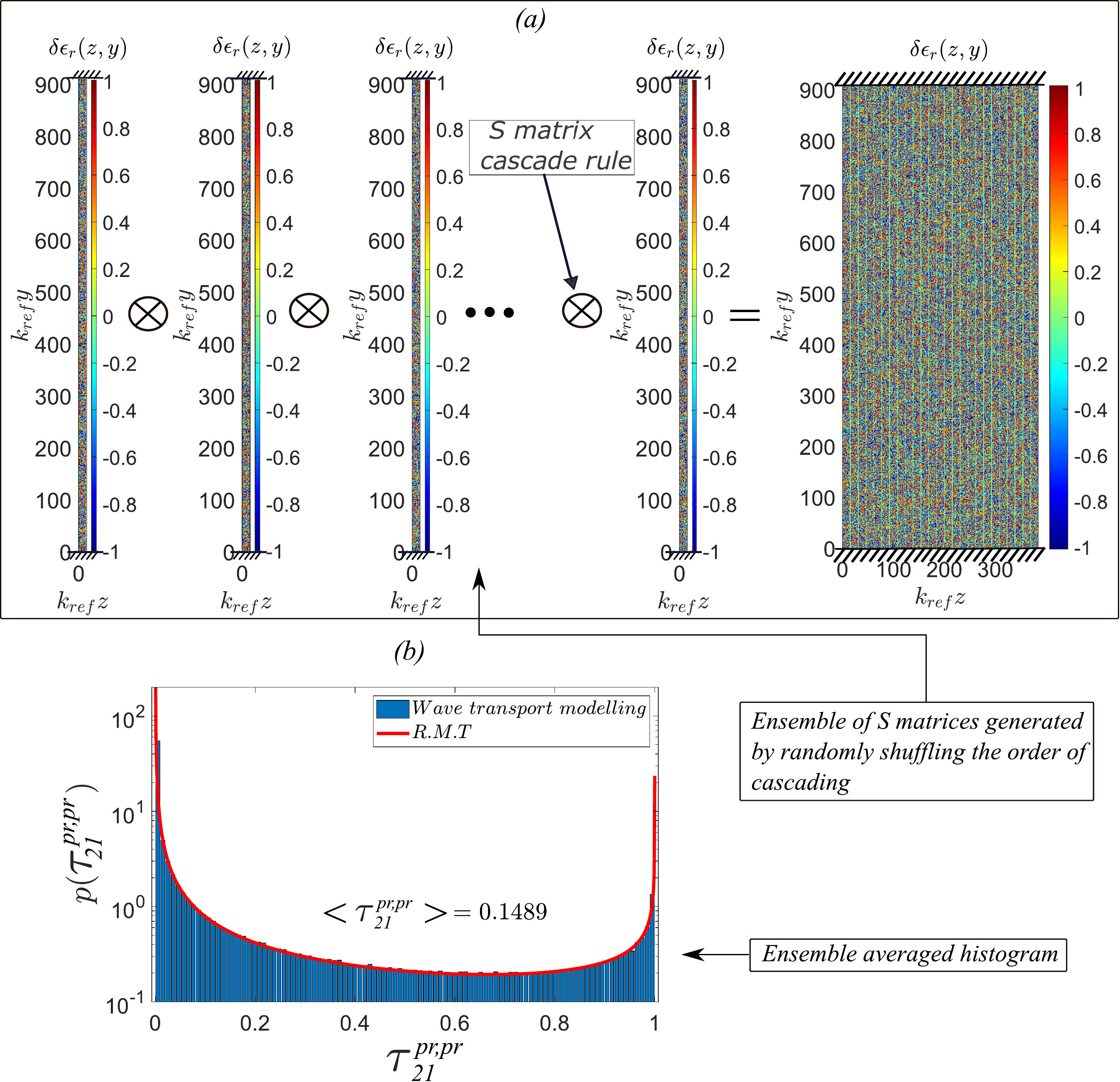}
	\caption{\textbf{Schematic to demonstrate the $S$ matrix cascade rule and the ensemble averaged eigenchannel transmission coefficient distribution obtained through cascading.} \textbf{(a)} Schematic of cascading thinner slabs to form thicker slab is shown. Cascade rule given in Eq.~\ref{eq:cascading} is used to cascade the slabs. First, the first and the second thin slabs are cascaded. It is followed by cascading the third slab with the cascaded result obtained from the first and the second slab. This pattern is repeated, until the last thin slab is cascaded. \textbf{(b)} The ensemble averaged frequency distribution of $\tau^{pr,pr}_{21}$ values after cascading. It is compared with the bimodal distribution associated with the Random Matrix theory ($R.M.T$). Note that here the histogram is evaluated for individual cascaded disorders, followed by averaging the obtained histogram over many cascaded disorder realizations.}
	\label{fig:cascadingschematic}
\end{figure*}
This section explains the $S$ matrix cascading method by which, a larger disordered slab can be formed, by cascading an ensemble of thinner slabs. Cascading of the generalized $S$ matrices of the thinner slabs, builds the sample along the longitudinal dimension ($z$ axis) step-by-step, as shown in the schematic given in the supplementary FIG.~\ref{fig:cascadingschematic}(a).  Different realizations of thicker slabs (1000 in number) can be formed from the same $S$ matrices of the thinner slabs (25 in number), taken in random order. This is particularly advantageous from a computational point of view that, one need not model 1000 different large slabs, but could estimate the $S$ matrices of a collection of smaller slabs once and cascade them using random order of the slabs for obtaining different realizations of larger slabs. The well known cascading formula for $S$ matrices is given as the following.  
\begin{align}
	S^{(a)}=\left(
	\begin{array}{cc}
		S_{11}^{(a)} & S_{12}^{(a)} \\
		S_{21}^{(a)} & S_{22}^{(a)} \\
	\end{array}
	\right)
\end{align} and \begin{align}
	S^{(b)}=\left(
	\begin{array}{cc}
		S_{11}^{(b)} & S_{12}^{(b)} \\
		S_{21}^{(b)} & S_{22}^{(b)} \\
	\end{array}
	\right)
\end{align}

\begin{align}
	&S^{\text{cascade}}=S^{(a)}\bigotimes S^{(b)}=\left(
	\begin{array}{cc}
		S_{11}^{(a)} & S_{12}^{(a)} \\
		S_{21}^{(a)} & S_{22}^{(a)} \\
	\end{array}
	\right) \bigotimes \left(
	\begin{array}{cc}
		S_{11}^{(b)} & S_{12}^{(b)} \\
		S_{21}^{(b)} & S_{22}^{(b)} \\
	\end{array}
	\right)\nonumber\\
	&=\left(
	\begin{array}{cc}
		S_{11}^{(a)}+S_{12}^{(a)} \left(I-S_{11}^{(b)} S_{22}^{(a)}\right)^{-1} S_{11}^{(b)} S_{21}^{(a)}  & S_{12}^{(a)} \left(I-S_{11}^{(b)} S_{22}^{(a)}\right)^{-1} S_{12}^{(b)} \\
		S_{21}^{(b)} \left(I-S_{22}^{(a)} S_{11}^{(b)}\right)^{-1} S_{21}^{(a)} & S_{22}^{(b)} + S_{21}^{(b)}\left(I-S_{22}^{(a)} S_{11}^{(b)}\right)^{-1} S_{22}^{(a)} S_{12}^{(b)} \\
	\end{array}
	\right)\label{eq:cascading}
\end{align}
where $I$ is the identity matrix. The $S$ matrices for a collection of the smaller slabs, are estimated serially using the generalized Fisher-Lee relations, followed by the use of the cascading formula given in Eq.~\ref{eq:cascading}. This helps in the ensemble averaging of the focusing results in the main paper. The ensemble averaged eigenchannel transmission coefficients of $S^{pr,pr}_{21}$ after cascading is given in FIG.~\ref{fig:cascadingschematic}(b).  

\bibliography{bibliography.bib}